\documentclass[%
 aip,
 amsmath,amssymb,
 reprint,%
]{revtex4-1}

\usepackage{graphicx}
\usepackage{dcolumn}
\usepackage{bm}

\usepackage[utf8]{inputenc}
\usepackage[T1]{fontenc}
\usepackage{newtxtext}
\usepackage[varvw]{newtxmath} 
 
\usepackage{etoolbox}

\usepackage{hyperref}

\usepackage{color}
\newif\ifshownewtextblue

\shownewtextbluefalse

\ifshownewtextblue
\else
  \definecolor{blue}{rgb}{0,0,0}
\fi

\definecolor{DarkGreen}{rgb}{0.0, 0.4, 0.0}
           
\newcommand{\Alfven}{Alfv\'{e}n}

\newcommand{\V}[1]{\mathbf{#1}} 
\newcommand{\T}[1]{{\tt #1}}

\makeatletter
\def\@email#1#2{%
 \endgroup
 \patchcmd{\titleblock@produce}
  {\frontmatter@RRAPformat}
  {\frontmatter@RRAPformat{\produce@RRAP{*#1\href{mailto:#2}{#2}}}\frontmatter@RRAPformat}
  {}{}
}%
\makeatother
\begin{document}

\preprint{Accepted for publication in Physics of Plasmas}

\title[Phase-Space Energy Transfer of Wave-Particle Interactions]{Phase-Space Energy Transfer of Wave-Particle Interactions using the Field-Particle Correlation Technique and Linear Plasma Theory with \texttt{JET-PLUME}}
\author{Collin R. Brown}
\affiliation{Department of Physics and Astronomy, University of Iowa, Iowa City, Iowa 52242, USA}
\affiliation{Plasma Physics Division, Naval Research Laboratory, Washington, D.C. 20375, USA}
\author{Gregory G. Howes}%
\affiliation{Department of Physics and Astronomy, University of Iowa, Iowa City, Iowa 52242, USA}
\author{Kristopher G. Klein} 
\affiliation{Lunar and Planetary Laboratory, University of Arizona, Tucson, AZ 85719, USA}
\author{Jason M. TenBarge}
\affiliation{Department of Astrophysical Sciences, Princeton University, Princeton, NJ 08544, USA}

\date{Accepted June 18, 2026}

\begin{abstract}
The collisionless transfer of energy between fields and particles through wave-particle interactions is a fundamental process in space plasmas but remains incompletely characterized because many mechanisms operate across a wide parameter range and diverse plasma conditions. The Field-Particle Correlation (FPC) technique reveals velocity-space signatures of particle energization by correlating measured electric field fluctuations with changes in the velocity distribution. Fully mapping these signatures across plasma parameters requires an impractically large number of kinetic simulations or observations. To address this challenge, we introduce \texttt{JET-PLUME} (\textbf{J}udging \textbf{E}nergy \textbf{T}ransfer in a \textbf{P}lasma in a \textbf{L}inear \textbf{U}niform \textbf{M}agnetized \textbf{E}nvironment), an extension of the \texttt{PLUME} Vlasov-Maxwell dispersion solver. \texttt{JET-PLUME} uses \texttt{PLUME}'s ability to model parallel drifting bi-Maxwellian distributions to examine phase-space energy transfer by adding an analytic Fourier-space formulation of the FPC. This approach isolates the contribution of individual resonances, separates degenerate entropy mode components, and allows systematic analysis of unstable, growing modes. Dimensionless expressions extend the results across a broad parameter range and highlight the role of off-diagonal elements of the susceptibility tensor in coupling electric field and current response. We show that during kinetic \Alfven\ wave damping, the perpendicular field can drive parallel ion currents among particles with large $v_\perp$, reducing the net Landau damping. The resulting velocity-space signatures, accessible through \texttt{JET-PLUME}, demonstrate how analytic formulations of phase-space energy transfer can reveal novel physics of wave-particle interactions across diverse plasma environments.
\end{abstract}

\maketitle

\begin{center}
\small
The following article has been accepted by \textit{Physics of Plasmas}.
After it is published, it will be found at
\url{https://pubs.aip.org/aip/pop}.
\end{center}

\section{\label{sec:intro} Introduction}

Linear mechanisms for transferring energy between electromagnetic fields and particles without collisions through wave-particle interactions are abundant in space and astrophysical plasmas. Fundamentally, either an unstable velocity distribution of particles can release its free energy by generating waves, or waves can interact with charged particles, perturbing their distribution in velocity space and resonantly transferring energy. These wave-particle interactions can be described with linear theory by assuming weak perturbations, helical trajectories for the particles, and a plane-wave ansatz for the electromagnetic fields. From these assumptions, the susceptibility of the system can be computed, leading to a linear dispersion relation \citep[e.g.,][]{Stix:1992,swanson2003plasma,gary1993theory,brambilla1998kinetic} that defines the properties of the resulting linear wave modes and enables the determination of the eigenmodes of the electromagnetic fields. Energy transfer mechanisms in collisionless plasmas produce non-trivial phase-space signatures of energy exchange between the particles and fields, spanning both spatial and velocity coordinates, that provide insight into the fundamental behavior of plasmas. This phase-space energy exchange can be used to identify mechanisms when observed \emph{in situ} using spacecraft observations \citep{chen2019evidence,afshari2021importance,Afshari:2024,shuster2021structures}, in the laboratory \citep{schroeder2021laboratory}, or in simulations \citep{howes2017prospectus,howes2017diagnosing,brown2022isolation,mccubbin2022characterizing,juno2021field} using the Field-Particle Correlation (FPC) technique \citep{howes2022revolutionizing}. The FPC technique diagnoses the rate of energy exchange between the electric field and particles as a function of the full phase space.

The inherent limitations of obtaining accurate measurements of space and laboratory plasmas at multiple points in space, particularly for \emph{in situ} measurements, provide the key motivation for the development of the FPC diagnostic \citep{klein2016measuring}. 
Correlating the field and particle measurements over a long enough correlation interval or a large enough spatial volume \citep{chen2019evidence,brown2022isolation} significantly increases the number of particles employed to reconstruct the particle velocity distribution, averaging out noisy data and more accurately sampling the fluctuations to yield a better determination of the physics of the wave-particle energy transfer. This is particularly valuable for \emph{in situ} spacecraft measurements with relatively low particle counts or kinetic simulations that sacrifice resolution in velocity space (either by fewer sampling particles or less dense grids) for computational efficiency.

Understanding the energy exchange inherent to kinetic processes within phase space remains at the forefront of plasma physics research. Progress in this area is constrained by the scarcity of detailed \emph{in situ} measurements in many parameter domains, the computational demands of simulation, and the inherent difficulties in measuring laboratory plasmas \citep{gilbert2024improving,mclaughlin2025spatially}. By leveraging linear kinetic theory, we develop here a computational methodology that efficiently computes the phase-space behavior for linear wave-particle interactions.

{\color{blue} Collisionless resonant wave--particle interactions in plasmas have a long theoretical history, beginning with Landau's treatment of plasma oscillations and collisionless damping and continuing through the creation of the plasma dispersion function and the standard theory of waves in magnetized collisionless plasmas \cite{Landau:1946,Fried:1961,Stix:1992,swanson2003plasma}. However, identifying which energization mechanisms operate in collisionless plasmas remains an active problem for many plasmas, such as the solar wind, magnetosphere, and astrophysical plasmas, where turbulent energy reaches kinetic scales and is partitioned among particle species through resonant damping, phase mixing, finite-Larmor-radius effects, cyclotron interactions, instability-driven processes, and other phase-space mechanisms \cite{howes2017prospectus,Howes:2024}. Although the basic linear mechanisms of collisionless wave--particle interaction are well established, their role in turbulent weakly collisional plasmas remains difficult to diagnose. In such systems, turbulent energy can cascade through both physical space and velocity space, with collisionless wave--particle interactions and phase mixing transferring field energy into particle distribution-function structure before weak collisions ultimately thermalize it \cite{schekochihin2009astrophysical}. These mechanisms can produce similar features in field spectra or low-order fluid moments, making it difficult to determine from these quantities alone the channels through which fields and particles exchange energy. Through the advancement of modern \emph{in situ} measurements, kinetic simulations, and laboratory diagnostics, there now exist increasingly high-quality particle velocity-distribution measurements that can overcome this challenge if the corresponding phase-space response is sufficiently well understood \cite{chen2019evidence,Afshari:2024,schroeder2021laboratory,howes2022revolutionizing}.}

In this work, we analytically compute the energy transfer in phase space using the FPC technique for wave–particle interactions within the framework of linear theory. 
We adopt the standard assumptions of plane-wave solutions for the electromagnetic fields and helical unperturbed particle trajectories, as in classical linear theory \citep{Stix:1992,swanson2003plasma}, to derive the FPC in terms of solutions to the Vlasov–Maxwell dispersion relation. 
Our methods apply to all solutions obtained from the linear dispersion solver \texttt{PLUME} (Section \ref{sec:PLUMEintro}), which permits parallel drifting bi-Maxwellian components.

To enable these calculations, we develop a routine within \texttt{PLUME}, \texttt{JET-PLUME} (\textbf{J}udging \textbf{E}nergy \textbf{T}ransfer in a \textbf{P}lasma in a \textbf{L}inear \textbf{U}niform \textbf{M}agnetized \textbf{E}nvironment), that predicts the linear velocity-distribution-function response and associated phase-space energy transfer for any admissible solution. 
This approach captures signatures of energization for both damped modes and growing instabilities. 
We compare \texttt{JET-PLUME} predictions to simulations of several key mechanisms, including ion and electron Landau damping, the Weibel instability with an external guide field, and ion cyclotron damping. 
We find that the velocity-space energy transfer computed from \texttt{JET-PLUME} agrees with that obtained from these simulations of collisionless plasmas. 
For unstable modes, agreement is strongest during the initial exponential growth phase, while deviations at later times reflect nonlinear modifications to the particle velocity distribution.

The code employed in this study is publicly available \citep{Brown_JETPLUME_2026_misc} and is formulated in dimensionless units, enabling broad applicability across plasma regimes, simplifying parameter scans, and emphasizing the underlying physics by removing specific scale dependencies.

{\color{blue}
The open-source code, \texttt{JET-PLUME}, introduced in this paper enables systematic studies of linear wave--particle energy transfer. By building on \texttt{PLUME}, \texttt{JET-PLUME} computes Vlasov--Maxwell eigenmodes for plasmas represented using an arbitrary number of drifting bi-Maxwellian components and species. It also adopts standard ingredients such as dimensionless variables, which facilitate parameter scans and comparison across plasma regimes. \texttt{JET-PLUME} evaluates the FPC from the complex Fourier eigenfunctions returned by \texttt{PLUME}. It computes the velocity-space energization signature of a selected monochromatic linear mode, or linear superposition of modes, directly from analytic expressions. 

The practical value of the FPC is that it converts distribution-function information into a velocity-space map of field--particle work. A perturbed distribution function contains resonant response, reversible oscillatory motion, sampling noise, and other phase-space structure that may not contribute to secular energization. By correlating the distribution-function response with the electric field, the FPC isolates the coherent contribution to the net field--particle energy exchange and is constructed so that its velocity-space integral gives the corresponding net-work channel, \(\langle j_{j,s}E_j\rangle\). This is especially useful when access to spatial information is limited but velocity-distribution measurements are more readily available, as in spacecraft measurements along a single space-time trajectory or laboratory diagnostics at a finite set of selected locations and viewing geometries. Even in kinetic simulations, where \(f(\mathbf{x},\mathbf{v},t)\) may be available in principle, the full seven-dimensional phase space is difficult to interpret directly. The FPC therefore provides a compact, physically interpretable diagnostic showing where in velocity space that energy is exchanged between fields and particles.

In this work, the use of analytic expressions for phase-space energization makes it possible to determine which components of the linear susceptibility response contribute to a given field--particle work channel and where in velocity space those contributions occur. Although Fourier-space formulations of the FPC have been developed previously \cite{li2019collisionless}, \texttt{JET-PLUME} applies this approach to analytic linear Vlasov--Maxwell eigenmodes while retaining the intermediate electric-field-driven pieces of the perturbed distribution function before they are summed into the total response. That is, \texttt{JET-PLUME} provides access to channel-resolved phase-space energization, separating diagonal and off-diagonal susceptibility-channel contributions that would otherwise be combined in the total perturbed distribution function, total current, or net damping rate. This illustrates the cross-channel structure that can underlie even familiar linear wave--particle interactions, in which the current response driven by one electric-field component contributes to the field--particle work associated with another. 

Specifically, here we identify the contribution of the \(E_{\perp,2}\)-driven response to the parallel current, \(\chi_{\parallel,\perp2}\), and show how this off-diagonal channel modifies the velocity-space signature of ion energization during kinetic \Alfven\ wave damping, including the reversal of the bipolar resonant structure at larger \(v_\perp\). The resulting extension provides a controlled way to predict phase-space energization signatures, separate distinct electric-field-driven response channels, and interpret off-diagonal susceptibility contributions that may be difficult to isolate in simulations or observations alone. We show that the full velocity-space behavior can be more complicated than suggested by single-parameter diagnostics such as the linear damping rate, \(\gamma\): particles at the same resonant parallel velocity can experience opposite signs of energy transfer at different perpendicular velocities because multiple electric-field-driven susceptibility channels contribute to the same component of the current response, a distinction that is hidden in the net rate but important for interpreting diagnostics and predicting wave-based heating, transport, or stability control.
}

Other applications of this diagnostic related to wave-particle interactions have analyzed simulations or observations, often using either time-integration over an integral number of wave periods, spatial integration over periodic domains, or high-pass filtering of the electric fields to eliminate the oscillating energy exchange by undamped, large-scale waves. The analytic formulation presented here ensures that only the selected modes are present in the computed signature. 
In the Appendix, we specify the conditions under which our analytic form for computing phase-space energy transfer is exact, as well as the conditions for it to be approximately correct.  Ultimately, this paper serves as the main reference for the open-source code \texttt{JET-PLUME} and advances techniques for analyzing phase-space energy transfer, a critical problem in kinetic plasma physics.

The paper is structured as follows.
We provide a brief overview of both the \texttt{PLUME} dispersion solver and the FPC method in Section~\ref{sec:methods}, and derive the methods used for \texttt{JET-PLUME} in Section~\ref{sec:linfpcintro}.
We demonstrate the accuracy of our solver through comparisons with numerical simulations in Section~\ref{sec:Comparisons}.
Discussion of the broader implications of the new diagnostics and insights enabled by this framework are given in Sections~\ref{sec:discussion} and ~\ref{sec:conclusion}.
We provide technical details regarding the implementation of \texttt{JET-PLUME} in the appendices. 
Appendix~\ref{app:vmlindisp} tabulates the variable definitions and normalizations used by the linear dispersion solver. 
Appendix~\ref{appendix:linFPCderiv} details the derivation of the phase-space energy transfer diagnostic using Fourier transforms, establishing the reality conditions and the treatment of linear superpositions of degenerate modes. 
Appendix~\ref{app:dimnormJetPlume} outlines the dimensionless framework employed by the code and describes the procedure for recovering dimensional physical quantities from the dimensionless output. {\color{blue}Appendix~\ref{app:workflow} summarizes the numerical workflow implemented in \texttt{JET-PLUME}.}


\section{Method Overview}
\label{sec:methods}
Prior to introducing \texttt{JET-PLUME}, we provide a brief introduction to both the \texttt{PLUME} dispersion solver and the FPC technique.

\subsection{\label{sec:PLUMEintro}\texttt{PLUME} (\textbf{P}lasma in a \textbf{L}inear
\textbf{U}niform \textbf{M}agnetized \textbf{E}nvironment)}

\texttt{PLUME} (\textbf{P}lasma in a \textbf{L}inear
\textbf{U}niform \textbf{M}agnetized \textbf{E}nvironment)\citep{Klein:PlumeCodePaper:2025} is a numerical Vlasov-Maxwell linear dispersion relation solver that solves for an arbitrary number of bi-Maxwellian species with drifts along the magnetic field for quasi-neutral plasmas. The main applications of \texttt{PLUME} have been to study turbulence and instability mechanisms in space plasmas such as the solar wind. Some example works using \texttt{PLUME} are as follows: to study the potential effects of anisotropic proton temperature distributions on the evolution of plasma fluctuations in the solar wind \citep{klein2015predicted}, to use Nyquist’s method when evaluating the instabilities that govern the evolution of the solar wind \citep{klein2017applying}, to perform a statistical assessment of solar wind stability considering multiple sources of ion free energy, analyzed at 1 AU using Wind observations \citep{klein2018majority}, between 0.3 and 1 AU using Helios observations  \citep{klein2018majority,martinovic2021ion}, and below 0.3 AU using PSP/SWEAP measurements \citep{klein2021inferred,mcmanus2024proton}.

{\color{blue}The linear response theory implemented in \texttt{PLUME} follows the standard hot-plasma Vlasov--Maxwell derivation presented in standard works such as \citet{Stix:1992} and \citet{swanson2003plasma}. We summarize the relevant expressions here for notation and self-containment, rather than as a new derivation of the theory. 

\texttt{PLUME} computes the linear plasma response of a collection of electrons and ions described by a linear superposition of hot, drifting, bi-Maxwellian velocity distributions coupled to electromagnetic perturbations using expressions derived from the wave equation \citep{Stix:1992},}
\begin{equation}
\V{n}^\prime \times (\V{n}^\prime \times \hat{\V{E}}) + \underline{\underline{\epsilon}}\cdot \hat{\V{E}} = 0,
\label{eqn:wave.hp}
\end{equation}
where the hat denotes the complex coefficient arising from the Fourier transform in space and Laplace transform in time, e.g. for the electric field $\hat{\V{E}}(\V{k},\omega)$.
Without loss of generality, we rotate the coordinates about the magnetic field such that $\mathbf{n}^\prime=n_{\perp,1}^\prime \hat{e}_{\perp,1} + n_{\parallel}^\prime \hat{e}_{\parallel}$, for a right-handed coordinate system ($\hat{E}_{\perp,1}$, $\hat{E}_{\perp,2}$, $\hat{E}_{\parallel}$), where $\hat{E}_{\parallel}$ is parallel to the equilibrium magnetic field, $\hat{E}_{\perp,1}$ is in the plane of $\mathbf{n}^\prime$, and $\hat{E}_{\perp,2}$ is out of the plane\footnote{Here, we denote components with $\parallel$, $\perp_1$, and $\perp_2$, where $\perp_1$ is the component in the plane perpendicular to the magnetic field and in the plane of the wavevector, $\mathbf{k}$. Note, it is common to denote $\perp_1$, $\perp_2$ and $\parallel$ as $x$, $y$, and $z$ respectively.} of $\mathbf{n}^\prime$. The index of refraction, $\mathbf{n}^\prime$, is related to the wavevector, $\mathbf{k}$, and complex frequency, $\omega$, via
\begin{equation}
\V{n}^\prime = \V{k} c/\omega,
\label{eq:indexofrefraction}
\end{equation}
and the dielectric tensor, $\underline{\underline{\epsilon}}$, trivially related to the susceptibility tensor, $\underline{\underline{\chi_s}}$, is solved for a Fourier mode $\V{k}$ within \texttt{PLUME} using
\begin{equation}
\underline{\underline{\epsilon}}=\underline{\underline{1}} + \sum_s \underline{\underline{\chi_s}}.
\label{eqn:dielectric.hp}
\end{equation}

Writing \eqref{eqn:wave.hp} in matrix form, we obtain an expression\footnotemark[\value{footnote}] 
that relates the wavevector, $\mathbf{k}$, to frequency, $\omega$,
\begin{equation}
\left( 
\begin{array}{ccc}
\epsilon_{xx}-n_z^{\prime \, 2} & \epsilon_{xy} & \epsilon_{xz}+n_x^\prime n_z^\prime\\
\epsilon_{yx} & \epsilon_{yy}-n_x^{\prime \, 2}-n_z^{\prime \, 2} & \epsilon_{yz}\\
\epsilon_{zx}+n_x^\prime n_z^\prime & \epsilon_{zy} & \epsilon_{zz}-n_x^{\prime \, 2}
\end{array}
\right)
\left(
\begin{array}{c}
\hat{E}_{\perp,1}\\
\hat{E}_{\perp,2}\\
\hat{E}_{\parallel}
\end{array}
\right)
=0.
\label{eqn:waveMatrix.hp}
\end{equation}
Here, for compactness in the matrix form, we identify $(x,y,z)\equiv(\perp_1,\perp_2,\parallel)$, so that $n_x^\prime\equiv n_{\perp,1}^\prime$ and $n_z^\prime\equiv n_\parallel^\prime$.

The code assumes a gyrotropic equilibrium velocity distribution about the equilibrium magnetic field $\V{B}_0=B_0 \hat{e}_{\parallel}$, with perpendicular and parallel thermal velocities $w_\perp$ and $w_{\parallel}$ respectively, \footnote{Here, the distribution function,  $f_{0s}(v_\perp,v_\parallel)$ is a function of both position and velocity, $f_{0s}(\mathbf{x};v_\perp,v_\parallel)$. However, we assume spatial homogeneity for background quantities, so the notation of the spatial dependence is suppressed. This subtle difference invokes an additional normalization factor of $n_{0,s}$, the equilibrium number density, in our definition.} 
\begin{equation}
f_{0s}(v_\perp,v_\parallel) = n_{0,s} h_s(v_\parallel) \exp
\left( -v_\perp^2/w_{\perp s}^2\right)/\pi w_{\perp s}^2.
\label{eq:maxwellian}
\end{equation}
Here, $h(v_{\parallel})$ includes an arbitrary drift $V_s$ of the species parallel to the magnetic field and is defined by, 
\begin{equation}
    h_s(v_\parallel) = \exp \left(
-(v_\parallel-V_s)^2/w_{\parallel s}^2\right)/\sqrt{\pi}
w_{\parallel s}.
\label{eq:drift_maxwellian}
\end{equation} 
Note that definitions for these variables can be found in Table~\ref{tab:defs}. One can analytically compute all terms in \eqref{eqn:waveMatrix.hp} by linearizing the Vlasov equation to first order, assuming plane-wave solutions for the perturbed fields and helical equilibrium particle trajectories, as presented in \citet{Stix:1992} and \citet{swanson2003plasma}. 
The dielectric tensor is a function of the wavevector $\V{k}$ and the complex wave frequency $\omega$, and can be expressed in terms of the sum of the tensor susceptibilities $\underline{\underline{\chi_s}}$ for each species $s$ using \eqref{eqn:dielectric.hp}. We compute the contribution to the total current by species $s$, denoted $\V{j}_s$, which is related to the electric field by $\hat{\V{j}}_s = -i \omega \underline{\underline{\chi_s}}(\V{k},\omega) \cdot \hat{\V{E}}/4 \pi$. 

\texttt{PLUME} solves for complex $\omega$ by finding solutions to the determinant of the matrix in \eqref{eqn:waveMatrix.hp} for specified $\V{k}$, also computing the associated electromagnetic, density, and velocity eigenfunctions for each species analytically at the solved frequency for the given wavevector. \texttt{PLUME} uses root-finding via the secant method to solve for the roots of the determinant of the matrix in \eqref{eqn:waveMatrix.hp}, starting with either a list of specified initial guesses or a 2D grid of the values of the determinant over a specified domain in complex omega space $(\omega_r,\gamma)$. \texttt{PLUME} can be used to compute the dispersion relation as a function of variations of a specified dimensionless input parameter from Table~\ref{tab:norm}. When doing so, \texttt{PLUME} uses the previous solution as the initial guess for the next step, improving numerical stability and performance while ensuring that the solver stays on the appropriate wave mode for each curve. The details for analytically computing $\underline{\underline{\chi_s}}$ with relevant normalizations are presented in \citet{Klein:PlumeCodePaper:2025}.

It should be noted that modeling particle velocity distributions with the many drifting bi-Maxwellians as done by \texttt{PLUME} is not without its limitations when applied to solar wind and space plasmas. The velocity distributions observed in the solar wind often deviate from a bi-Maxwellian (e.g. \citet{coburn2024regulation,bowen2022situ,bowen2023data,Klein:2026}) and have the \emph{potential} to significantly modify the dispersion relation, but we note that this bi-Maxwellian approximation is typically sufficient to compute an accurate dispersion relation for small to moderate non-Maxwellianity\citep{klein2015predicted,verscharen2018alps,walters2023effects}. Furthermore, using more accurate representations of the distribution function is computationally intense \citep{verscharen2018alps,Klein_Verscharen_2025}. Using the bi-Maxwellian approximation has the advantage of enabling computationally efficient solutions to the dispersion relation.

\subsection{\label{sec:FPCintro}Field-Particle Correlation Technique}
The Field-Particle Correlation (FPC) technique \citep{klein2016measuring,howes2017diagnosing} is a diagnostic that reveals energy transfer between fields and particles in a plasma as a function of phase space. The FPC is defined by multiplying the Vlasov equation for species $s$ by $\tfrac{1}{2} m_s v^2$ to obtain an equation for the evolution of the phase-space energy density $\mathcal{E}_s(\mathbf{x}, \mathbf{v}, t) = \tfrac{1}{2} m_s v^2 f_s(\mathbf{x}, \mathbf{v}, t)$, yielding
\begin{equation}
\frac{\partial \mathcal{E}_s}{\partial t} = - \V{v} \cdot \nabla \mathcal{E}_s -q_s \frac{v^2}{2} \V{E} \cdot \frac{\partial f_s}{\partial \V{v}} - \frac{q_s}{c} \frac{v^2}{2} (\V{v} \times \V{B}) \cdot \frac{\partial f_s}{\partial \V{v}}.
\label{eq:desdt}
\end{equation}
Integrating this equation over all velocity space and over a periodic or infinite spatial domain leaves only the term involving the electric field, because the advective term does not change the total particle energy and magnetic fields do no net work \citep{klein2017diagnosing}. The electric-field term is broken into components, as work done by each component can often be related to a specific mechanism (e.g., Landau damping is done by the parallel electric field, while cyclotron damping is done by the perpendicular components of the electric field). Following this procedure, the FPC due to electric field component $j$ on species $s$ is defined as \citep{klein2016measuring},
\begin{eqnarray}
C_{E_j,s}(\mathbf{x}_0,\mathbf{v},t_0)&\equiv&
\left\langle C_{E_j,s}(\mathbf{x},\mathbf{v},t)\right\rangle_{\mathbf{x}\in\mathcal{N}(\mathbf{x}_0),\,t\in[t_0,t_1]} \nonumber \\
&=&\bigg < -q_s \frac{v_j^2}{2} \frac{\partial f_s(\V{x},\V{v},t)}{\partial v_j} E_j(\V{x},t) \bigg >,
\label{eq:FPCdefinition}
\end{eqnarray}
where $\langle\cdot\rangle$ denotes an average over time at fixed $\mathbf{x}$,
over space at fixed $t$, or over the combined spatiotemporal region
$\mathbf{x}\in\mathcal{N}(\mathbf{x}_0)$ and $t\in[t_0,t_1]$, depending on the
application and reference frame. {\color{blue} The FPC is used here as a diagnostic of net field--particle energization. It does not evolve \(f_s\) or \(f_{1s}\). The replacement \(v^2\rightarrow v_j^2\) in the term involving the electric field in Equation~\ref{eq:desdt} is a choice of diagnostic representation, chosen because the velocity-space integral with this replacement gives the net work channel \(\langle j_{j,s}E_j\rangle\), as the remaining velocity components perpendicular to \(\hat{e}_j\) yield no net contribution after integration.}

In principle one may {\color{blue}analyze} an instantaneous (unaveraged) correlation $C_{E_j,s}(\mathbf{x},\mathbf{v},t)\equiv -q_s \frac{v_j^2}{2}\,\frac{\partial f_s(\mathbf{x},\mathbf{v},t)}{\partial v_j}\,E_j(\mathbf{x},t)$, but throughout this work we use the averaged form since the relevant energization processes are distributed in space and time and the net (cumulative) transfer is most robustly characterized by spatiotemporal averaging. Note that the $v^2$ factor in the electric field term of \eqref{eq:desdt} is replaced here by $v_j^2$ because the contributions from the other components yield zero net change in the particle energy when integrated over all velocity space \citep{klein2017diagnosing}.  For wave-particle interactions or other oscillatory particle energization mechanisms, the correlation is optimally taken over an integral number of wave periods or a periodic spatial domain.  Furthermore, the correlation may be taken with the perturbed velocity distribution, $f_{1,s}$, rather than total velocity distribution, $f_s=f_{0,s}+f_{1,s}$, as long as there is zero net flow in the equilibrium velocity distribution. {\color{blue} For the averaged linear-wave correlations considered here, the nonzero contribution to the FPC arises from the correlation between the first-order perturbed distribution function and the first-order electric field, and is therefore second order in the fluctuation amplitude. Note that this ordering is closely related to quasilinear theory, where averaged products of first-order fluctuations drive the slow evolution of the background distribution function \citep{Stix:1992}. However, here the same ordering is used to compute the velocity-space distribution of field--particle work for a specified linear eigenmode on a fixed equilibrium distribution, rather than to evolve the distribution function quasilinearly in time.}

The FPC technique serves as a valuable diagnostic for net energy transfer in collisionless systems. In systems governed by the Vlasov–Maxwell equations, all net energy transfer occurs through collisionless field–particle interactions and is therefore fully captured by the correlation. This relationship is illustrated by the identity
\begin{equation}
    \int C_{E_{j,s}}(\V{x},\V{v}) \, d\V{v} = \Big < j_{j,s}(\V{x}) E_j(\V{x}) \Big >_{\mathbf{x}\in\mathcal{N}(\mathbf{x}_0),\,t\in[t_0,t_1]}.
\label{eq:CEiequivjdotE}
\end{equation}

The FPC can also be expressed as $C_{E_j,s}(\V{x}_0,\V{v},t) = -(v_j/2)\partial C_{E_j,s}^\prime(\V{v})/\partial v_j + C_{E_j,s}^\prime(\V{v})/2$, where $C^\prime_{E_j,s}=\big<q_s v_j E_j f_s \big>$  is known as the \emph{alternative} form of the Field-Particle Correlation\citep{chen2019evidence}. As it is necessary to have a sufficient sampling of the distribution function to take the velocity derivative in \eqref{eq:FPCdefinition}, the alternative form is preferred when the distribution function is computed using an integration volume over which the electric field significantly varies \citep{brown2022isolation} or when the velocity gradient in \eqref{eq:FPCdefinition} has significant uncertainty. This $C_{E_j,s}^\prime$ form was the primary form used in the work by \citet{zhao2022quantifying}.

If the correlation is integrated over all velocity space, $C_{E_j,s}^\prime(\V{v})$ can be obtained from $C_{E_j,s}(\V{v})$ using integration by parts, which critically relies on the fact that the distributions vanish at the boundaries of the integration. 
The alternative FPC $C_{E_j,s}^\prime(\V{v})$ represents the energy transport in the Lagrangian frame moving with particles \citep{montag2022field}, whereas the standard FPC $C_{E_j,s}(\V{v})$ defines the rate of change of the phase-space energy density $\mathcal{E}_s(\mathbf{x}, \mathbf{v}, t)$ in the Eulerian frame of 3D-3V phase space $(\mathbf{x}, \mathbf{v})$.
Both the standard and alternative forms yield the same net rate of particle energization when integrated over velocity space, meaning they provide equivalent measures of total energy transfer. However, the standard form $C_{E_j,s}(\V{v})$ is generally preferred for velocity-space visualization. Because its signatures cross through zero exactly at the resonant phase velocity in the Eulerian frame, it provides a clearer, more direct visual connection to the underlying wave-particle resonance.

The seemingly chaotic behavior of collisionless mechanisms can be locally misleading, and thus the FPC technique was developed to diagnose mechanisms of energy transfer that lead to net plasma heating or particle acceleration. One of the notable advantages of this diagnostic is the use of velocity-space information rather than spatial information. This capability enables one to identify, isolate, and measure energy transfer mechanisms from measurements at a single point, typical of most \emph{in situ} satellite measurements, and to separate secular energy transfer from mechanisms that induce oscillatory energy transfer. Consequently, the FPC technique has found wide application across space and astrophysical plasmas, being used to study a variety of mechanisms including auroral electron acceleration \citep{schroeder2021laboratory}, turbulence in the heliosphere \citep{afshari2021importance,chen2019evidence,horvath2020electron,horvath2022observing,klein2017diagnosing,klein2020diagnosing,li2019collisionless,Afshari:2024,Huang:2024}, collisionless shocks \citep{juno2021field,brown2022isolation,juno2022phase,Montag:2025,Howes:2025}, magnetic reconnection \citep{mccubbin2022characterizing}, and magnetic pumping \citep{montag2022field}.

The FPC diagnostic has been successful in identifying the particle energization mechanisms and quantifying their rates of energy transfer, even in situations where multiple energization mechanisms are acting simultaneously at the same position in space \citep{klein2020diagnosing,Afshari:2024}.  Carefully selected approaches, such as filtering in frequency \citep{afshari2021importance,Afshari:2024}, exploiting  particular system symmetries such as periodicity in certain spatial dimensions \citep{brown2022isolation}, or deliberate tuning of the time integration window \citep{klein2017diagnosing,verniero2021determining} have enabled the separation of the different energization mechanisms.  Yet, if the regions of velocity space corresponding to two different energization mechanisms overlap, it is challenging to isolate the energization rates due to each mechanism.  Particularly, if the wavelength and frequency associated with two different mechanisms are too similar in scale, it can be difficult to unambiguously associate a signature with its respective process. Here, we address these challenges by employing a formulation of the FPC that leverages analytic eigenmode solutions of the dispersion relation for both the electromagnetic fields and the particle velocity distribution, enabling the specification and analysis of a single wave mode in isolation.


While invaluable for understanding the energy transfer associated with wave–particle interactions, implementation of the FPC diagnostic for use with experimental measurements or numerical simulations can present significant challenges. The experimental determination of the FPC often requires sophisticated and costly particle diagnostics, and the numerical investigation of specific modes or mechanisms over the full relevant range of system parameters is often not feasible due to the computational cost of kinetic simulations. Particle-in-cell simulations require a large number of particles to accurately resolve the system's particle velocity distributions in six-dimensional phase space \citep{brown2022isolation,juno2020noise}, while direct solutions of the Vlasov–Maxwell equations, such as those employed in \texttt{Gkeyll} \citep{juno2018discontinuous}, are computationally demanding to simulate all six dimensions of phase space.  The analytical methodology developed here offers a significant advantage in computational efficiency and control. By directly calculating the FPC from linear theory, we circumvent these issues, ensuring a focus on the fundamental linear wave–particle interactions. 

\section{Analytical Formulations of Phase-Space Field-Particle Energy Exchange}
\label{sec:linfpcintro}
Here, we extend \texttt{PLUME} to predict \textit{velocity-space signatures} of the energy transfer between an arbitrary number of drifting bi-Maxwellian particle species, $f_s(v_{\parallel},v_{\perp})$, and the linear response of the electric fields using the susceptibility of a hot plasma in a uniform magnetic field \citep{swanson2003plasma,Stix:1992} to a single perturbing wave mode. That is, we extend our solver to predict the wave–particle interactions between an arbitrary number of species and a single wave with wavevector $\mathbf{k}$ and frequency $\omega$, determined by the linear dispersion relation computed by \texttt{PLUME}. We name this extension \texttt{JET-PLUME} (\textbf{J}udging \textbf{E}nergy \textbf{T}ransfer in a \textbf{P}lasma in a
\textbf{L}inear \textbf{U}niform \textbf{M}agnetized \textbf{E}nvironment). \texttt{JET-PLUME} correlates the perturbed velocity distribution with the electromagnetic eigenfunctions for a selected solution of the dispersion relation for a single monochromatic linear wave to compute the FPC of wave–particle interactions averaged over a correlation interval of one wave period. In Appendix~\ref{appendix:linFPCderiv}, we show that the form of the FPC, Equation~\eqref{eq:FPCdefinition},  in the weak growth rate limit $|\gamma|\ll |\omega_r|$ at position $\V{x}_0$ is given by
\begin{equation}
  C_{E_j,s}(\V{x}_0,\V{v},t)=
   -q_s \frac{v_j^2}{2}
  \frac{1}{4}\left( \frac{\partial \hat{f}_{1,s}}{\partial v_j}  \hat{E}_j^* +  \frac{\partial \hat{f}^*_{1,s}}{\partial v_j}  \hat{E}_j \right)
 \label{eq:ceipredict}
\end{equation}
where the complex Fourier coefficient  is denoted by the hat, \emph{e.g.}, $\hat{E}_j(\V{k},\omega)$, and $*$ denotes the complex conjugate.
The complex Fourier coefficient 
of the perturbed distribution function is given by\citep{swanson2003plasma}
\begin{align}
    \hat{f}_{1,s}(\V{k},\V{v},&\omega) = - \frac{iq_s}{m_s} \sum_{m=-\infty}^{\infty} \sum_{n=-\infty}^{\infty} \frac{J_m(b_s) e^{i(m-n)\phi}}{\omega - n \Omega_s - k_{\parallel} v_{\parallel}} \nonumber \\
    & \times \left\{ \frac{n J_n(b_s)}{b_s} U_s \hat{E}_{\perp,1} + i J'_n(b_s) U_s \hat{E}_{\perp,2} + J_n(b_s) W_s \hat{E}_{\parallel} \right\}
\label{eq:f1a}
\end{align}
with
\begin{equation}
  U_s=\frac{\partial f_{0,s}}{\partial v_\perp}+  \frac{k_{\parallel}}{\omega}\left(v_\perp \frac{\partial f_{0,s}}{\partial v_{\parallel}} - v_{\parallel} \frac{\partial f_{0,s}}{\partial v_\perp} \right),
  \label{eq:u}
\end{equation}
\begin{equation}
  W_s=\frac{\partial f_{0,s}}{\partial v_{\parallel}}-  \frac{n \Omega_s}{v_\perp \omega}\left(v_\perp \frac{\partial f_{0,s}}{\partial v_{\parallel}} - v_{\parallel} \frac{\partial f_{0,s}}{\partial v_\perp} \right),
  \label{eq:w}
\end{equation}
\begin{equation}
  b_s=\frac{ k_\perp v_\perp}{\Omega_{s}}.
\end{equation}
The variable definitions above can be found in Tables \ref{tab:defs} and \ref{tab:norm}. Note that, without loss of generality, we choose coordinates such that the wavevector lies in the $\hat{e}_{\parallel}$ and $\hat{e}_{\perp,1}$ plane, eliminating the $\phi$ dependence in Equation~(4.180) in the text by \citet{swanson2003plasma}. Additional details about variable definitions and the dimensionless normalization used by \texttt{JET-PLUME} can be found in \citet{Klein:PlumeCodePaper:2025}.

We present the correlation in two forms in this work: (i) as \textit{gyrotropic signatures},
\begin{align}
C_{E_{\parallel},s}(v_{\parallel},v_{\perp}) &\equiv \int_0^{2 \pi} C_{E_{\parallel},s}(v_{\parallel},v_{\perp},\phi) \, v_\perp d\phi,  \label{eq:gyrocorr}\\
C_{E_{\perp},s}(v_{\parallel},v_{\perp}) &\equiv \int_0^{2 \pi} \bigg[ C_{E_{\perp,1},s}(v_{\parallel},v_{\perp},\phi) \\ &+ C_{E_{\perp,2},s}(v_{\parallel},v_{\perp},\phi) \bigg] \, v_\perp d\phi, \nonumber
\end{align}
with $(v_{\perp,1},v_{\perp,2}) = (v_{\perp} \cos{\phi}, v_{\perp} \sin{\phi})$; or (ii) as \textit{Cartesian signatures}, which are reductions from 3V to 2V of the correlation $C_{E_j}$ by integration over one of the dimensions in Cartesian velocity space. For example, the integration over the parallel velocity component to obtain velocity-space signatures in the perpendicular plane would be
\begin{equation}
C_{E_{\perp,1,s}}(v_{\perp,1},v_{\perp,2}) \equiv \int C_{E_{\perp,1,s}}(v_{\parallel},v_{\perp,1},v_{\perp,2}) \, \, dv_\parallel.
\end{equation}
The motivation behind these reductions to a 2V velocity space is two-fold. First, particle counts from \emph{in situ} observations and in kinetic simulations are often limited, so reducing the dimensionality increases the number of counts in each 2V bin, effectively increasing the signal-to-noise ratio. Second, a three-dimensional (3V) structure in velocity space is unwieldy to plot, more data-intensive, and not always necessary to identify mechanisms using the FPC technique (for example, see \citet{chen2019evidence,li2019collisionless,klein2020diagnosing}).

With \texttt{JET-PLUME}, one can analytically separate the contributions to the perturbed velocity distribution \( \hat{f}_{1,s} \) in Equation~\eqref{eq:f1a} according to their dependence on each component of the electric field,
\begin{equation}
\hat{f}_{1,s} = \hat{f}_{1,s}^{(\perp,1)} + \hat{f}_{1,s}^{(\perp,2)} + \hat{f}_{1,s}^{(\parallel)},
\label{eq:fdecomptotal}
\end{equation}
where
\begin{align}
\hat{f}_{1,s}^{(\perp,1)} &= - \frac{iq_s}{m_s} \sum_{m,n} \frac{J_m(b_s) e^{i(m-n)\phi}}{\omega - n  \Omega_s - k_{\parallel} v_{\parallel}}  \frac{n J_n(b_s)}{b_s} \, U_s \, \hat{E}_{\perp,1}, \label{eq:fseparateddefs_justperp1}\\
\hat{f}_{1,s}^{(\perp,2)} &= - \frac{iq_s}{m_s} \sum_{m,n} \frac{J_m(b_s) e^{i(m-n)\phi}}{\omega - n  \Omega_s - k_{\parallel} v_{\parallel}}  i J'_n(b_s) \, U_s \, \hat{E}_{\perp,2},
\label{eq:fseparateddefs_justperp2}\\
\hat{f}_{1,s}^{(\parallel)} &= - \frac{iq_s}{m_s} \sum_{m,n} \frac{J_m(b_s) e^{i(m-n)\phi}}{\omega - n  \Omega_s - k_{\parallel} v_{\parallel}} J_n(b_s) \, W_s \, \hat{E}_{\parallel},
\label{eq:fseparateddefs}
\end{align}
such that each term isolates the particle response driven by the corresponding electric field component.

These terms are related to the velocity moment integrals that define the elements of the susceptibility tensor, $\chi_{ij}$, such that multiplying each isolated term $\hat{f}_{1,s}^{(i)}$ by $q_s v_j$ and integrating over velocity space yields the induced current density component, which defines the susceptibility tensor\citep{Stix:1992}. Similarly, we decompose the FPC into directional contributions to isolate the velocity-space energy transfer associated with each field component with
\begin{equation}
C_{E_j,s}(\mathbf{v}) = C_{E_j,s}^{(\perp,1)}(\mathbf{v}) + C_{E_j,s}^{(\perp,2)}(\mathbf{v}) + C_{E_j,s}^{(\parallel)}(\mathbf{v}),
\end{equation}
where each subscript denotes the component of the electric field driving the correlation, and the superscript indicates the velocity component over which the distribution is perturbed, i.e.
\begin{eqnarray}
C_{E_j,s}^{(\perp,1)}(\mathbf{v}) &=&
-q_s \frac{v_j^2}{2}
\frac{1}{4}\left( \frac{\partial \hat{f}_{1,s}^{(\perp,1)}}{\partial v_j}  \hat{E}_j^* +  \frac{\partial \hat{f}^{*,(\perp,1)}_{1,s}}{\partial v_j}  \hat{E}_j \right), \nonumber \\
C_{E_j,s}^{(\perp,2)}(\mathbf{v}) &=&
-q_s \frac{v_j^2}{2}
\frac{1}{4}\left( \frac{\partial \hat{f}_{1,s}^{(\perp,2)}}{\partial v_j}  \hat{E}_j^* +  \frac{\partial \hat{f}^{*,(\perp,2)}_{1,s}}{\partial v_j}  \hat{E}_j \right),  \label{eq:sepCEpar}\\
C_{E_j,s}^{(\parallel)}(\mathbf{v}) &=&
-q_s \frac{v_j^2}{2}
\frac{1}{4}\left( \frac{\partial \hat{f}_{1,s}^{(\parallel)}}{\partial v_j}  \hat{E}_j^* +  \frac{\partial \hat{f}^{*,{(\parallel)}}_{1,s}}{\partial v_j}  \hat{E}_j \right). \nonumber 
\end{eqnarray}

{\color{blue}
This decomposition is used to construct component-resolved FPC terms from the electric-field-driven pieces of \(\hat f_{1s}\), so that individual response channels can be analyzed before they are summed into the total FPC signature.
} Often, the off-diagonal elements of both the susceptibility tensor and their contributions to wave–particle energy exchange are small and neglected for simplicity. Expressed mathematically, $C_{E_i}^{(j)}$, with $j \neq i$, is often small. However, as demonstrated in Section~\ref{sec:ikldcomp} and the study of transit-time damping in \citet{Huang:2024}, off-diagonal contributions can play a significant role in the net energization of the particle distribution and can be clearly diagnosed with the FPC. This motivates the analysis of these off-diagonal terms in advanced diagnostics of wave–particle energy transfer.

\section{\label{sec:Comparisons}Comparisons with Simulations}

In this section, we compare the analytic computations of \textit{velocity-space signatures} produced by \texttt{JET-PLUME} against those derived from numerical simulations using the gyrokinetic code \texttt{AstroGK} \citep{numata2010astrogk} and the Vlasov-Maxwell solver in \texttt{Gkeyll} \citep{juno2018discontinuous}. Simulations are initialized such that energy transfer due to wave-particle interactions will be dominated by a single mode to facilitate comparison to the \texttt{JET-PLUME} predictions.

\subsection{Ion Landau Damping}
\label{sec:ikldcomp}

\subsubsection{Kinetic \Alfven\ Wave Simulation Setup and FPC Comparison}

We start by comparing the signatures computed with \texttt{JET-PLUME} to the signatures from a kinetic \Alfven\ wave simulation using \texttt{AstroGK}, described in depth in \citet{howes2017prospectus}. We observe significant energy exchange related to the off-diagonal elements of the susceptibility tensor, i.e., energization between the parallel electric field and the distribution function's response to the out-of-plane perpendicular electric field.

\texttt{AstroGK} solves the gyrokinetic equation and gyroaveraged Maxwell's equations, evolving the gyroaveraged distribution function of each species, the scalar potential, the parallel vector potential, and the parallel magnetic field perturbation \citep{howes2006astrophysical}. The code uses an upwind finite-difference scheme in the parallel direction and represents the perpendicular spatial dimensions pseudospectrally. The domain is periodic with size $L_{\parallel} \times L_{\perp}^2$, where $L_{\parallel}/L_{\perp} \gg 1$. We initialize the simulation with uniform Maxwellian equilibria for a system with ions and electrons of realistic mass ratio, $m_i/m_e = 1836$. Here, as in \citet{howes2017prospectus}, a single kinetic \Alfven\ wave is initialized using the eigenfunctions for the perturbed distribution functions and electromagnetic fields computed from the linear, collisionless gyrokinetic dispersion relation for a kinetic \Alfven\ wave \citep{howes2006astrophysical} with $k_{\perp} \rho_i = 1.3$ and plasma parameters $\beta_i = 1$, and $T_i/T_e = 1$, assuming an isotropic Maxwellian for each species, in a simulation domain of size $L_{\perp} = 2\pi \rho_i / 1.3$ and $L_{\parallel} = L_{\perp} / \epsilon$, where $\epsilon \ll 1$ is the gyrokinetic expansion parameter. The eigenfunctions for this mode, which are equal to the gyrokinetic solutions in the low frequency, wavevector anisotropic limit, computed by \texttt{PLUME} with $k_\parallel \rho_i = 0.001$ (corresponding to specifying $\epsilon= 7.69 \times 10^{-4}$) are provided in Table~\ref{tab:ikld}. Note, the initialized kinetic \Alfven\ wave has a small field amplitude, with \(q_i\phi/T_i \lesssim 3\times10^{-4}\). {\color{blue} This corresponds to an \(E\times B\)-velocity estimate \(\epsilon_i\sim\delta v_E/v_{\mathrm{th},i}\lesssim2\times10^{-4}\), orders of magnitude below the \(\epsilon_i\sim0.1\)--\(0.3\) values typically associated with the onset of strong stochastic ion heating in finite-amplitude kinetic \Alfven\ wave turbulence \citep{Chandran:2010a,Cerri:2021}.}

{Similarly, for \texttt{JET-PLUME}, we adopt a proton-electron plasma with equal number densities $n_i = n_e$ and a realistic mass ratio $m_i/m_e = 1836$. The ion beta is set to $\beta_i = 1$, with isotropic temperatures $(T_\perp/T_\parallel)_s = 1$ for all species, and an ion--electron temperature ratio of $T_i/T_e = 1$. The ion thermal speed is specified directly as $v_{ti}/c = 10^{-4}$. Wavenumbers are normalized to the ion gyroradius, and in the example shown we take $k_\parallel \rho_i = 0.001$ and $k_\perp \rho_i = 1.3$. No background drifts are imposed in this system. 
The input parameters used in the \texttt{JET-PLUME} and \texttt{PLUME} calculations, for all sections of this work, are available in the Data Availability section.}


To better approximate a collisionless system, the collisionality is decreased by two orders of magnitude from the original simulation in \citet{howes2017prospectus}, taking $\nu_i = \nu_e = 2.0 \times 10^{-4} \, \omega_A$, where $\omega_A=k_\parallel v_A$ is the characteristic \Alfven\ wave frequency. We compute the velocity-space signature at the start of the simulation, correlating over the first wave period, $\omega_A T = 2\pi/(\omega_r/\omega_A) = (2\pi)/(1.237)$, at a single spatial point. 
\begin{table}
 \begin{center}
 \begin{tabular}{|l||c|c|c|}
    \hline
    & $\hat{E}_{\perp,1}/\hat{E}_{\perp,1}$ & $\hat{E}_{\perp,2}/\hat{E}_{\perp,1}$ & $\hat{E}_\parallel/\hat{E}_{\perp,1}$ \\
    \hline
    Re. & $1.0$ & $6.3 \times 10^{-6}$ & $-2.79 \times 10^{-4}$ \\
    Im. & $0$ & $5.6 \times 10^{-4}$ & $6.63 \times 10^{-5}$ \\
    \hline
    & $\hat{B}_{\perp,1}/\hat{E}_{\perp,1}$ & $\hat{B}_{\perp,2}/\hat{E}_{\perp,1}$ & $\hat{B}_\parallel/\hat{E}_{\perp,1}$ \\
    \hline
    Re. & $1.11 \times 10^{-1}$ & $1.1 \times 10^{4}$ & $-1.45 \times 10^{2}$ \\
    Im. & $-4.53$ & $-3.01 \times 10^{2}$ & $5.88 \times 10^{3}$ \\
    \hline
    & $\hat{U}_{\perp,1,i}/(c\hat{E}_{\perp,1}/B_0)$ & $\hat{U}_{\perp,2,i}/(c\hat{E}_{\perp,1}/B_0)$ & $\hat{U}_{\parallel,i}/(c\hat{E}_{\perp,1}/B_0)$ \\
    \hline
    Re. & $-2.38 \times 10^{-5}$ & $-6 \times 10^{-1}$ & $-8.53 \times 10^{-2}$ \\
    Im. & $-5.43 \times 10^{-4}$ & $4.72 \times 10^{-2}$ & $1.33 \times 10^{-2}$ \\
    \hline
    & $\hat{U}_{\perp,1,e}/(c\hat{E}_{\perp,1}/B_0)$ & $\hat{U}_{\perp,2,e}/(c\hat{E}_{\perp,1}/B_0)$ & $\hat{U}_{\parallel,e}/(c\hat{E}_{\perp,1}/B_0)$ \\
    \hline
    Re. & $6.32 \times 10^{-6}$ & $-1.37$ & $-1.24 \times 10^{-1}$ \\
    Im. & $5.6 \times 10^{-4}$ & $2.84 \times 10^{-2}$ & $-1.42$ \\
    \hline
    & $\hat{n}_{i}/(n_{0,i}\hat{E}_{\perp,1}/B_0)$ & $\hat{n}_{e}/(n_{0,e}\hat{E}_{\perp,1}/B_0)$ &  $\omega/\Omega_R$\\
    \hline
    Re. & $-7.38 \times 10^{2}$ & $-7.38 \times 10^{2}$ & $1.24 \times 10^{-3}$ \\
    Im. & $-5.63 \times 10^{3}$ & $-5.63 \times 10^{3}$ & $-4.44 \times 10^{-5}$ \\
    \hline
 \end{tabular}
 \end{center}
 \caption{Selected complex Fourier coefficients of the eigenfunctions for the ion Landau damping of the kinetic \Alfven\ wave described in Section \ref{sec:ikldcomp} using \texttt{PLUME}.}
 \label{tab:ikld}
\end{table}

\begin{figure}
 \begin{center}
    \includegraphics[width=.45\textwidth]{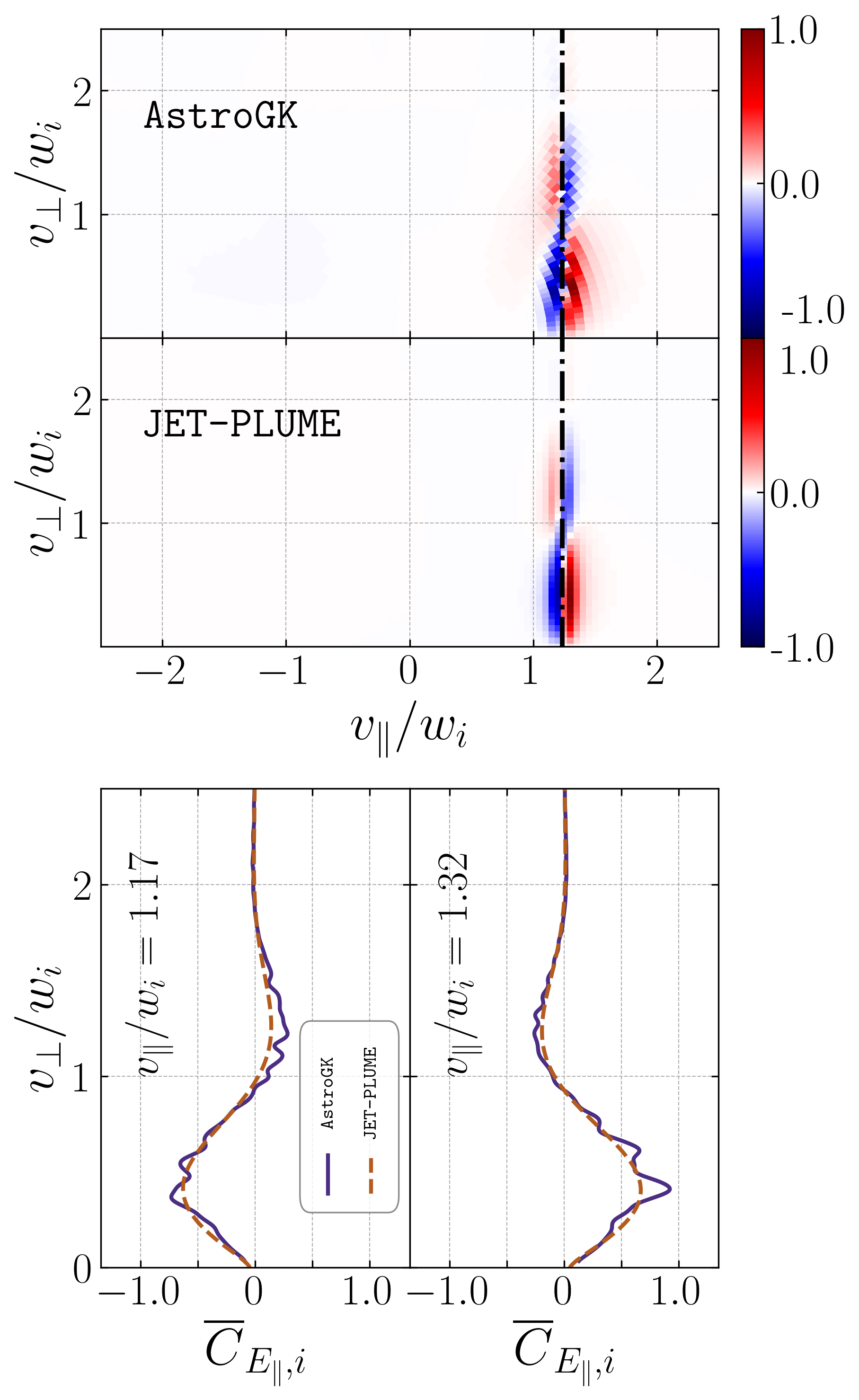}
   \end{center}
 \caption{{\color{blue}Comparison between \textit{gyrotropic velocity-space signatures}, $C_{E_{\parallel},i}(v_{\parallel},v_\perp)$, for ion Landau damping computed from the \texttt{AstroGK} simulation of a kinetic \Alfven\ wave with $\beta_i=1$ and computed using \texttt{JET-PLUME}. The upper two panels show the full velocity-space signatures from \texttt{AstroGK} and \texttt{JET-PLUME}, respectively. Both are separately normalized so that $\max|C_{E_{\parallel},i}(v_{\parallel},v_\perp)| = 1$. The vertical black dash-dotted line indicates the resonant parallel phase velocity, $\omega_r/k_\parallel$. The lower panels show cuts through the same normalized signatures at fixed values $v_\parallel/w_i=1.17$ and $v_\parallel/w_i=1.32$, plotted as functions of $v_\perp/w_i$, with the \texttt{AstroGK} result shown by the solid curve and the \texttt{JET-PLUME} result shown by the dashed curve.}}
    \label{fig:howes2017compar}
\end{figure}

We compare the gyrotropic velocity-space signatures of the simulation to those computed by \texttt{JET-PLUME} for the ions in Figure~\ref{fig:howes2017compar}, showing the parallel correlation, $C_{E_{\parallel},i}(v_{\parallel},v_{\perp})$. \texttt{JET-PLUME} accurately computes the location of the signature at the resonant velocity, the `twist' in the features as the bipolar structure inverts at about $v_{\perp}/w_{i} \simeq 1$, and the overall magnitude of the signature. {\color{blue}The one-dimensional cuts in the lower panels provide a direct comparison of the normalized profile height and shape at selected values of \(v_\parallel/w_i\), showing that the agreement is not limited to the visual appearance of the two-dimensional color plots.} We interpret the features of Figure~\ref{fig:howes2017compar} as follows: there exists a bipolar structure surrounding the resonant parallel phase velocity $\omega_r/k_\parallel$ (vertical dot-dashed line), indicating the resonant interaction of particles with velocities similar to that of the wave. Integrating over velocity space yields a positive quantity, indicating the net transfer of energy from the wave to the particles, as expected for a damping wave. The \texttt{JET-PLUME} signature appears marginally thinner in $v_\parallel$ than the \texttt{AstroGK} signature; this is likely due to the effect of weak collisionality in the simulation relative to the collisionless \texttt{JET-PLUME} calculation.  In addition, small differences in the visual appearance of the plots may arise from 
different discretizations of the gyrotropic velocity space: \texttt{AstroGK} employs a polar grid in $(v_\parallel, v_\perp)$, while \texttt{JET-PLUME} uses a cylindrical grid in 3V aligned along the magnetic field, which reduces to a rectangular grid after integration over the gyrophase angle.


\begin{figure}
 \begin{center}
    \includegraphics[width=.45\textwidth]{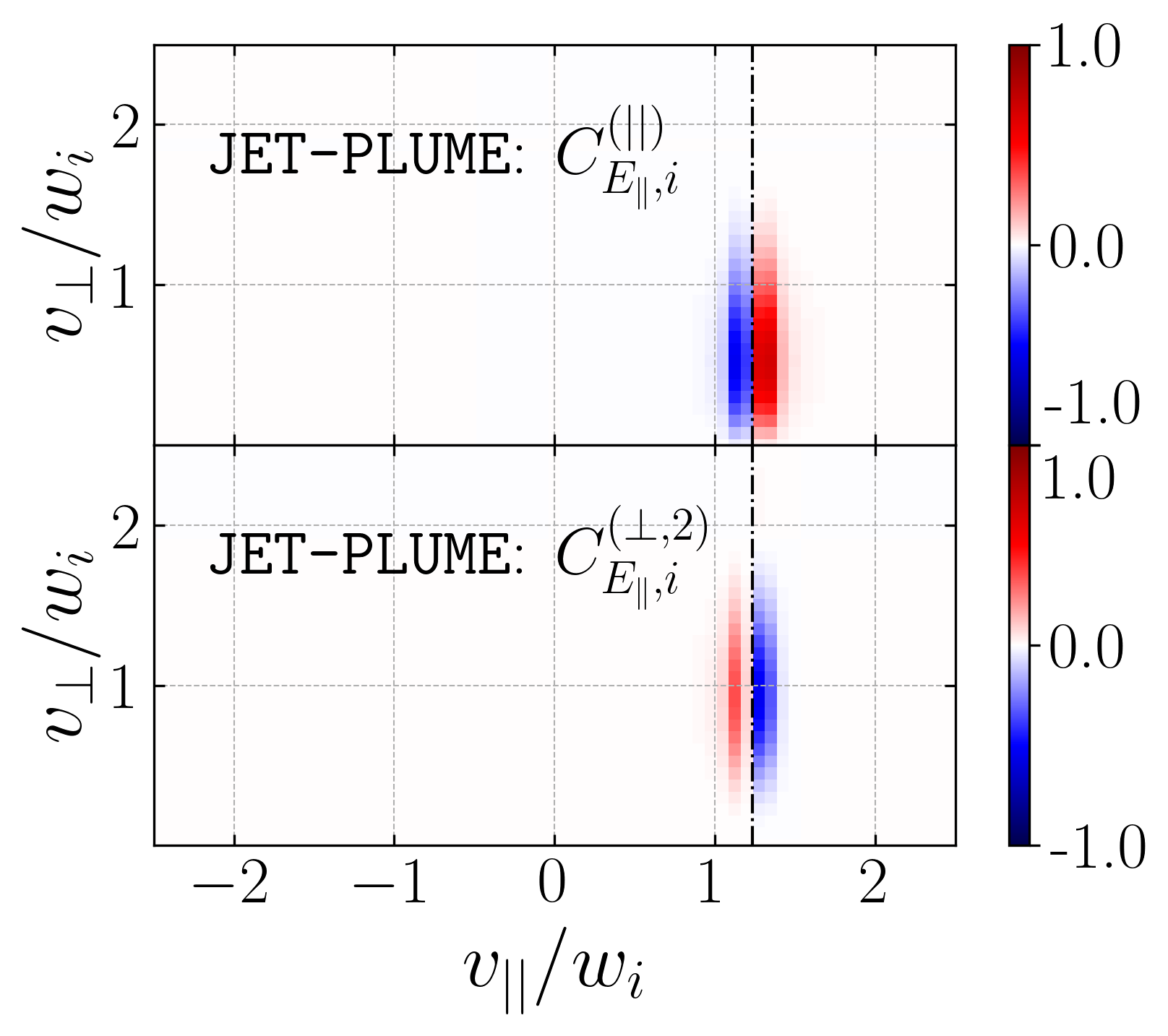}
   \end{center}
 \caption{Comparison between the separated contributions to the \emph{gyrotropic velocity-space signature}  from ion Landau damping of a kinetic \Alfven\ wave computed using \texttt{JET-PLUME}. (\emph{Top}) The contribution $C_{E_\parallel,i}^{(\parallel)}$ to the energization by the parallel electric field due to contribution to the parallel current caused by $f_{1,i}^{(\parallel)}$. (\emph{Bottom}) The contribution $C_{E_\parallel,i}^{(\perp,2)}$ to the energization by the parallel electric field due to contribution to the parallel current caused by $f_{1,i}^{(\perp,2)}$. The panels are normalized separately. 
 }
    \label{fig:howes2017comparSeparated}
\end{figure}

Figure~\ref{fig:howes2017compar} exhibits a `twist' as the sign of the bipolar structure about the resonant parallel phase velocity flips as $v_\perp$ increases. To explain this twist, we plot in Figure~\ref{fig:howes2017comparSeparated} the separated contributions to the energization by the parallel electric field, $C_{E_{\parallel},i}^{(\parallel)}$ and $C_{E_{\parallel},i}^{(\perp,2)}$ given by \eqref{eq:sepCEpar}. In the top panel, $C_{E_{\parallel},i}^{(\parallel)}$ yields the expected signature for the Landau damping of a wave \citep{howes2017diagnosing}: a bipolar signature showing a loss of phase-space energy density (blue) below the resonant parallel phase velocity at $v_\parallel <\omega_r/k_\parallel$ and a gain (red) above at  $v_\parallel > \omega_r/k_\parallel$.
Physically, this is interpreted as ions being accelerated from the blue to the red region by $E_\parallel$, leading to a net acceleration of the resonant ions and corresponding damping of the kinetic \Alfven\ wave. 

In the bottom panel, however, $C_{E_{\parallel},i}^{(\perp,2)}$ shows a reversal of the bipolar signature associated with $C_{E_{\parallel},i}^{(\parallel)}$. (Note that the third potential contribution to the parallel energization, $C_{E_{\parallel},i}^{(\perp,1)}$, is identically zero for this resonance, as detailed below.) This reversal physically represents ions that are decelerated from parallel velocities above the resonant velocity to parallel velocities below the resonant velocity. Thus, the contribution from this term represents a loss of energy in the ions and a corresponding increase of the energy in the parallel electric field.

The capabilities of \texttt{JET-PLUME} enable us to identify the reason for the `twist' in the bipolar signature as $v_\perp$ increases.  First, note that the Landau resonance is the $n=0$ contribution to the summation over Bessel functions in \eqref{eq:f1a}.  For the $n=0$ contribution, the perturbation to the ion velocity distribution driven by $E_{\perp,1}$ from \eqref{eq:fseparateddefs_justperp1} is zero, $\hat{f}_{1,i}^{(\perp,1)}=0$, leaving only the perturbations $\hat{f}_{1,i}^{(\perp,2)}$ caused by $E_{\perp,2}$ given by \eqref{eq:fseparateddefs_justperp2} and  $\hat{f}_{1,i}^{(\parallel)}$ caused by $E_{\parallel}$ given by \eqref{eq:fseparateddefs}.  These perturbations to the ion velocity distribution ultimately lead to the ion current density of the wave, $\V{j}_i$, determined mathematically through the ion susceptibility tensor, $\underline{\underline{\chi_i}}$.  

The velocity-space signature of ion energization through the Landau resonance by $E_{\parallel}$ contains the two contributions $C_{E_{\parallel},i}^{(\perp,2)}$ and $C_{E_{\parallel},i}^{(\parallel)}$ given by \eqref{eq:sepCEpar}. The mathematical forms of  $\hat{f}_{1,i}^{(\perp,2)}$ and $\hat{f}_{1,i}^{(\parallel)}$ explain the qualitative appearance of $C_{E_{\parallel},i}^{(\perp,2)}$ and $C_{E_{\parallel},i}^{(\parallel)}$.  The perturbation $\hat{f}_{1,i}^{(\parallel)}$ involves the zeroth-order Bessel function $J_0$ in $v_\perp$ and has an overall envelope of $\exp(-v_\perp^2/v_{ti}^2)$ through the $W$ factor given by \eqref{eq:w}.  Therefore the peak of this pattern will occur at $v_\perp=0$; note, however, that integration of the 3V correlation over gyrophase to obtain the correlation in 2V gyrotropic velocity space $(v_\parallel,v_\perp)$, which incorporates the factor of $v_\perp$ in \eqref{eq:gyrocorr}, pushes the peak above $v_\perp=0$.  In contrast, the perturbation $\hat{f}_{1,i}^{(\perp,2)}$ involves the derivative of the zeroth-order Bessel function $J'_0$ in $v_\perp$ and also has an overall envelope of $\exp(-v_\perp^2/v_{ti}^2)$ through the $U$ factor given by \eqref{eq:u}. Therefore the peak of this curve will occur at $v_\perp>0$, pushing the pattern of $C_{E_{\parallel},i}^{(\perp,2)}$  to higher $v_\perp$ than that of $C_{E_{\parallel},i}^{(\parallel)}$.  

Furthermore, the sign of the bipolar pattern will depend on the relative phase of parallel electric field $E_\parallel$ to the phases of the contributions to the parallel current driven by $E_{\perp,2}$ and by $E_{\parallel}$ through the ion susceptibility tensor.  In this case, the contribution to the parallel current driven by $E_{\parallel}$ has a component in-phase with $E_{\parallel}$, leading to energy transfer to the ions; the contribution to the parallel current driven by $E_{\perp,2}$ has a component out-of-phase with $E_{\parallel}$, leading to energy transfer from the ions, thus reversing the bipolar pattern. The contributions of $C_{E_{\parallel},i}^{(\parallel)}$ and $C_{E_{\parallel},i}^{(\perp,2)}$ are comparable in magnitude, with $\int C_{E_{\parallel},i}^{(\parallel)} \, d^3v/\int C_{\mathbf{E},i} \, d^3v = 2.57$ and $\int C_{E_{\parallel},i}^{(\perp,2)} \, d^3v/\int C_{\mathbf{E},i} \, d^3v = -1.82$.  The sum of these contributions leads to a net energization of the ions that is positive, leading to the Landau damping of the wave.

{\color{blue}
\subsubsection{Susceptibility-channel Interpretation}

The channel decomposition of \(C_{E_\parallel,i}\) defined in Eq.~\eqref{eq:sepCEpar} can be related directly to susceptibility-channel contributions to the field--particle power. We start with the separation of the perturbed distribution function in Eqs.~\eqref{eq:fdecomptotal}--\eqref{eq:fseparateddefs}. Each term \(\hat f_{1,s}^{(\ell)}\), with \(\ell\in\{\perp,1;\perp,2;\parallel\}\), is the part of the linear distribution-function response driven by the corresponding electric-field component \(\hat E_\ell\). Taking a velocity moment of this same separated response gives the corresponding channel-resolved current,
\begin{equation}
\hat{j}^{(\ell)}_{j,s}
\equiv
q_s\int v_j \hat{f}^{(\ell)}_{1,s}\,d^3v ,
\label{eq:channel_current_def}
\end{equation}
which is related to the total current response via
\begin{equation}
\hat{j}_{j,s}
=
\sum_{\ell}
\hat{j}^{(\ell)}_{j,s}.
\end{equation}
This current contribution is the susceptibility-channel response
\begin{equation}
\hat{j}^{(\ell)}_{j,s}
=
q_s\int v_j \hat{f}^{(\ell)}_{1,s}\,d^3v
=
-\frac{i\omega}{4\pi}\chi_{j\ell,s}\hat{E}_\ell .
\label{eq:channel_current}
\end{equation}
Here, \(j\) labels the current component and \(\ell\) labels the electric-field component that drives that piece of the response.

The same separation carries over to the channel-resolved FPC. Substituting \(\hat f_{1,s}^{(\ell)}\) into Eq.~\eqref{eq:sepCEpar} and integrating over velocity space yields
\begin{eqnarray}
\int C^{(\ell)}_{E_j,s}(\mathbf{v})\,d^3v
&=&
-q_s\frac{1}{4}
\int
\frac{v_j^2}{2}
\left[
\frac{\partial \hat f^{(\ell)}_{1,s}}{\partial v_j}\hat E_j^*
+
\frac{\partial \hat f^{(\ell)*}_{1,s}}{\partial v_j}\hat E_j
\right]d^3v \nonumber \\
&=&
\frac{1}{4}
\left[
\hat E_j^* q_s\int v_j\hat f^{(\ell)}_{1,s}\,d^3v
+
\hat E_j q_s\int v_j\hat f^{(\ell)*}_{1,s}\,d^3v
\right] \nonumber \\
&=&
\frac{1}{4}
\left[
\hat E_j^*\hat j^{(\ell)}_{j,s}
+
\hat E_j\hat j^{(\ell)*}_{j,s}
\right].
\label{eq:fpc_channel_integral}
\end{eqnarray}
We therefore define
\begin{equation}
\int C^{(\ell)}_{E_j,s}(\mathbf{v})\,d^3v
=
\left\langle E_j j^{(\ell)}_{j,s}\right\rangle
\equiv
P^{\rm fp}_{s;j\leftarrow \ell},
\label{eq:fpc_channel_power}
\end{equation}
where \(P^{\rm fp}_{s;j\leftarrow \ell}\) is the power transferred to species \(s\) by the \(j\)-component of the electric field acting on the \(j\)-directed current induced by the \(\ell\)-component of the electric field.

Using Eq.~\eqref{eq:channel_current}, the same power can be written in terms of the susceptibility channel as
\begin{eqnarray}
P^{\rm fp}_{s;j\leftarrow \ell}
&=&
\frac{1}{4}
\left[
\hat{E}_j^*\hat{j}^{(\ell)}_{j,s}
+
\hat{E}_j\hat{j}^{(\ell)*}_{j,s}
\right]_{\omega=\omega_r} \nonumber \\
&=&
\frac{i\omega_r}{16\pi}
\left[
\chi^*_{j\ell,s}\hat{E}_\ell^*\hat{E}_j
-
\chi_{j\ell,s}\hat{E}_\ell\hat{E}_j^*
\right] \nonumber \\
&=&
\frac{\omega_r}{8\pi}
\operatorname{Im}
\left[
\chi_{j\ell,s}\hat{E}_\ell\hat{E}_j^*
\right].
\label{eq:channel_power_susc}
\end{eqnarray}
Here, positive \(P^{\rm fp}_{s;j\leftarrow \ell}\) corresponds to net energy transfer from the fields to species \(s\), while negative \(P^{\rm fp}_{s;j\leftarrow \ell}\) corresponds to energy transfer from the particles to the wave.

For the ion kinetic \Alfven\ wave example, the FPC shown in Fig.~\ref{fig:howes2017compar} is \(C_{E_\parallel,i}\), so \(j=z\). The velocity-space integral of the total signature can therefore be decomposed as
\begin{eqnarray}
\int C_{E_\parallel,i}\,d^3v
&=&
\int
\left[
C^{(\perp,1)}_{E_\parallel,i}
+
C^{(\perp,2)}_{E_\parallel,i}
+
C^{(\parallel)}_{E_\parallel,i}
\right]d^3v \nonumber \\
&=&
P^{\rm fp}_{i;z\leftarrow x}
+
P^{\rm fp}_{i;z\leftarrow y}
+
P^{\rm fp}_{i;z\leftarrow z}.
\label{eq:cepar_channel_sum}
\end{eqnarray}
For the \(n=0\) Landau/transit-time response relevant to the specific kinetic \Alfven\ wave damping example considered here in this section, the \(E_{\perp,1}\)-driven contribution to \(C_{E_\parallel,i}\) is negligible. The relevant contributions to net energization here are therefore the direct 
$E_{\parallel}$-driven Landau channel,
\begin{eqnarray}
P^{\rm fp}_{i;z\leftarrow z}
&=&
\int C_{E_\parallel,i}^{(\parallel)}\,d^3v \nonumber \\
&=&
\frac{\omega_r}{8\pi}
\operatorname{Im}\left[\chi_{zz,i}\right]|\hat{E}_z|^2 ,
\label{eq:landau_channel_power}
\end{eqnarray}
and the off-diagonal $E_{\perp,2}\to j_{\parallel}$ channel,
\begin{eqnarray}
P^{\rm fp}_{i;z\leftarrow y}
&=&
\int C_{E_\parallel,i}^{(\perp,2)}\,d^3v \nonumber \\
&=&
\frac{\omega_r}{8\pi}
\operatorname{Im}
\left[
\chi_{zy,i}\hat{E}_y\hat{E}_z^*
\right].
\label{eq:offdiag_channel_power}
\end{eqnarray}
This latter term is the susceptibility-channel contribution associated with the transit-time-damping-like part of the \(n=0\) response that appears in the parallel FPC. In the notation of Eq.~\eqref{eq:fpc_channel_power},
\begin{equation}
P^{\rm TTD}_{i;z\leftarrow y}
\equiv
P^{\rm fp}_{i;z\leftarrow y}
=
\int C_{E_\parallel,i}^{(\perp,2)}\,d^3v .
\label{eq:ttd_fpc_equiv}
\end{equation}

These channel powers can also be related to contributions to the linear damping or growth rate. Let \(\mathcal{W}\) denote the linear wave energy density of the eigenmode. Since the wave energy evolves as \(d\mathcal{W}/dt=2\gamma\mathcal{W}\), a field--particle power \(P^{\rm fp}_{s;j\leftarrow \ell}\), defined as positive for particle energization, contributes
\begin{equation}
\gamma_{s;j\leftarrow \ell}
=
-\frac{P^{\rm fp}_{s;j\leftarrow \ell}}{2\mathcal{W}},
\label{eq:gamma_channel_def}
\end{equation}
or equivalently
\begin{equation}
-\frac{\gamma_{s;j\leftarrow \ell}}{|\omega_r|}
=
\frac{P^{\rm fp}_{s;j\leftarrow \ell}}
{2\mathcal{W}|\omega_r|}.
\label{eq:gamma_channel_norm}
\end{equation}

For the two dominant ion channels in Fig.~\ref{fig:howes2017compar}, the direct Landau-like contribution is damping-like,
\begin{equation}
-\frac{\gamma_{i,\mathrm{LD}}}{|\omega_r|}
=
\frac{1}{2\mathcal{W}|\omega_r|}
\int C_{E_\parallel,i}^{(\parallel)}\,d^3v
=
\frac{P^{\rm fp}_{i;z\leftarrow z}}
{2\mathcal{W}|\omega_r|},
\label{eq:gamma_ild_fpc}
\end{equation}
whereas the off-diagonal \(E_{\perp,2}\rightarrow j_\parallel\) channel is growth-like for this mode,
\begin{equation}
\frac{\gamma_{i,\mathrm{TTD},z\leftarrow y}}{|\omega_r|}
=
-\frac{1}{2\mathcal{W}|\omega_r|}
\int C_{E_\parallel,i}^{(\perp,2)}\,d^3v
=
-\frac{P^{\rm fp}_{i;z\leftarrow y}}
{2\mathcal{W}|\omega_r|}.
\label{eq:gamma_ittd_fpc}
\end{equation}
In the presented kinetic \Alfven\ wave case, the \(E_\parallel\)-driven Landau channel is damping-like, while the off-diagonal \(E_{\perp,2}\rightarrow j_\parallel\) channel has the opposite sign and \emph{partially} returns energy to the wave. The total ion contribution remains damping-like because the direct Landau channel is larger in magnitude than the off-diagonal growth-like contribution.

\begin{figure}
\begin{center}
   \includegraphics[width=.45\textwidth]{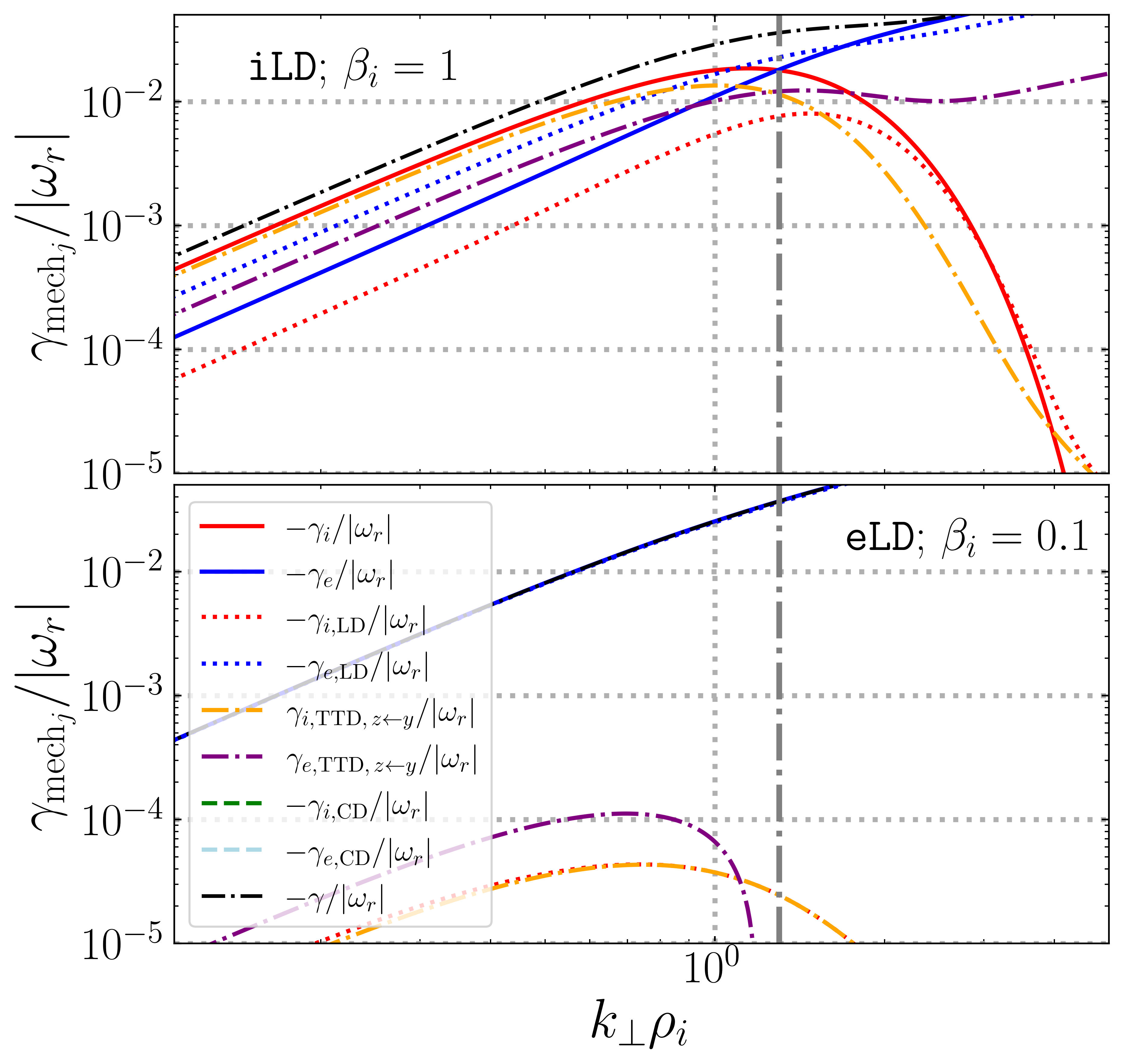}
  \end{center}
\caption{{\color{blue}Comparison of selected normalized damping- and growth-rate contributions inferred from susceptibility-channel powers. The upper panel corresponds to the ion Landau damping case with \(\beta_i=1\) shown in Fig.~\ref{fig:howes2017compar}, and the lower panel corresponds to the electron Landau damping case with \(\beta_i=0.1\) shown in Fig.~\ref{fig:ekdeleccomp}. The black dash-dotted curve shows the total normalized damping rate, \(-\gamma/|\omega_r|\). The red and blue solid curves show the total ion and electron contributions, \(-\gamma_i/|\omega_r|\) and \(-\gamma_e/|\omega_r|\). The dotted curves show the Landau-like contributions, \(-\gamma_{i,\mathrm{LD}}/|\omega_r|\) and \(-\gamma_{e,\mathrm{LD}}/|\omega_r|\), the dashed curves show the \(n=1\) cyclotron damping contributions, \(-\gamma_{i,\mathrm{CD}}/|\omega_r|\) and \(-\gamma_{e,\mathrm{CD}}/|\omega_r|\), and the dot-dashed orange and purple curves show the growth-like off-diagonal transit-time-damping-associated contributions, \(\gamma_{i,\mathrm{TTD},z\leftarrow y}/|\omega_r|\) and \(\gamma_{e,\mathrm{TTD},z\leftarrow y}/|\omega_r|\). The vertical gray line marks \(k_\perp\rho_i=1.3\), corresponding to the kinetic \Alfven\ wave comparison in Fig.~\ref{fig:howes2017compar}. In the lower panel, the electron Landau damping contribution, the total electron contribution, and total wave damping-rate curves are nearly coincident.}}
   \label{fig:ratecomparison}
\end{figure}

Figure~\ref{fig:ratecomparison} provides the corresponding rate-level view of the same channel decomposition. At \(k_\perp\rho_i=1.3\), the off-diagonal \(E_{\perp,2}\rightarrow j_\parallel\) contribution has the opposite sign from the direct Landau channel, so it reduces the net ion damping rate but does not overcome it. This quantitative comparison supports the interpretation of the ``twist'' in Fig.~\ref{fig:howes2017compar}. The reversal of the bipolar structure at larger \(v_\perp\) is produced by a growth-like off-diagonal susceptibility channel superposed with a stronger damping-like direct Landau channel, which preferentially energizes particles with a higher perpendicular velocity.

The apparent reversal for particular parameter regimes modifies but does not discredit the usual Landau-damping picture. The direct \(E_\parallel\)-driven channel, \(C_{E_\parallel,i}^{(\parallel)}\), retains the familiar Landau-damping structure. Particles on one side of \(v_\parallel=\omega_r/k_\parallel\) gain energy while particles on the other side lose energy, producing net damping. The ``twist'' appears only in the summed \(C_{E_\parallel,i}\) after the off-diagonal \(E_{\perp,2}\rightarrow j_\parallel\) response is included. This off-diagonal contribution is part of the same linear kinetic \Alfven\ wave eigenmode, but it is not the direct parallel Landau response. Instead, it represents parallel electric-field work on a parallel current contribution driven by the perpendicular wave polarization. Around \(v_\perp/w_i\sim1\), the direct Landau and off-diagonal contributions become comparable and partially cancel. At larger \(v_\perp\), the off-diagonal contribution becomes more prominent because of the finite-Larmor-radius dependence through \(J_0'(b_i)\) from Equation~\ref{eq:f1a}, with \(b_i=k_\perp v_\perp/\Omega_i\).
}


\subsection{Electron Landau Damping}
\label{sec:ekldcomp}

\begin{figure}
 \begin{center}
    \includegraphics[width=.45\textwidth]{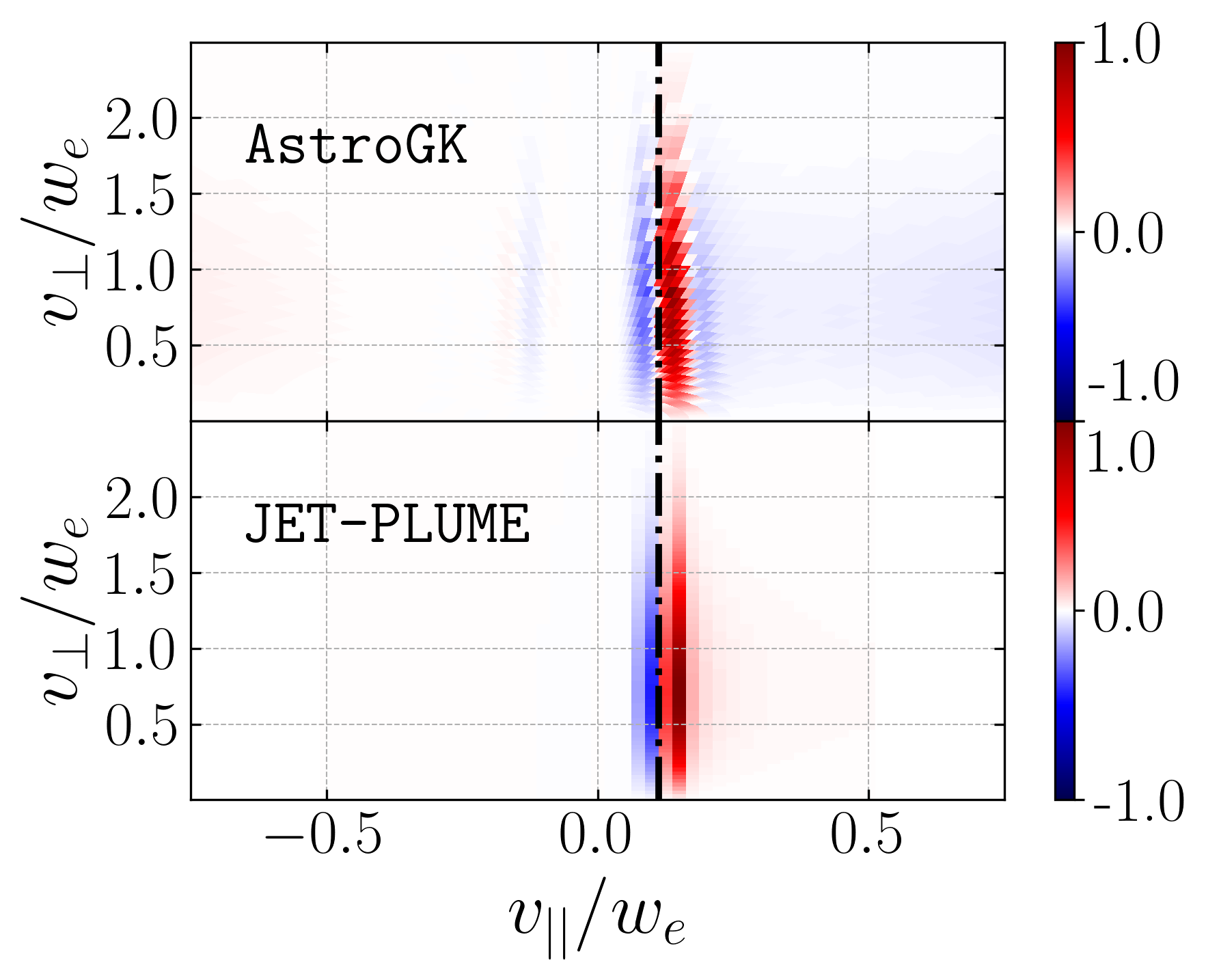}
   \end{center}
 \caption{Comparison between \textit{gyrotropic velocity-space signatures} for the electrons, $C_{E_{\parallel},e}(v_{\parallel},v_\perp)$, of the electron Landau damping computed from the \texttt{AstroGK} simulation of a kinetic \Alfven\ wave with $\beta_i=0.1$ (\emph{Top}), and computed using \texttt{JET-PLUME} (\emph{Bottom}) using the parallel correlation, $C_{E_{\parallel}}$. Both figures are separately normalized so that $\max|C_{E_{\parallel},e}(v_{\parallel},v_\perp)| = 1$. The vertical black dash-dotted line indicates the resonant parallel velocity, $\omega_r/k_\parallel$.}
    \label{fig:ekdeleccomp}
\end{figure}

{We repeat the simulation of Landau damping from Section~\ref{sec:ikldcomp}, {holding all parameters and wavevectors fixed except for a reduced plasma beta of $\beta_i = 0.1$}. The corresponding eigenmodes, computed by \texttt{PLUME}, are listed in Table~\ref{tab:ekld}. 
Lowering $\beta_i$ diminishes the net ion contribution to power exchange, leaving it negligible compared to the dominant electron Landau resonance. For the parameters in this case, only one mechanism transfers a significant amount of energy from the wave to the particles through work by the parallel component of the electric field: electron Landau damping. We compute the  electron Landau damping signature in the \texttt{AstroGK} simulation using the FPC. 
Specifically, we compute the parallel correlation over three wave periods in the simulation and compare it to the prediction for the same mode obtained with \texttt{JET-PLUME} in Figure~\ref{fig:ekdeleccomp}.}

\begin{table}
 \begin{center}
 \begin{tabular}{|l||c|c|c|}
    \hline
    & $\hat{E}_{\perp,1}/\hat{E}_{\perp,1}$ & $\hat{E}_{\perp,2}/\hat{E}_{\perp,1}$ & $\hat{E}_\parallel/\hat{E}_{\perp,1}$ \\
    \hline
    Re. & $1.0$ & $9.25 \times 10^{-6}$ & $-3.59 \times 10^{-4}$ \\
    Im. & $0$ & $3.24 \times 10^{-4}$ & $8.26 \times 10^{-5}$ \\
    \hline
    & $\hat{B}_{\perp,1}/\hat{E}_{\perp,1}$ & $\hat{B}_{\perp,2}/\hat{E}_{\perp,1}$ & $\hat{B}_\parallel/\hat{E}_{\perp,1}$ \\
    \hline
    Re. & $5.57 \times 10^{-3}$ & $3.03 \times 10^{3}$ & $-7.25$ \\
    Im. & $-6.7 \times 10^{-1}$ & $-1.1 \times 10^{2}$ & $8.71 \times 10^{2}$ \\
    \hline
    & $\hat{U}_{\perp,1,i}/(c\hat{E}_{\perp,1}/B_0)$ & $\hat{U}_{\perp,2,i}/(c\hat{E}_{\perp,1}/B_0)$ & $\hat{U}_{\parallel,i}/(c\hat{E}_{\perp,1}/B_0)$ \\
    \hline
    Re. & $-1.01 \times 10^{-4}$ & $-3.44 \times 10^{-1}$ & $-7.74 \times 10^{-3}$ \\
    Im. & $-2.71 \times 10^{-3}$ & $-8.62 \times 10^{-4}$ & $-3.78 \times 10^{-2}$ \\
    \hline
    & $\hat{U}_{\perp,1,e}/(c\hat{E}_{\perp,1}/B_0)$ & $\hat{U}_{\perp,2,e}/(c\hat{E}_{\perp,1}/B_0)$ & $\hat{U}_{\parallel,e}/(c\hat{E}_{\perp,1}/B_0)$ \\
    \hline
    Re. & $9.34 \times 10^{-6}$ & $-1.48$ & $-1.51 \times 10^{-1}$ \\
    Im. & $3.27 \times 10^{-4}$ & $-1.03 \times 10^{-2}$ & $-3.98$ \\
    \hline
    & $\hat{n}_{i}/(n_{0,i}\hat{E}_{\perp,1}/B_0)$ & $\hat{n}_{e}/(n_{0,e}\hat{E}_{\perp,1}/B_0)$ &  $\omega/\Omega_R$\\
    \hline
    Re. & $-1.54 \times 10^{1}$ & $-1.54 \times 10^{1}$ & $4.84 \times 10^{-3}$ \\
    Im. & $-7.35 \times 10^{3}$ & $-7.35 \times 10^{3}$ & $-1.78 \times 10^{-4}$ \\
    \hline
 \end{tabular}
 \end{center}
 \caption{Selected eigenfunctions for the electron Landau damping of the kinetic \Alfven\ wave described in Section \ref{sec:ekldcomp} using \texttt{PLUME}.}
 \label{tab:ekld}
\end{table}


{\color{blue}
Figure~\ref{fig:ekdeleccomp} shows strong agreement between the resonant electron Landau damping signature predicted by \texttt{JET-PLUME} and that computed from the \texttt{AstroGK} simulation. Both exhibit a bipolar structure centered near \(v_\parallel=\omega_r/k_\parallel\), with net positive energy transfer to the electrons. The weaker off-resonant structures in the simulation are not reproduced by the single-mode calculation and are likely residual oscillatory or finite-window contributions rather than part of the secular Landau damping signature. Unlike the ion kinetic \Alfven\ wave case, no comparable twist is expected here because \(k_\perp\rho_e\ll1\), which suppresses the \(J_0'(b_e)\) off-diagonal \(E_{\perp,2}\rightarrow j_\parallel\) response, leaving the direct \(E_\parallel\rightarrow j_\parallel\) Landau response dominant.
}


\subsection{Weibel Instability}
\label{sec:weibelcomp}

To compare the results of \texttt{JET-PLUME} to a simulation for an unstable mode, we use the Vlasov-Maxwell solver in the \texttt{Gkeyll} framework \citep{juno2018discontinuous,Hakim:2020b} to examine the Weibel instability, performing a simulation based on the “low-temperature” Weibel instability from \citet{cagas2017nonlinear}\footnote{Specifically, we perform a simulation similar to the `low temperature' simulation in \emph{Section 4}.}. \texttt{Gkeyll} uses the discontinuous Galerkin finite element method to directly discretize the Vlasov-Maxwell system on a phase-space grid. The simulation is performed in 2D-2V (two spatial and two velocity-space dimensions) with both ions and electrons. We employ quadratic Serendipity basis elements \citep{Arnold:2011} with $n_\mathbf{x} = 32^2$ spatial elements and $n_\mathbf{v} = 64^2$ velocity-space elements for the discontinuous Galerkin expansion.

The simulation is performed by initializing two counter-propagating electron beams with drift velocities of $\pm 5 w_{e}$ in the $\hat{e}_{\parallel}$ direction and a Maxwellian background ion population (Note that the definition of thermal velocity is simulation-dependent. Because \texttt{Gkeyll} defines thermal velocity as $v_{th} = \sqrt{T/m}$ while \texttt{PLUME} uses $w = \sqrt{2T/m}$, the simulated drift of 5 in \texttt{Gkeyll} corresponds to a numerical value of $5/\sqrt{2} \approx 3.54$ in \texttt{JET-PLUME} to ensure identical physical parameters.) The configuration space domain size is $L_{\perp,1} = L_{\parallel} = 6\pi/|\mathbf{k}|$. The velocity-space domain extends from $-20 w_{i}$ to $+20 w_{i}$ for ions, and from $-50 \, w_{e}$ to $+50 \, w_{e}$ for electrons, in both $v_{\perp,1}$ and $v_{\parallel}$. We use a realistic mass ratio of $m_i/m_e = 1836$, and assume equal temperatures, $T_i = T_e$, for both the ion background population and the two electron beams. The plasma beta is set to $\beta_e = \beta_i = 0.05$, with an in-plane background magnetic field oriented in the $\hat{e}_{\parallel}$ direction of the simulation. This setup excites both the forward and backward branches of the mode, since perturbing only $E_{\parallel}$ necessarily drives fluctuations at both $+\mathbf{k}$ and $-\mathbf{k}$.

{The parameters used by \texttt{JET-PLUME} are as follows. We use a proton-electron plasma with equal total number densities ($n_i = n_e$) and a realistic mass ratio $m_i/m_e = 1836$. The ion beta is $\beta_i = 0.05$. Temperatures are isotropic for all species, $(T_\perp/T_\parallel)_s = 1$, with an ion-electron temperature ratio $T_i/T_e = 1$. The ion thermal speed is specified directly as $v_{ti}/c = 1.0437\times 10^{-3}$. Wavenumbers are normalized to the ion gyroradius; in this case we take $k_\perp \rho_i = 3.8325$ and $k_\parallel \rho_i = 0.10$. The electron population is represented by two counter-streaming subpopulations of equal density, each representing $1/2$ of the total electron density, and equal temperature. The electrons are drifting parallel and antiparallel to the background field with equal magnitude and opposite directions, specifically with drift velocities of $\pm 3.54 w_{e}$ (equivalent to $\pm 47.91 v_{AR}$), corresponding to the $\pm 5 v_{th,e}$ drift in \texttt{Gkeyll} once the differing thermal velocity conventions are taken into account.
}

We note three key differences from the `low-temperature' simulation presented in \citet{cagas2017nonlinear}. First, we introduce an external guiding magnetic field, which is necessary to match the assumptions of \texttt{JET-PLUME}, as it requires the presence of such a field. While this guiding field slightly reduces the growth rate of the instability \citep{bret2011robustness}, it does not suppress it in this regime \citep{grassi2017electron,stockem2007relativistic,stockem2008suppression}. Second, we simulate the instability in 2D-2V. Ideally, we would perform the simulation in full 3D-3V, but this is not computationally feasible at present. However, for the parameters of this system, \texttt{PLUME} predicts weaker relative growth in the out-of-plane components of both the electric field and bulk velocities, as shown in Table~\ref{tab:weibeltable}\footnote{In this particular setup, the determinant of the dispersion tensor is very shallow surrounding the fast-growing entropy solution and remains approximately zero along the line $\omega = 0$ for sufficiently large $\gamma$. Many nearby large-$\gamma$ values along this line appear nearly equivalent solutions, making automated solvers unreliable. In practice, we therefore insert the approximate $\gamma$ observed in the simulation, which yields eigenmodes consistent with those across all order-unity values of $\gamma$ observed in the simulation for this case.}. This result suggests that restricting the simulation to 2D-2V likely has little impact on the behavior of the system. Third, we introduce a perturbation in the electric field in the simulation of the form $\mathbf{E}(x_{\perp,1}) = 1 \times 10^{-3} \sqrt{4 \pi} \frac{q_e n_0 d_e }{\epsilon_0} \sin(k_{\perp} x_{\perp,1}) \, \hat{e}_{\parallel}$, with $k_\perp \rho_i = 3.83$\footnote{Even with the discontinuous Galerkin discretization, finite-grid effects cause the nominally perpendicular mode to acquire a weak parallel component. We therefore set $k_{\parallel}\rho_i = 0.1$ within \texttt{PLUME}.}, where $d_e$ is the electron inertial length and $\epsilon_0$ is the permittivity of free space, to drive the growth of a single wavenumber.

{This perturbation excites unstable entropy modes in the system. Because of a degeneracy, the initial perturbation drives two entropy modes simultaneously. We present both sets of associated eigenmodes in Table~\ref{tab:weibeltable}: one corresponding to the mode with the smaller growth rate, $\gamma/\Omega_{R} = 39.4$, hereafter referred to as the slow-growing mode, and one corresponding to the mode with the larger growth rate, $\gamma/\Omega_{R} = 3750$, hereafter referred to as the fast-growing mode. We argue that both modes are present for the following reasons.

Evidence for the slow-growing mode lies in its stronger coupling to the parallel velocity of the particles, which drives the observed $\hat{U}_{\parallel,e1}$ and $\hat{U}_{\parallel,e2}$ eigenmodes (that is, the eigenmodes of the two streaming electron populations, $e1$ and $e2$) with a relative phase shift from the parallel electric field. This is visible directly in the system; equivalently, the characteristic blue-red-blue `tripolar' structure of both electron populations, in Figure~\ref{fig:weibelelecstackcomp}, indicates oscillatory motion~\citep{brown2022isolation} along the parallel axis, consistent with the orientation of the signature. In contrast, the fast-growing mode produces only a tiny in-phase perturbation, which is not observed in the system.
 
The fast-growing mode amplifies the fields more efficiently, yielding a growth rate in agreement with \citet{cagas2017nonlinear} (noting that the magnetic field minimally alters the growth rate for this system~\citep{bret2011robustness}), rather than the nearly two orders of magnitude slower rate of the slow-growing mode.  

{In principle, one would prefer to excite a single mode in the simulation. In practice, this is difficult because any initial perturbation decomposes into the system’s eigenmodes, and the fastest-growing mode typically dominates the evolution even if weaker modes are initially present~\citep{gary1993theory}. Moreover, many eigenmodes of the $+k$ slow-growing branch have field components (e.g., $E_{\parallel}$) that are nearly in phase with those of the $+k$ and $-k$ fast modes. As a result, perturbations intended to drive the slow mode can also excite the fast mode, making it challenging to isolate the slow-growing branch for comparison with \texttt{JET-PLUME}. In principle, one might suppress the fast-growing mode by introducing density perturbations aligned only with the slow-growing mode, but we do not pursue that approach here. }

{For the present comparison, it is sufficient to note that the slow-growing mode exhibits a much larger in magnitude $\hat{U}_{\parallel,s}$ response for all species, resulting in $j_{\parallel,s}E_{\parallel}$ being generated almost exclusively by the slow-growing mode for both counter-streaming electrons and the background ions.} Likewise, $\partial \hat{f}_{1,s}/\partial v_j$ is much larger for the slow-growing mode, producing the dominant contribution to $C_{E_{\parallel},s}$. This coupling allows the slow-growing mode to dominate the observed signature, and it is therefore the focus of the comparison in this section.
}
\begin{table*}
 \begin{center}
 \small
 \setlength{\tabcolsep}{6pt}
 \renewcommand{\arraystretch}{1.1}
\resizebox{\textwidth}{!}{%
\begin{tabular}{|l||c|c|c||c|c|c|}
    \hline
    & \multicolumn{3}{c||}{\textbf{slow-growing entropy mode}} & \multicolumn{3}{c|}{\textbf{fast-growing entropy mode}} \\
    \hline
    & $\hat{E}_{\perp,1}/\hat{E}_{\perp,1}$ & $\hat{E}_{\perp,2}/\hat{E}_{\perp,1}$ & $\hat{E}_\parallel/\hat{E}_{\perp,1}$
    & $\hat{E}_{\perp,1}/\hat{E}_{\perp,1}$ & $\hat{E}_{\perp,2}/\hat{E}_{\perp,1}$ & $\hat{E}_\parallel/\hat{E}_{\perp,1}$ \\
    \hline
    Re. & $1.0$ & $-1.32 \times 10^{-1}$ & $-7.16 \times 10^{-1}$
        & $1.0$ & $-3.47 \times 10^{-1}$ & $4.63 \times 10^{-3}$ \\
    Im. & $0$ & $3.02 \times 10^{-16}$ & $-8.62 \times 10^{-13}$
        & $0$ & $6.00 \times 10^{-18}$ & $3.28 \times 10^{-19}$ \\
    \hline
    & $\hat{B}_{\perp,1}/\hat{E}_{\perp,1}$ & $\hat{B}_{\perp,2}/\hat{E}_{\perp,1}$ & $\hat{B}_\parallel/\hat{E}_{\perp,1}$
    & $\hat{B}_{\perp,1}/\hat{E}_{\perp,1}$ & $\hat{B}_{\perp,2}/\hat{E}_{\perp,1}$ & $\hat{B}_\parallel/\hat{E}_{\perp,1}$ \\
    \hline
    Re. & $1.73 \times 10^{-14}$ & $8.41 \times 10^{-11}$ & $-6.63 \times 10^{-13}$
        & $-1.53 \times 10^{-19}$ & $-3.21 \times 10^{-19}$ & $5.87 \times 10^{-18}$ \\
    Im. & $-3.22 \times 10^{-1}$ & $-6.9 \times 10^{1}$ & $1.23 \times 10^{1}$
        & $-8.87 \times 10^{-3}$ & $-2.10 \times 10^{-2}$ & $3.40 \times 10^{-1}$ \\
    \hline
    & $\hat{U}_{\perp,1,i}/(c\hat{E}_{\perp,1}/B_0)$ & $\hat{U}_{\perp,2,i}/(c\hat{E}_{\perp,1}/B_0)$ & $\hat{U}_{\parallel,i}/(c\hat{E}_{\perp,1}/B_0)$
    & $\hat{U}_{\perp,1,i}/(c\hat{E}_{\perp,1}/B_0)$ & $\hat{U}_{\perp,2,i}/(c\hat{E}_{\perp,1}/B_0)$ & $\hat{U}_{\parallel,i}/(c\hat{E}_{\perp,1}/B_0)$ \\
    \hline
    Re. & $2.49 \times 10^{-2}$ & $-3.96 \times 10^{-3}$ & $-1.81 \times 10^{-2}$
        & $2.67 \times 10^{-4}$ & $4.27 \times 10^{-25}$ & $-9.26 \times 10^{-5}$ \\
    Im. & $1.42 \times 10^{-15}$ & $-2.47 \times 10^{-16}$ & $-3.04 \times 10^{-14}$
        & $1.60 \times 10^{-21}$ & $1.23 \times 10^{-6}$ & $8.74 \times 10^{-23}$ \\
    \hline
    & $\hat{U}_{\perp,1,e1}/(c\hat{E}_{\perp,1}/B_0)$ & $\hat{U}_{\perp,2,e1}/(c\hat{E}_{\perp,1}/B_0)$ & $\hat{U}_{\parallel,e1}/(c\hat{E}_{\perp,1}/B_0)$
    & $\hat{U}_{\perp,1,e1}/(c\hat{E}_{\perp,1}/B_0)$ & $\hat{U}_{\perp,2,e1}/(c\hat{E}_{\perp,1}/B_0)$ & $\hat{U}_{\parallel,e1}/(c\hat{E}_{\perp,1}/B_0)$ \\
    \hline
    Re. & $1.7 \times 10^{-1}$ & $-9.91 \times 10^{-1}$ & $2.18 \times 10^{1}$
        & $-4.61 \times 10^{-1}$ & $-3.24 \times 10^{-4}$ & $-5.57 \times 10^{-2}$ \\
    Im. & $-5.55 \times 10^{-1}$ & $-2.05 \times 10^{1}$ & $-1.55 \times 10^{1}$
        & $1.57 \times 10^{-3}$ & $-2.25 \times 10^{-3}$ & $1.71 \times 10^{-5}$ \\
    \hline
    & $\hat{U}_{\perp,1,e2}/(c\hat{E}_{\perp,1}/B_0)$ & $\hat{U}_{\perp,2,e2}/(c\hat{E}_{\perp,1}/B_0)$ & $\hat{U}_{\parallel,e2}/(c\hat{E}_{\perp,1}/B_0)$
    & $\hat{U}_{\perp,1,e2}/(c\hat{E}_{\perp,1}/B_0)$ & $\hat{U}_{\perp,2,e2}/(c\hat{E}_{\perp,1}/B_0)$ & $\hat{U}_{\parallel,e2}/(c\hat{E}_{\perp,1}/B_0)$ \\
    \hline
    Re. & $1.7 \times 10^{-1}$ & $-9.91 \times 10^{-1}$ & $2.18 \times 10^{1}$
        & $-4.61 \times 10^{-1}$ & $3.24 \times 10^{-4}$ & $-5.57 \times 10^{-2}$ \\
    Im. & $5.55 \times 10^{-1}$ & $2.05 \times 10^{1}$ & $1.55 \times 10^{1}$
        & $-1.57 \times 10^{-3}$ & $-2.25 \times 10^{-3}$ & $-1.71 \times 10^{-5}$ \\
    \hline
    & $\hat{n}_{i}/(n_{0,i}\hat{E}_{\perp,1}/B_0)$ & $\hat{n}_{e1}/(n_{0,e1}\hat{E}_{\perp,1}/B_0)$ & $\hat{n}_{e2}/(n_{0,e2}\hat{E}_{\perp,1}/B_0)$
    & $\hat{n}_{i}/(n_{0,i}\hat{E}_{\perp,1}/B_0)$ & $\hat{n}_{e1}/(n_{0,e1}\hat{E}_{\perp,1}/B_0)$ & $\hat{n}_{e2}/(n_{0,e2}\hat{E}_{\perp,1}/B_0)$ \\
    \hline
    Re. & $1.86 \times 10^{-13}$ & $-9.11 \times 10^{1}$ & $9.11 \times 10^{1}$
        & $2.65 \times 10^{-24}$ & $3.15 \times 10^{-3}$ & $-3.15 \times 10^{-3}$ \\
    Im. & $-2.27$ & $-2.35$ & $-2.35$
        & $-2.61 \times 10^{-4}$ & $4.52 \times 10^{-1}$ & $4.52 \times 10^{-1}$ \\
    \hline
    & \multicolumn{3}{c||}{$\omega/\Omega_R$} & \multicolumn{3}{c|}{$\omega/\Omega_R$} \\
    \hline
    Re. & \multicolumn{3}{c||}{$2.21 \times 10^{-12}$} & \multicolumn{3}{c|}{$0$} \\
    Im. & \multicolumn{3}{c||}{$3.94 \times 10^{1}$} & \multicolumn{3}{c|}{$3.75 \times 10^{3}$} \\
    \hline
 \end{tabular}%
}
 \end{center}
 \caption{Selected eigenfunctions for the $+k_{\perp}$ entropy modes from a single simulation using \texttt{PLUME}. The left block shows the slow-growing mode and the right block the fast-growing mode. }
 \label{tab:weibeltable}
\end{table*}

{Using the simulation, we compute the instantaneous correlation with the parallel component of the electric field at various frames during the interval 
$t \in [60 \, \omega_{p,e}^{-1}, \, 75.03 \, \omega_{p,e}^{-1}]$, 
using the full spatial domain but only the perturbed portion of the distribution function and applying a band-pass filter in space at fixed time to isolate the wave of interest:
\begin{equation}
C_{E_j,s}^{(k)}(\mathbf{v},t)
= \Big\langle -\,q_s \tfrac{v_j^2}{2}
\frac{\partial f_s^{(k)}(\mathbf{x},\mathbf{v},t)}{\partial v_j}
E_j^{(k)}(\mathbf{x},t)
\Big\rangle_{\mathbf{x}},
\end{equation}
where the superscript $(k)$ denotes the contribution from a single Fourier mode. 
Explicitly,
\begin{equation}
A^{(k)}(\mathbf{x},\ldots) \equiv 
\hat{A}(\mathbf{k},\ldots)\,e^{i\mathbf{k}\cdot\mathbf{x}}
+ \hat{A}^*(\mathbf{k},\ldots)\,e^{-i\mathbf{k}\cdot\mathbf{x}},
\end{equation}
with $\hat{A}(\mathbf{k})=\langle A(\mathbf{x})\,e^{-i\mathbf{k}\cdot\mathbf{x}}\rangle_{\mathbf{x}}$, which is the Fourier transform of $A$ at mode $\mathbf{k}$. 
In other words, $A^{(k)}$ is obtained by projecting $A$ onto the $(\mathbf{k},-\mathbf{k})$ pair and reconstructing only that contribution, so that this operation acts as a filter which isolates a single wavenumber mode from the full field.
This procedure isolates the selected mode from the background fluctuations and other channels of energy exchange. 
Note that this expression is equivalent to a direct computation of \eqref{eq:ceipredict} for a single mode, except the correlation is performed in real space rather than frequency space.}

{In Figure~\ref{fig:weibelelecstackcomp}, we compare the electron
velocity–space signatures obtained from \texttt{Gkeyll} \emph{(top three panels)} with
the total predictions from \texttt{JET-PLUME} \emph{(bottom)}, which include contributions from all modes, but again note that the contribution from the fast-growing mode (including cross terms) is negligible. For the total electron signatures computed with \texttt{JET-PLUME} in Figure~\ref{fig:weibelelecstackcomp} \emph{(bottom)}, we emphasize the need to carefully compute the linear superposition of the $+k$ and $-k$ modes using the expression derived in Appendix~\ref{app:linsupapp}, which uniquely arises for degenerate modes. Here, we do not isolate the $+k$ and $-k$ modes, as their signatures are practically identical.

During the early
linear growth phase of the Weibel instability (e.g., $t = 60.0\,\omega_{p,e}^{-1}$),
we find strong agreement between the simulation and analytic prediction. At
later times (e.g., $t = 67.52\,\omega_{p,e}^{-1}$ and $t = 75.03\,\omega_{p,e}^{-1}$), the
agreement degrades, reflecting the nonlinear evolution of the perturbed electron
velocity distribution, as expected given that the analytic calculation is linear. {\color{blue}
Note that for unstable modes, the \texttt{JET-PLUME} prediction should be interpreted as the velocity-space energization signature of the linear eigenmode on the specified initial particle velocity distribution. The complex frequency obtained by \texttt{PLUME} determines whether the mode is damped or growing through the sign of \(\gamma=\mathrm{Im}(\omega)\). Stability is therefore not assumed a priori for a specified particle velocity distribution and wavevector. In the simulation, as the unstable mode grows, nonlinear saturation and quasilinear modification of the distribution function eventually invalidate the fixed-equilibrium linear response assumed in the analytic theory used here. Agreement with simulation is expected only during the linear growth phase, before the background distribution has been significantly modified.
} 

We also note a slight tilt in the simulated signature that is not captured by the linear calculation. We interpret this as a small contribution from finite $k_\parallel$ coupling appearing even in a nominally $k_\perp$ mode. Such a contribution may arise from a physical nonlinear process, from the restrictions of the 2D2V configuration, or from the assumption that $\gamma/\omega_r$ is sufficiently small (see Appendix~\ref{appendix:linFPCderiv}). Further investigation is warranted but lies beyond the scope of this work, as the overall agreement between the simulated and calculated signatures from \texttt{JET-PLUME} is sufficient for the core purposes of the FPC: measuring phase-space energy transfer and identifying mechanisms \emph{in situ}.

\begin{figure}
 \begin{center}
    \includegraphics[width=.45\textwidth]{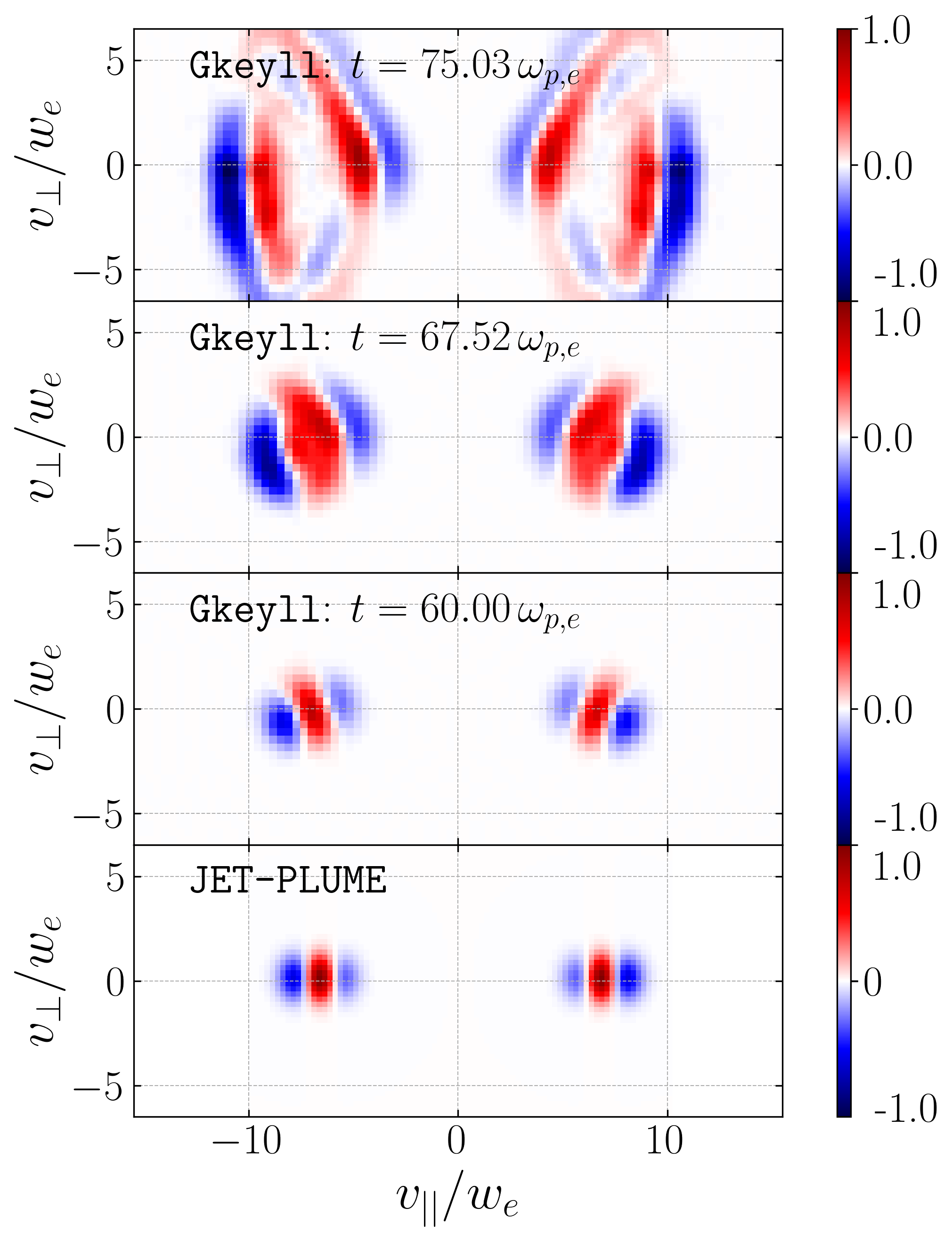}
   \end{center}
 \caption{Comparison between \textit{velocity-space signatures} of the electrons in a \texttt{Gkeyll} simulation at times $t=60.0 \omega_{p,e}^{-1}$, $t=67.52 \omega_{p,e}^{-1}$, and $t=75.03 \omega_{p,e}^{-1}$ (\emph{Top three panels}), and total prediction using \texttt{JET-PLUME} (\emph{Bottom}) of the Weibel instability with guiding field using the parallel correlation, $C_{E_{\parallel},e}$. As the \emph{fast-growing entropy} mode's contribution (including cross-terms) is weak, we do not plot separately the \emph{slow-growing entropy} mode, as it is practically equivalent to the total signature. The simulation and calculation using \texttt{JET-PLUME} are normalized separately so that $\max|C_{E_{\parallel},e}(v_{\parallel},v_\perp)| = 1$. Note that the simulation is two-dimensional, so “$\perp$” refers only to the in-plane perpendicular direction and does not imply any gyro-averaging.}
    \label{fig:weibelelecstackcomp}
\end{figure}
}

\begin{figure}
 \begin{center}
    \includegraphics[width=.45\textwidth]{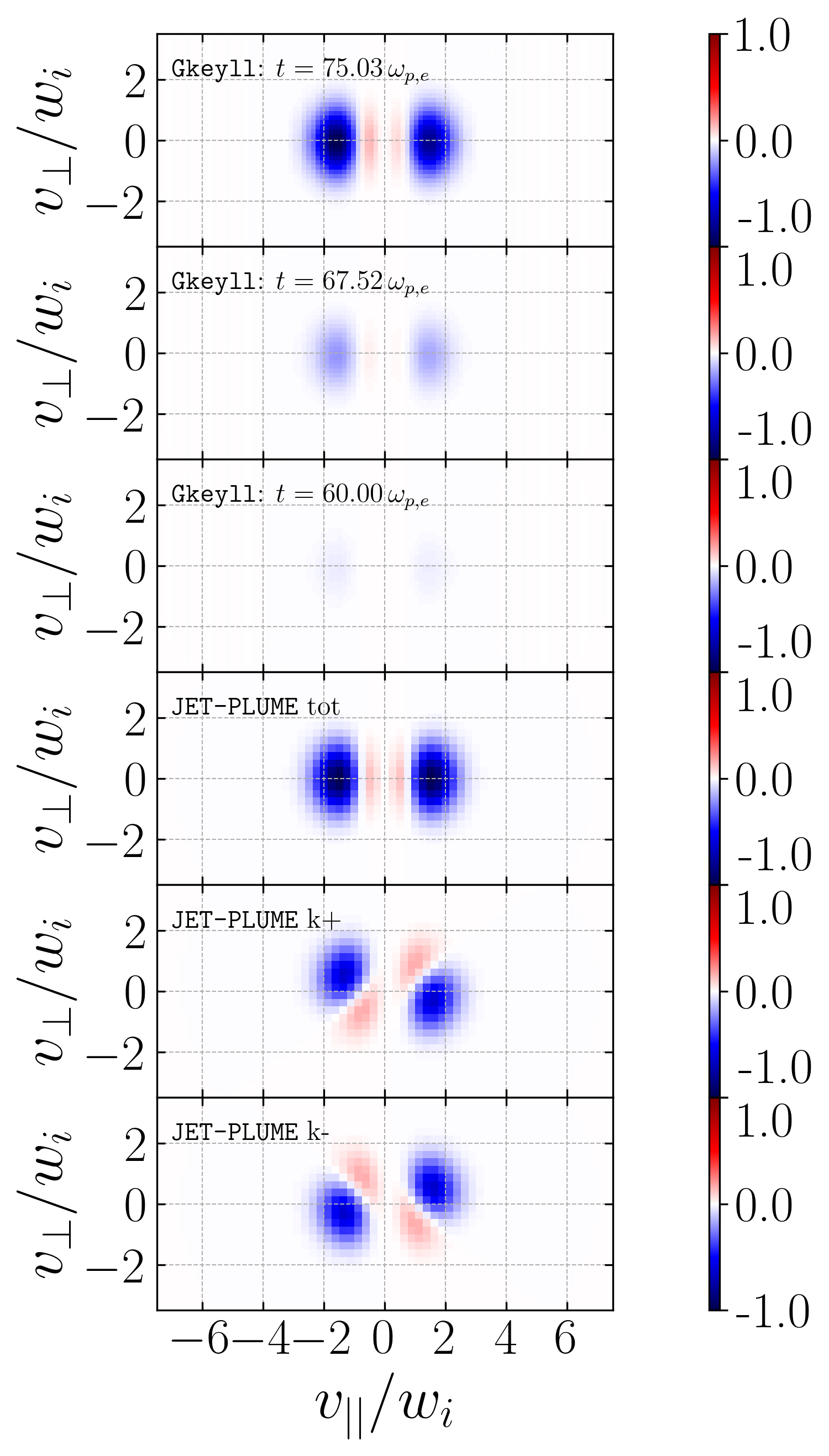}
   \end{center}
 \caption{Comparison between \textit{velocity-space signatures} of the ions in a \texttt{Gkeyll} simulation at times $t=60.0 \omega_{p,e}^{-1}$, $t=67.52 \omega_{p,e}^{-1}$, and $t=75.03 \omega_{p,e}^{-1}$ (\emph{Top three panels}), and prediction using \texttt{JET-PLUME} (\emph{Bottom three panels}) of the Weibel instability with guiding field using the parallel correlation, $C_{E_{\parallel},i}$. The simulation and calculation using \texttt{JET-PLUME} are normalized separately so that $\max|C_{E_{\parallel},i}(v_{\parallel},v_\perp)| = 1$. Note that the simulation is two-dimensional, so “$\perp$” refers only to the in-plane perpendicular direction and does not imply any gyro-averaging.}
    \label{fig:weibelionstackcomp}
\end{figure}

Figure~\ref{fig:weibelionstackcomp} compares the predicted and simulated ion energy exchange. 
The figure illustrates the development of the parallel velocity-space signature of energy transfer between the electric field and the ions during the growth of the Weibel instability with an external magnetic guide field. 
We plot both the total signature, which combines contributions from the $+k$ and $-k$ modes, as well as the signatures of each mode individually. Again, the fast-growing mode produces negligible contributions to the signature relative to the slow-growing mode, and is therefore treated as absent in the following discussion. We find strong agreement between the predicted total signature and the observed signatures at all times during the linear growth of the instability. In the simulation, the shape of the signature remains constant in this time range, but its amplitude increases as the wave grows. This effect may appear visually exaggerated due to the normalization to $\max|C_{E_{\parallel},i}(v_{\parallel},v_\perp)|$ at $t=75.03 \omega_{p,e}^{-1}$ across all times presented.

In the bottom two panels of Figure~\ref{fig:weibelionstackcomp}, we plot the $+k$ and $-k$ modes separately. The signatures of the isolated modes display a clover-like pattern, similar to that reported in \citet{Afshari:2024}, which likely arises from the phase difference between $\hat{E}_{\parallel}$ and $\hat{U}_{\parallel,i}$ listed in Table~\ref{tab:weibeltable}. Without separation, one might mistakenly infer that $\hat{\mathbf{U}}_{i}$ is linearly polarized and corresponds to a single non-resonant mode, resembling the behavior and associated signatures of non-resonant features described in \citet{conley2023characterizing}. The more accurate interpretation is that the observed response reflects the superposition of two oppositely directed, circularly polarized bulk flows of the plasma, which cancel when combined.


\subsection{Ion Cyclotron Damping}
\label{sec:icdcomp}
We again employ the Vlasov–Maxwell solver in the \texttt{Gkeyll} framework \citep{juno2018discontinuous,Hakim:2020b} to examine the ion cyclotron damping of an \Alfven-mode wave. The simulations are performed in 1D-3V (one spatial and three velocity-space dimensions) with protons and electrons, using $n_{\text{cells}} = 16$ spatial and $n_v = 16^3$ cells for each species. We again employ quadratic Serendipity basis elements \citep{Arnold:2011} for the discontinuous Galerkin expansion. In configuration space, we use $L = 2\pi / k$, where $k = \sqrt{k_\perp^2 + k_\parallel^2}$ is the wavevector of the mode of interest, and in velocity space we define $|v_{\text{max},s}| = 6 w_{s}$ for each species\footnote{\label{fn:thermroot2}Note that the inclusion of the $\sqrt{2}$ factor in the definition of thermal velocity is simulation-dependent in this work. \texttt{JET-PLUME} includes $\sqrt{2}$ in the definition of the thermal velocity, and comparisons appropriately account for this.}. Periodic boundary conditions are used in configuration space and zero-flux boundary conditions are applied in velocity space. We choose $m_i / m_e = 100$, $T_i = T_e$, and $\beta_i = 1$, with $v_{A,e} / c = 0.1$, where $v_{A,e} = B_0 / \sqrt{\mu_0 m_e n_e}$ is the electron \Alfven\ speed based on the background magnetic field $B_0$. A small amount of collisions is included to regularize velocity space, with collision frequencies $\nu_{ee} = 0.001 \, \Omega_i$ and $\nu_{ii} = 0.0001 \, \Omega_i$ for electron–electron and ion–ion collisions, respectively.

To initialize an ion cyclotron wave, we choose $k_\perp \rho_i = 0.01$ and $k_\parallel \rho_i = 0.4$, resulting in a small angle from the direction of the external magnetic field of $\theta_{kB} = 1.4^\circ$. We solve the dispersion relation in \texttt{PLUME} to obtain the appropriate eigenvectors for the mode, which are used to initialize the wave in \texttt{Gkeyll}. The solution to the dispersion relation yields a normalized real frequency of $\omega_r/\Omega_i = 0.244$ and a normalized damping rate of $\gamma/\Omega_i = -0.022$. An overall amplitude scaling, $\delta B/B_0 = 1 \times 10^{-3}$, is used to initialize a linear wave. The amplitude and phase of the eigenvectors for this mode computed by \texttt{PLUME} are supplied in Table~\ref{tab:icw}.

{Again, with the parameters used by \texttt{JET-PLUME}, we are able to match those used by the simulation. We use a proton-electron plasma with equal number densities, $n_i = n_e$, and a reduced mass ratio $m_i/m_e = 100$. The ion beta is $\beta_i = 1.0$, and both species are isotropic, $(T_\perp/T_\parallel)_s = 1$. The temperature ratio is $T_i/T_e = 1$. The ion thermal speed is $v_{ti}/c = 7.071\times 10^{-3}$. Wavenumbers are normalized to the ion gyroradius, with $k_\perp \rho_i = 0.01$ and $k_\parallel \rho_i = 0.40$. No background drifts are imposed for either species.}

\begin{table}
 \begin{center}
 \begin{tabular}{|l||c|c|c|}
    \hline
    & $\hat{E}_{\perp,1}/\hat{E}_{\perp,1}$ & $\hat{E}_{\perp,2}/\hat{E}_{\perp,1}$ & $\hat{E}_\parallel/\hat{E}_{\perp,1}$ \\
    \hline
    Re. & $1.0$ & $8.58 \times 10^{-4}$ & $-2.43 \times 10^{-3}$ \\
    Im. & $0$ & $-9.99 \times 10^{-1}$ & $5.3 \times 10^{-3}$ \\
    \hline
    & $\hat{B}_{\perp,1}/\hat{E}_{\perp,1}$ & $\hat{B}_{\perp,2}/\hat{E}_{\perp,1}$ & $\hat{B}_\parallel/\hat{E}_{\perp,1}$ \\
    \hline
    Re. & $-2.1 \times 10^{1}$ & $2.3 \times 10^{2}$ & $5.24 \times 10^{-1}$ \\
    Im. & $2.3 \times 10^{2}$ & $2.08 \times 10^{1}$ & $-5.75$ \\
    \hline
    & $\hat{U}_{\perp,1,i}/(c\hat{E}_{\perp,1}/B_0)$ & $\hat{U}_{\perp,2,i}/(c\hat{E}_{\perp,1}/B_0)$ & $\hat{U}_{\parallel,i}/(c\hat{E}_{\perp,1}/B_0)$ \\
    \hline
    Re. & $5.98 \times 10^{-2}$ & $-1.65$ & $-2.04 \times 10^{-2}$ \\
    Im. & $-1.65$ & $-5.94 \times 10^{-2}$ & $3.35 \times 10^{-2}$ \\
    \hline
    & $\hat{U}_{\perp,1,e}/(c\hat{E}_{\perp,1}/B_0)$ & $\hat{U}_{\perp,2,e}/(c\hat{E}_{\perp,1}/B_0)$ & $\hat{U}_{\parallel,e}/(c\hat{E}_{\perp,1}/B_0)$ \\
    \hline
    Re. & $1.08 \times 10^{-3}$ & $-9.98 \times 10^{-1}$ & $-1.9 \times 10^{-2}$ \\
    Im. & $-9.97 \times 10^{-1}$ & $-4.66 \times 10^{-5}$ & $1.73 \times 10^{-2}$ \\
    \hline
    & $\hat{n}_{i}/(n_{0,i}\hat{E}_{\perp,1}/B_0)$ & $\hat{n}_{e}/(n_{0,e}\hat{E}_{\perp,1}/B_0)$ &  $\omega/\Omega_R$\\
    \hline
    Re. & $-4.20$ & $-4.20$ & $2.44 \times 10^{-1}$ \\
    Im. & $-2.16$ & $-2.16$ & $-2.2 \times 10^{-2}$ \\
    \hline
 \end{tabular}
 \end{center}
 \caption{Selected eigenfunctions for the ion cyclotron mode described in Section~\ref{sec:icdcomp} using \texttt{PLUME}.} \label{tab:icw} \end{table}

\begin{figure}
 \begin{center}
    \includegraphics[width=.499\textwidth]{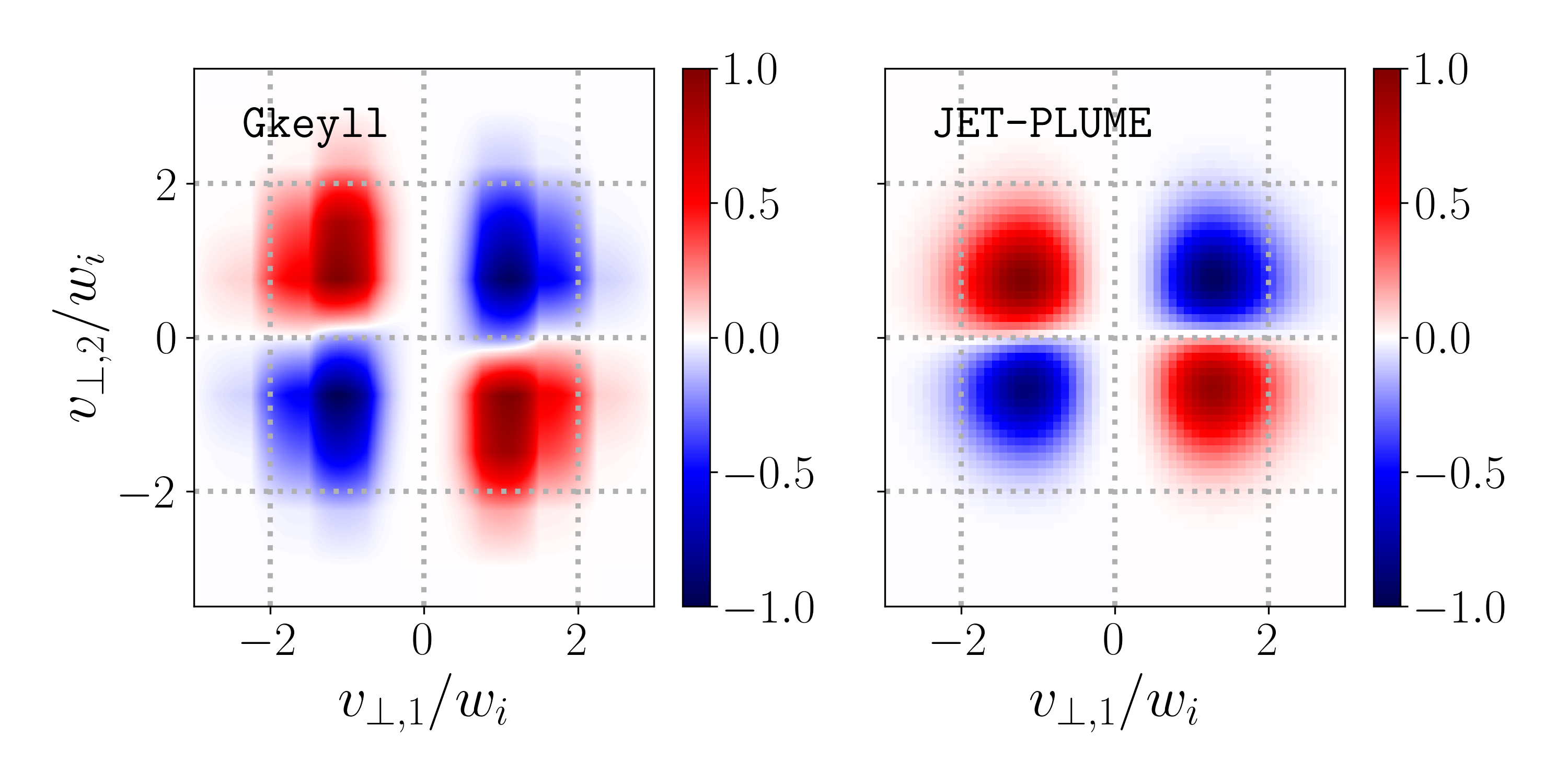}
    \includegraphics[width=.499\textwidth]{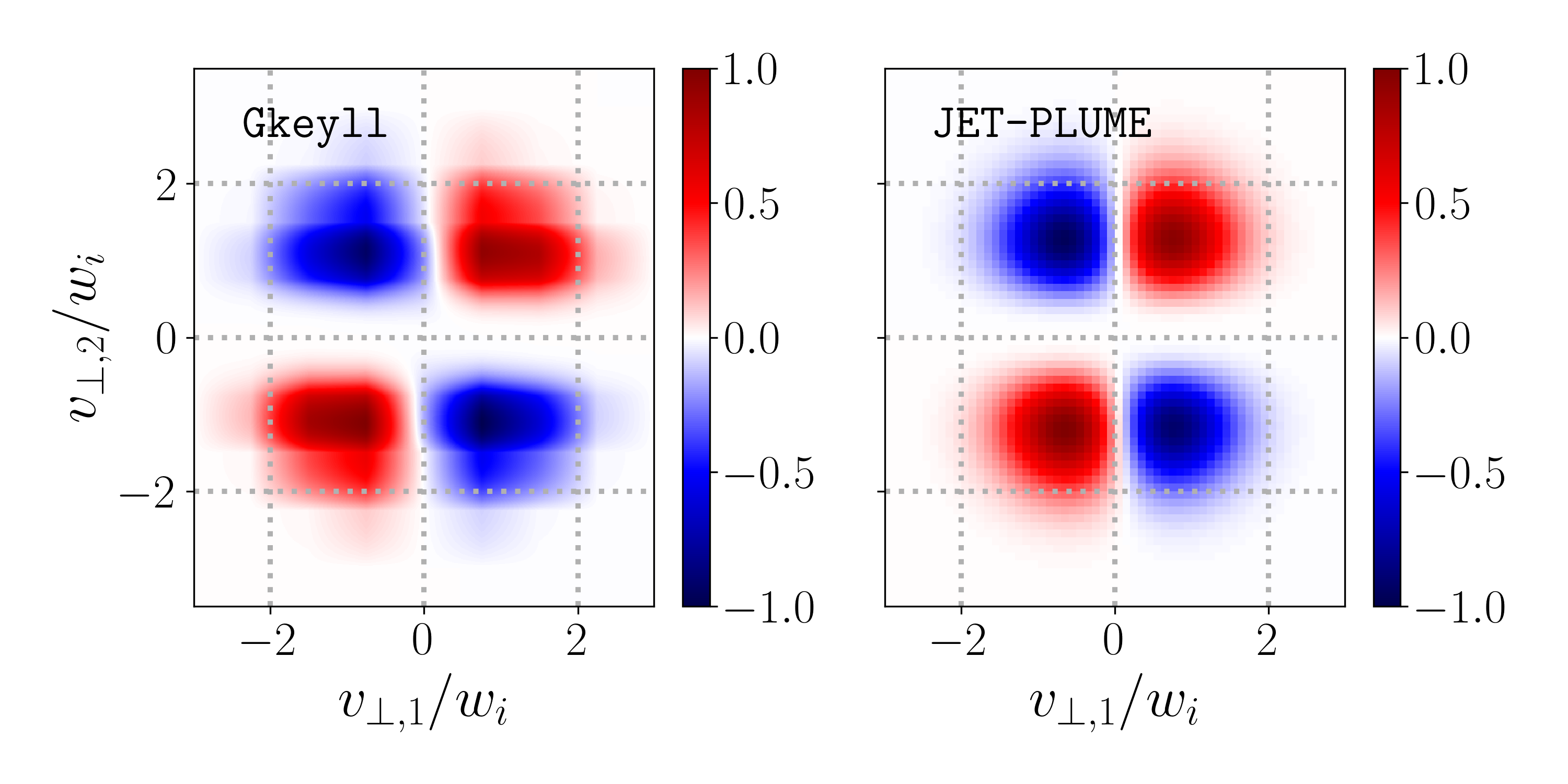}
   \end{center}
\caption{Comparisons between the \texttt{JET-PLUME} (\emph{Right}) and \texttt{Gkeyll} (\emph{Left}) \textit{Cartesian velocity-space signatures} of the ions for an \Alfven\ wave that is damped by ion-cyclotron damping using the correlation with both perpendicular coordinates, $C_{E_{\perp,1},i}(v_{\perp,1},v_{\perp,2})=\int C_{E_{\perp,1}} \, \, dv_{\parallel}$ (\emph{Top}) and $C_{E_{\perp,2}}(v_{\perp,1},v_{\perp,2})=\int C_{E_{\perp,2},i} \, \, dv_{\parallel}$ (\emph{Bottom}). The signature from \texttt{Gkeyll} is generated using a correlation over space over the entire periodic domain of the simulation. The pixelated appearance in the \texttt{Gkeyll} signatures corresponds to the simulation's discrete mesh elements, which are spaced at intervals of $0.75 \,w_{i}$.}
    \label{fig:icwcomp}
\end{figure}

We compare the \textit{Cartesian velocity-space signatures} of the ions, in the perpendicular plane, $C_{E_{\perp,1},i}(v_{\perp,1},v_{\perp,2})$ and  $C_{E_{\perp,2},i}(v_{\perp,1},v_{\perp,2})$, between the simulation and \texttt{JET-PLUME} in Figure~\ref{fig:icwcomp}\footnote{The apparent nonzero value of $C_{E_{\perp1}}$ at $v_{\perp1} = 0$ and of $C_{E_{\perp2}}$ at $v_{\perp2} = 0$ that is expected to arise due to the $v_{j}^2$ factor in the definition of the FPC, \eqref{eq:FPCdefinition} in the \texttt{Gkeyll} results arises because the simulation does not sample exactly at $v_{\perp 1,2} = 0$. Instead, the lowest grid point lies at $v_{\perp1,2} \approx \Delta v/2$, so interpolation by the plotting routine produces the spurious nonzero values.}. For an ion cyclotron wave, the velocity-space signatures exhibit a quadrupolar pattern due to the phase relations between the perpendicular electric field and the bulk flow of the ion species \citep{Afshari:2024}. We find good agreement for the $C_{E_{\perp,1},i}$ and $C_{E_{\perp,2},i}$ signatures, accurately capturing the quadrupolar structure, the skew between positive and negative features, and the relative strength of those features. For a full interpretation of the quadrupolar structure, we refer the reader to the discussion in \citet{Afshari:2024}. At present, we attribute the observed `smearing' of the phase-space signature in the simulation to finite amplitude effects; however, this interpretation warrants a more detailed investigation.

\section{Discussion}
\label{sec:discussion}

The novel routine developed alongside this publication, \texttt{JET-PLUME}, extends \texttt{PLUME} and is capable of accurately predicting velocity-space energy exchange due to wave–particle interactions using the FPC technique. {\color{blue}For a specified linear wave mode, \texttt{JET-PLUME} captures the corresponding collisionless net field--particle energy exchange through the FPC, including both the total work and its velocity-space structure.} The resulting diagnostic produces signatures with both qualitative and quantitative features that enhance our understanding of energy transfer in collisionless plasmas. These signatures can now be efficiently and accurately computed analytically for linear wave–particle interactions using this new routine.

Efficiently predicting the velocity-space energy exchange of a system has many applications. First, in systems with complex and sensitive dependence on their parameters, this routine can isolate the effect of a single parameter by varying it while holding others constant, revealing its influence on energy exchange. Second, when both distribution function and field measurements are available—such as in \emph{in situ} spacecraft observations or simulations—the tools presented in this work are a viable approach to identifying specific modes, instabilities, or mechanisms. Comparing the FPC of the observed system to predictions from the routine developed here for candidate modes provides a basis for confirming the presence of linear wave–particle interactions and for quantifying the associated properties. Third, this tool identifies the most energetic regions of velocity space for many plasma mechanisms, which is valuable when selecting laser wavelengths for probing plasmas using laser-induced fluorescence (LIF). To maximize the signal-to-noise ratio when measuring the perturbed distribution with LIF, it is optimal to probe these high-activity regions \citep{chu2018lagrangian}. This targeted approach is especially useful for verifying the successful excitation of a specific mode. Fourth, in principle, one could predict the energy exchange of heavier species that are present but not directly measurable due to instrumental limitations. This is possible because \texttt{PLUME} can model an arbitrary number of drifting bi-Maxwellian species, enabling the inclusion of heavy ions even when direct measurements are unavailable. Finally, the analytic formulation is essential because it provides a guaranteed means of disentangling overlapping processes—such as separating the degenerate $+k$ and $-k$ components of entropy modes or isolating specific damping channels—that cannot be uniquely resolved through direct measurement or Fourier analysis alone.

The linear plasma theory used here has several critical limitations, most notably of course, its inability to capture nonlinear regimes. For instance, processes such as nonlinear scattering \citep{ganguli2010three,soto20232d}, trapping-induced nonlinearity \citep{bernstein1957exact}, and stochastic heating \citep{johnson2001stochastic} are observed in both space and laboratory plasmas but are, by definition, not captured by linear theory. In cases where these mechanisms are the dominant energy transfer mechanisms, one will not be able to produce the velocity-space energization signatures produced with direct computation of \eqref{eq:FPCdefinition}. However, one may reduce—or in some cases practically eliminate—the contributions of these nonlinear mechanisms through careful choice of the integration window, Fourier filtering of the velocity distribution or electric fields to a single $(\mathbf{k},\omega)$ and by exploiting system geometry to isolate the linear mode\citep{brown2022isolation}, either individually or in combination.

Another fundamental issue of note is the challenge of accurately representing the perturbed distribution function, $\hat{f}_{s,1}$, on a finite velocity-space grid in a manner consistent with the density and velocity moments. Ideally, one would obtain these eigenfunctions directly by taking the relevant moments of the computed $\hat{f}_{s,1}$. However, this is often not feasible due to the form of $\hat{f}_{s,1}$, which involves a resonant denominator (see Equation \eqref{eq:f1a}). This denominator introduces asymptotic behavior in velocity space, leading to singularities that complicate numerical evaluation of the principal value integral. In practice, this issue can be circumvented by using an alternative representation of $\hat{f}_{s,1}$ that avoids the asymptotic behavior, such as the `algebraic form' presented in \citet{verscharen2018alps}. However, since such approaches require redundant computation of specific eigenmodes, we have not implemented similar approximations in our code. Instead, we urge caution when interpreting or computing eigenfunctions from $\hat{f}_{s,1}$ and emphasize the importance of careful selection of the velocity grid used to represent $\hat{f}_{s,1}$ whenever relevant.

Characterizing particle energization as a function of phase space is crucial for determining field and particle energy exchange in plasmas, particularly in kinetic regimes involving mechanisms including—but not limited to—particle acceleration, collisionless damping, turbulence, heating, particle reflection, magnetic pumping\citep{montag2022field}, and magnetic reconnection. Mechanisms such as these require kinetic descriptions, which track the full spatial and velocity distribution of particles, unlike fluid models that only consider macroscopic spatial quantities like density and bulk species velocity, which may be complicated by spatially coincident energy transfer by other waves or mechanisms. In principle, one can use more sophisticated closures to capture relevant effects, but constructing such closures without a comprehensive understanding of the underlying kinetic physics is challenging—and, at best, these approximations are valid only within limited parameter regimes. Thus, fluid descriptions can be inaccurate in the kinetic regime and difficult to apply to \emph{in situ} measurements, where spatial information is limited to the trajectory of the spacecraft. Using a kinetic description, it becomes possible to study distinct particle subpopulations—differentiated by their velocities—that experience disparate energization at the same spatial location. While it is conceptually possible to incorporate kinetic effects into a fluid framework through, for example, closure approximations or higher-order moments, such approaches inevitably yield incomplete descriptions. This limitation arises from the fundamental inability of a fluid model to fully resolve the detailed velocity-space dynamics that drive kinetic processes.

In this work, we focus on studying phase-space energy exchange using the FPC technique, which highlights net energization in phase space via the only channel exchanging net energy between fields and particles in collisionless plasma systems: $\mathbf{j} \cdot \mathbf{E}$. However, diagnostics analogous to the FPC for other channels of energy exchange in phase space are also critical for advancing our understanding. Recently, \citet{conley2024kinetic} developed methods for studying pressure-strain interactions \citep{cassak2022pressure,yang2022pressure} arising from the ballistic term in the Vlasov equation using phase-space diagnostics analogous to the field-particle correlation in simulations of fundamental plasma systems. The diagnostic introduced in that work would similarly benefit from linear analytical formulations like those presented here.

\section{Conclusion}
\label{sec:conclusion}

We present the novel code \texttt{JET-PLUME}, which computes the Field-Particle Correlation (FPC) diagnostic showing the phase-space energy exchange between fields and particles in response to a small wave-like disturbance in a neutral, magnetized, hot plasma with an arbitrary number of drifting bi-Maxwellian species. \texttt{JET-PLUME} is an extension of \texttt{PLUME}, a solver that provides insight into collisionless plasma systems like the solar wind by computing the dispersion relation and eigenfunction response of the system. \texttt{JET-PLUME} enables the efficient calculation of the perturbed distribution and average phase-space energy density rate of change due to linear wave-particle interactions across a wide range of parameters, facilitating detailed analysis of these ubiquitous processes in laboratory, space, and astrophysical plasmas. Analysis using the FPC diagnostic continues to demonstrate success in identifying mechanisms \emph{in situ}, and despite the well-established nature of linear theory, the specific calculations enabled by \texttt{JET-PLUME} across the full parameter domain of collisionless plasmas reveal surprisingly complex and insightful results, highlighting the continued relevance of this fundamental framework. 

{Analytic formulations of phase-space energy transfer offer several advantages over \emph{in situ} observational, laboratory, and simulation-based approaches, including computational efficiency, precise control of parameters, and, critically, the ability to systematically isolate specific transfer channels. This capability of isolation enables the examination of both well-known processes, such as Landau damping or cyclotron damping, and it enables the examination of the less accessible contributions, including off-diagonal terms and degenerate superpositions of modes, by selectively manipulating the governing equations to include only the desired contribution or by evaluating a single wavevector mode in isolation, which might be difficult to selectively excite in the laboratory or simulation.}

We compare the diagnostic computed from simulation with that from \texttt{JET-PLUME} to verify our routine and to explore the limits of linear theory in describing energization in velocity space. \texttt{JET-PLUME} successfully reproduces the measured phase-space energy exchange rate between fields and particles for both damping and growing modes, accurately capturing the non-trivial structures in velocity space associated with each tested mechanism. We make comparisons to four simulations of distinct mechanisms using two simulation codes: \texttt{AstroGK} and \texttt{Gkeyll}. We show agreement between predictions from \texttt{JET-PLUME} and simulation results for ion Landau damping, electron Landau damping, the Weibel instability with a magnetic guide field, and ion cyclotron damping. For damping modes, we find strong agreement. For growing modes, linear theory reproduces the features of the velocity-space diagnostic during the linear regime of the instability. Overall, these tests demonstrate that \texttt{JET-PLUME} can generate accurate phase-space energy signatures that aid in the identification and statistical characterization of relevant modes in collisionless plasmas across a broad range of parameters. 

Linear plasma theory is capable of explaining a wide range of phenomena but remains underexplored. {For example, we demonstrate in Figure \ref{fig:howes2017comparSeparated} significant energy exchange arising from distinct responses of the perturbed velocity distribution function $\hat{f}_{s1}$ to both perpendicular and parallel components of the electric field during the Landau damping of kinetic Alfvén waves. In this case, the resulting structures in the perturbed velocity distribution $\hat{f}_{s1}$  and the velocity-space signatures of energy exchange associated with the parallel component of the electric field, $E_\parallel$, given by $C_{E_\parallel}(v_\parallel,v_\perp)$, show substantial contributions from the response of $\hat{f}_{s1}$ to both $E_\parallel$ and $E_\perp$. 
} 
The velocity-space signatures predicted by linear theory thus provide a powerful diagnostic for identifying underlying energization mechanisms and their constituent components, and for understanding particle energization in a wide variety of laboratory, space, and astrophysical plasma environments.

\begin{acknowledgments}
The authors thank Sarah Conley for helpful discussions and advice during this project, and Chris Crabtree for his feedback during the editing of this paper. \texttt{Gkeyll} simulations were performed on the Argon High-Performance Computing Cluster at the University of Iowa. \\ 
\end{acknowledgments}

\section*{Funding}
C.R.B. and G.G.H were supported in part by NASA grant 80NSSC20K1273. C.R.B. was supported in part by an appointment to the NRC Research Associateship Program at the Plasma Physics Division in the Naval Research Laboratory, administered by the Fellowships Office of the National Academies of Sciences, Engineering, and Medicine. K.G.K. was supported in part by NASA grants 80NSSC19K0912 and 80NSSC24K0171. J.M.T. was supported by NASA Grant 80NSSC23K0099.

\section*{Data Availability Statement}

\texttt{PLUME} and \texttt{JET-PLUME} are open source codes available in the repository at \url{https://github.com/kgklein/PLUME}. Inputs can be found at \url{https://doi.org/10.5281/zenodo.19225024}. The working branch for \texttt{JET-PLUME} can be found at \url{https://github.com/kgklein/PLUME/tree/JETPLUME}. \texttt{AstroGK} is available at \url{https://homepage.physics.uiowa.edu/~ghowes/astrogk/index.html}. Information for obtaining, installing, and running \texttt{Gkeyll} may be found on the documentation site (\url{https://gkeyll.readthedocs.io}). The input files for the simulations used to produce the results in this paper may be acquired from the repository at \url{https://github.com/ammarhakim/gkyl-paper-inp/}.

\appendix
\begin{widetext}
\section{Variable Definitions used by the Linear Vlasov-Maxwell Dispersion Relation, \texttt{PLUME}}
\label{app:vmlindisp}

The \texttt{PLUME} implementation largely follows the derivation of the linear Vlasov-Maxwell dispersion relation presented in Chapter 10 of \citet{Stix:1992}. Details can be found in \citet{Klein:PlumeCodePaper:2025}, and variable definitions can be found in Table \ref{tab:defs}. The input and output parameters relevant here are found in Table \ref{tab:norm}. Given all other values as in Table \ref{tab:norm} as inputs, \texttt{PLUME} solves for the normalized Eigenfrequency, $\omega/\Omega_r$. \texttt{JET-PLUME} introduces additional input and output parameters found in Table \ref{tab:normJP}.

{\color{blue}
Because a linear eigenmode is defined only up to an arbitrary complex multiplicative constant, we fix the eigenmode normalization by choosing the phase such that \(\hat E_{\perp,1}\) is real and positive, and then normalizing all eigenfunction coefficients by this reference Fourier coefficient. Thus, the code computes normalized quantities such as \(\hat E_j/\hat E_{\perp,1}\), \(\hat B_j/\hat E_{\perp,1}\), \(\hat U_{j,s}/(c\hat E_{\perp,1}/B_0)\), and \(\hat n_{1,s}/(n_{0,s}\hat E_{\perp,1}/B_0)\) for the complex Fourier eigenfunction coefficients.
}

\begin{table}[h]
\begin{tabular}{|l|c|l|c|}
\hline
Variable & Definition & Variable & Definition \\
\hline
Equilibrium Magnetic Field & $\V{B}_0=B_0 \hat{e}_{\parallel}$ & Wavevector & $\V{k} = k_\perp \hat{e}_{\perp,1} + k_\parallel \hat{e}_{\parallel}$ \\
Species Charge&  $ q_s $ & Species Mass & $m_s$ \\
Species Perpendicular Temperature & $ T_{\perp s}$ & 
Species Parallel Temperature  &$ T_{\parallel s}$ \\
Species Number Density (Equilibrium; Pert.) & $n_s=n_{0,s}+n_{1,s}$ & Speed of Light & $c$ \\
Perpendicular Particle Velocity & $v_\perp=( v_{\perp,1}^2 + v_{\perp,2}^2)^{1/2}$ &  Parallel Particle Velocity & $v_\parallel$ \\
Gyrophase (vel. space) & $\phi$ & Azimuth (conf. space) & $\theta$ \\
Spec. Thermal Velocity   & $w_{s} =(2 T_{s}/m_s)^{1/2}\,$& 
Perp./Parallel Spec. Thermal Velocity  &$w_{\{\parallel/\perp\} s} = (2 T_{\{\parallel/\perp\} s}/m_s)^{1/2}\,$\\
Species Cyclotron Frequency & $\Omega_s= q_s B_0/(m_s c)$ &  Species Alfv\'en Velocity & $v_{A s}= B_0/(4 \pi n_{0,s} m_s)^{1/2}$ \\
Modified Bessel Function Argument & $\lambda_s=k_\perp^2 w_{\perp s}^2/(2 \Omega_s^2)$  &
Perpendicular Thermal Larmor Radius & $\rho_s=w_{\perp s}/\Omega_s$  \\
Complex Wave Frequency &$\omega = \omega_r + i \gamma $ &Index of Refraction & $\V{n}^\prime=c \V{k}/\omega$ \\
Real Component of Wave Freq. &$\omega_r $ &Imag./Growth Comp. of Wave Freq.& $\gamma$ \\
Perturbed Electric Field Component &$\hat{E}_{j}(\mathbf{k},\omega) $ &Perturbed Magnetic Field Component &$\hat{B}_{j}(\mathbf{k},\omega)$\\
Species Phase-Space Density Factor &$f_{00s} = \frac{n_{0,s}}{\pi^{3/2}w_{\parallel,s} w_{\perp,s}^2}$ & Field-Particle Correlation & $C_{E_j,s}$ ; ( $\int C_{E_j} d\mathbf{v} = \big <j_{j,s} E_j$ \big > )\\
\hline
\end{tabular}
\caption{ \label{tab:defs} Definitions. Note that, in our notation, the species cyclotron frequency carries the sign of the species charge. Boltzmann's constant $\kappa$ is absorbed in the temperature, giving temperature in units of energy. In place of the commonly used $(\hat{x},\hat{y},\hat{z})$, we use coordinates $(\hat{e}_{\perp,1},\hat{e}_{\perp,2}, 
\hat{e}_{\parallel})$, where $\hat{e}_{\perp,1}$ is in the direction of the perpendicular component of the wavevector, $\hat{e}_{\perp,2}$ is perpendicular to the plane of the wavevector and the equilibrium magnetic field, and $\hat{e}_{\parallel}$ is parallel to the equilibrium magnetic field.}
\end{table}

\begin{table}[h]
\begin{tabular}{|l|c|}
\hline
\multicolumn{2}{|l|}{Global \texttt{PLUME} Input Parameters$^\dagger$} \\
\hline
Dimensionless Parameter & Definition  \\
\hline
Reference Parallel Plasma Beta & $\beta_{\parallel R}= 8 \pi n_R T_{\parallel R}/B_0^2$\\
Reference Perpendicular Wavenumber & $\overline{k}_\perp = k_\perp \rho_R$ \\ 
Reference Parallel Wavenumber & $\overline{k}_\parallel = k_\parallel \rho_R$ \\ 
Reference Parallel Thermal Velocity & $\overline{w}_{\parallel R} = w_{\parallel R}/c$\\
\hline
\multicolumn{2}{|l|}{Species Input Parameters}\\
\hline
Dimensionless Parameter & Definition  \\
\hline
Reference-to-Species Parallel Temperature Ratio & $\tau_s = T_{\parallel R}/T_{\parallel s}$ \\
Species Temperature Anisotropy & $\aleph_s = T_{\perp s}/T_{\parallel s} $ \\
Reference-to-Species Mass Ratio & $\mu_s = m_R/m_s $ \\
Reference-to-Species Charge Ratio & $Q_s = q_R/q_s $ \\
Species-to-Reference Number Density Ratio & $D_s = n_s/n_R$ \\
Parallel Species Flow & $\overline{V}_s = V_s/v_{A R}$\\
\hline
\multicolumn{2}{|l|}{Output}\\
\hline
Normalized Eigenfrequency & $\widetilde{\omega} = \omega/\Omega_R$ \\
Normalized Electric Eigenmodes & $\hat{E}_j/\hat{E}_{\perp,1}$ \\
Normalized Magnetic Eigenmodes & $\hat{B}_j/\hat{E}_{\perp,1}$ \\
Normalized Species Bulk Velocity Eigenmode & $\hat{U}_{j,s}/(c \hat{E}_{\perp,1}/B_0)$ \\
Normalized Species Density Eigenmode & $\hat{n}_{1,s}/(n_{0,s} \hat{E}_{\perp,1}/B_0)$ \\
\hline
\end{tabular}
\caption{Dimensionless normalization of \T{PLUME} parameters.  Boltzmann's constant $\kappa$ is absorbed in the temperature, giving temperature in units of energy.  $^\dagger$Note that Global quantities are defined based on the chosen reference species, $s=R$, typically taken to be the dominant ion species. This choice is not followed by \texttt{AstroGK} and may not be followed by \texttt{Gkeyll} based on user input, thus conversion may be required.  {\color{blue}
Because a linear eigenmode is defined only up to an arbitrary complex multiplicative constant, we normalize the eigenfunction coefficients by \(\hat E_{\perp,1}\). This choice fixes the arbitrary phase and amplitude of the eigenmode so that the normalized reference component satisfies \(\hat E_{\perp,1}/\hat E_{\perp,1}=1\).
}
\label{tab:norm}}
\end{table}

\begin{table}[h]
\begin{tabular}{|l|c|}
\hline
\multicolumn{2}{|l|}{Global \texttt{JET-PLUME} Parameters} \\
\multicolumn{2}{|l|}{Species Parameters}\\
\hline
Dimensionless Parameter & Definition  \\
\hline
Species Velocity Spacing & $\Delta \hat{v}_j = \Delta v_j/w_{\parallel,s}$ \\
Species Velocity Domain & $[\hat{v}_{i,\min}, \hat{v}_{i,\max}] = [v_{i,\min}/w_{\parallel,s}, v_{i,\max}/w_{\parallel,s}]$ \\
\hline
\multicolumn{2}{|l|}{Output}\\
\hline
Species Perturbed Distribution & $\check{f}_{s,1} = \hat{f}_{s,1}/(f_{00s} \hat{E}_{\perp,1}/B_0)$ \\
Field-Particle Correlation & $\overline{C}_{E_j} = C_{E_j}/\big(q_R w_{\parallel s} f_{00s} \hat{E}_{\perp,1}^2/B_0\big)$\\
\hline
\end{tabular}
\caption{Dimensionless normalization of \T{JET-PLUME} parameters. These variables, in addition to those in Table \ref{tab:norm}, are used in \T{JET-PLUME}.}
\label{tab:normJP}
\end{table}

\section{Using Fourier Transforms and Reality Conditions to Compute the Field-Particle Correlation}
\label{appendix:linFPCderiv}
For a given choice of global parameters $(\beta_{\parallel
  R},w_{\parallel R}/c)$, sets of parameters for each species $s$
$(\tau_s,\aleph_s,\mu_s,Q_s,D_s,\overline{V}_s)$, and a specified
normalized wavevector $\V{k} \rho_R = k_\perp \rho_R \hat{e}_{\perp,1}
+ k_\parallel \rho_R \hat{e}_{\parallel}$, \T{PLUME} computes the
solutions for the complex frequency $\omega_\alpha=\omega_{r \alpha} +
i \gamma_{\alpha}$ using the Laplace-Fourier solution of the
Vlasov-Maxwell linear dispersion relation.  In general, there are
multiple possible solutions for the complex frequency $\omega_\alpha$
that satisfy the linear dispersion relation (yielding the normal modes
of the system), so we specify the solution representing the wave mode
of interest---\emph{e.g.}, the \Alfven\ wave---by the subscript
$\alpha$. For typical applications, we solve for the complex frequency
with fixed global and species parameters as the wavevector is varied,
yielding a solution that depends on the wavevector for the wave mode
$\alpha$ of interest, $\omega_\alpha(\V{k})$.  In addition to the
complex frequency, \T{PLUME} can also be directed to compute the
complex spatial Fourier components of the eigenfunction for the chosen
wave mode \citep{Klein:PlumeCodePaper:2025}, \emph{e.g.}, the electric field $
\hat{E}[\V{k},\omega_\alpha(\V{k})]$, which is a function of the
wavevector $\V{k}$, and the hat symbol indicates the Fourier coefficient.

\subsection{Reality Conditions}
Let us consider the expression for the electric field as a function of
space and time $\V{E}(\V{x},t)$ as the sum of the Fourier components of
all possible wavevectors and all wave modes $\alpha$ allowed by the
linear dispersion relation,
\begin{equation}
\V{E}(\V{x},t) =\sum_\alpha \sum_{k_{\perp,1}=-\infty}^{+\infty} \sum_{k_{\perp,2}=-\infty}^{+\infty}\sum_{k_{\parallel}=-\infty}^{+\infty}  \tilde{\V{E}}[\V{k},\omega_\alpha(\V{k})] \exp[i \V{k}\cdot \V{x} - \omega_\alpha(\V{k})t],
\end{equation}
where the general wavevector is given by $\V{k}= k_{\perp,1}
\hat{e}_{\perp,1} +k_{\perp,2} \hat{e}_{\perp,2}+ k_\parallel \hat{e}_{\parallel}$.  Here the $\tilde{\V{E}}[\V{k},\omega_\alpha(\V{k})]$ are the complex Fourier coefficients, to be replaced below for consistency with the conventions adopted in \T{PLUME} for the output of the Fourier coefficients of the eigenfunctions.
Note that we are \emph{not} performing a transform in time, but rather we
incorporate the harmonic time variation of each contributing linear wave
mode $\alpha$ due to its real frequency $\omega_{r \alpha}$ and its growth or
damping rate in time due to the imaginary component of the frequency
$\gamma_{r \alpha}$.  Since the application of \T{JET-PLUME} is generally
to determine the velocity-space signature of the damping or growth of
a single wave mode using the FPC technique, we will restrict the
analysis to the single wave mode of interest $\alpha$ and thereby drop
the summation over this subscript.

Since the electric field must be real, we can determine the reality conditions on the complex Fourier coefficients $\hat{\V{E}}[\V{k},\omega_\alpha(\V{k})]$ and complex frequency $\omega_\alpha(\V{k})$ by rewriting the summation over all wavevectors as follows,
\begin{equation}
  \V{E}(\V{x},t) = \sum_{k_{\perp,1}=-\infty}^{+\infty} \sum_{k_{\perp,2}=-\infty}^{+\infty}
  \sum_{k_{\parallel}=k_{\parallel,\text{min}}}^{+\infty} \frac{1}{2}
  \left[ \hat{\V{E}}[\V{k},\omega_\alpha(\V{k})] e^{i[ \V{k}\cdot \V{x} - \omega_\alpha(\V{k})t]} +
 \hat{\V{E}}[-\V{k},\omega_\alpha(-\V{k})] e^{-i[ \V{k}\cdot \V{x} -\omega_\alpha(-\V{k})t]} \right] ,
\end{equation}
where for simplicity in this expression we take the $k_\parallel=0$ Fourier coefficients to be zero, $\hat{\V{E}}[k_{\perp,1}, k_{\perp,2},k_\parallel=0,\omega_\alpha(\V{k})]=0$.  We have also defined the Fourier coefficients with $k_\parallel \ne 0$ with a factor of $1/2$, giving $\tilde{\V{E}}[\V{k},\omega_\alpha(\V{k})]= \hat{\V{E}}[\V{k},\omega_\alpha(\V{k})]/2$, to be consistent with the convention used in \texttt{PLUME} for the output of the Fourier coefficients of the eigenfunctions.  In order for this expression to be purely real, we must satisfy the reality conditions
\begin{equation}
  \hat{\V{\V{E}}}[-\V{k},\omega_\alpha(-\V{k})] = \hat{\V{E}}^*[\V{k},\omega_\alpha(\V{k})]
  \quad  \quad  ; \quad  \quad
  \omega_{r \alpha}(-\V{k})=\omega_{r \alpha}(\V{k})
  \quad  \quad  ; \quad  \quad
  \gamma_{\alpha}(-\V{k})=-\gamma_{\alpha}(\V{k}),
\end{equation}
where $*$ indicates the complex conjugate.
Applying the reality conditions yields the result
\begin{equation}
  \V{E}(\V{x},t) = \sum_{k_{\perp,1}=-\infty}^{+\infty} \sum_{k_{\perp,2}=-\infty}^{+\infty}
  \sum_{k_{\parallel}=k_{\parallel,min}}^{+\infty} \frac{1}{2}
  \left[ \hat{\V{E}}[\V{k},\omega_\alpha(\V{k})] e^{i[ \V{k}\cdot \V{x} - \omega_{r \alpha}(\V{k})t]} +
    \hat{\V{E}}^*[\V{k},\omega_\alpha(\V{k})] e^{-i[ \V{k}\cdot \V{x} -\omega_{r \alpha}(\V{k})t]} \right] e^{\gamma_\alpha(\V{k}) t},
  \label{eq:ftreal}
\end{equation}
which is purely real as required.

\subsection{Calculating the Field-Particle Correlation for a Single Wave Mode}
Note that \eqref{eq:ftreal} demonstrates that real electromagnetic field eigenfunctions require contributions from  both the  $\V{k}$ and  $-\V{k}$ modes, so the electric field in $(\V{x},t)$ space for a specified single  wavevector $\V{k}$ can be expressed by
\begin{equation}
  \V{E}(\V{x},t) =
  \frac{1}{2}\left(\hat{E}[\V{k},\omega_\alpha(\V{k})] e^{i[ \V{k}\cdot \V{x} - \omega_{r \alpha}(\V{k})t]} +
    \hat{E}^*[\V{k},\omega_\alpha(\V{k})] e^{-i[ \V{k}\cdot \V{x} -\omega_{r \alpha}(\V{k})t]} \right) e^{ \gamma_\alpha(\V{k}) t} ,
  \label{eq:one_ek}
\end{equation}
where we have eliminated the sums in  \eqref{eq:ftreal} since we are incorporating only a single plane-wave mode.
The perturbed velocity distribution function for species $s$, $f_{1,s}(\V{x},\V{v},t)$, for this single wave mode can be expressed as a sum of the  $\V{k}$ and  $-\V{k}$ modes but with the additional dependence on the particle velocity $\V{v}$,
\begin{equation}
  f_{1,s}(\V{x},\V{v},t) =
  \frac{1}{2}\left(\hat{f}_{1,s}[\V{k},\V{v},\omega_\alpha(\V{k})] e^{i[ \V{k}\cdot \V{x} - \omega_{r \alpha}(\V{k})t]} +
    \hat{f}_{1,s}^*[\V{k},\V{v},\omega_\alpha(\V{k})] e^{-i[ \V{k}\cdot \V{x} -\omega_{r \alpha}(\V{k})t]} \right) e^{ \gamma_\alpha(\V{k}) t} .
  \label{eq:one_fk}
\end{equation}

Next, without loss of generality, in place of a full space–time correlation, we calculate the Field–Particle Correlation \eqref{eq:FPCdefinition} as a time average over a correlation interval equal to one wave period, $\tau=T=2\pi/\omega_{r\alpha}$, at a single spatial position $\mathbf{x}_0$ due to a specified component of the electric field $E_i$,
\begin{equation}
  C_{E_j,s}(\mathbf{x}_0,\mathbf{v},t)
  = \frac{1}{T} \int_{t-T/2}^{t+T/2}
    \left(-q_s \frac{v_j^2}{2} \frac{\partial f_{1,s}(\mathbf{x}_0,\mathbf{v},t')}{\partial v_j}
    E_j(\mathbf{x}_0,t')\right)\, dt' .
  \label{eq:cei}
\end{equation}
Substituting \eqref{eq:one_ek} and \eqref{eq:one_fk} into \eqref{eq:cei}, and simplifying yields
\begin{eqnarray}
  C_{E_j,s}(\V{x}_0,\V{v},t)& =& -q_s \frac{v_j^2}{2} \frac{1}{4}
  \left( \frac{\partial \hat{f}_{1,s}}{\partial v_j}  \hat{E}_j^* +  \frac{\partial \hat{f}^*_{1,s}}{\partial v_j}  \hat{E}_j \right)  \frac{1}{T} \int dt' \   e^{ \gamma_\alpha(\V{k}) t'}   \label{eq:cei2} \\
  &-&q_s \frac{v_j^2}{2} \frac{1}{4} \left[ \left( \frac{\partial \hat{f}_{1,s}}{\partial v_j}  \hat{E}_j \right)
  \frac{1}{T} \int dt' \ e^{2i[ \V{k}\cdot \V{x}_0 - \omega_{r \alpha}t']} e^{ \gamma_\alpha(\V{k}) t'}-\left(    \frac{\partial \hat{f}^*_{1,s}}{\partial v_j}  \hat{E}^*_j \right)
   \frac{1}{T} \int dt' \ e^{-2i[ \V{k}\cdot \V{x}_0 - \omega_{r \alpha}t']} e^{ \gamma_\alpha(\V{k}) t'} \right]. \nonumber
\end{eqnarray}
Although the time integrals can be calculated for finite $\gamma_\alpha$, the latter two terms in \eqref{eq:cei2} depend on the phase of the wave at the beginning of the correlation interval $t'=t-T/2$, representing the lack of cancellation of oscillatory energy transfer due to the finite damping of the wave over one wave period.  The first term represents the dominant, wave-phase independent contribution to the velocity-space signature of the energy transfer.  In the weak growth or damping rate limit, $\gamma_\alpha/\omega_{r \alpha} \rightarrow 0$,
the latter two terms in  \eqref{eq:cei2} integrate to zero over the wave period and the normalized time integral in the first term simply yields a factor of one, leading to the final result
\begin{equation}
  \lim_{\gamma_\alpha/ \omega_{r \alpha} \rightarrow 0}  C_{E_j}(\V{x}_0,\V{v},t)=
   -q_s \frac{v_j^2}{2}
  \frac{1}{4}\left( \frac{\partial \hat{f}_{1,s}}{\partial v_j}  \hat{E}_j^* +  \frac{\partial \hat{f}^*_{1,s}}{\partial v_j}  \hat{E}_j \right).
  \label{eq:cei_final}
\end{equation}
In practice, this limit is typically adequately satisfied whenever $ |\gamma|/\omega_r < 1/e$. \texttt{JET-PLUME} computes 
$C_{E_j,s}(\V{x}_0,\V{v},t)$ normalized to $q_R w_{\parallel,s} \frac{n_{0,s}}{\pi^{3/2}w_{\parallel,s}w_{\perp,s}^2} \hat{E}_{\perp,1}^2/B_0$. Details on the normalization of the velocity distribution function $f_s$ can be found in the following section, Appendix \ref{app:dimnormJetPlume}.

{\subsection{Calculating the Field-Particle Correlation for Multiple Wave Modes}
\label{app:linsupapp}
When two linear modes with plane–wave dependence 
$\exp[i(\mathbf{k}\cdot\mathbf{x}-\omega t)]$ for their respective currents and electric fields have distinct $(\mathbf{k}_1,\omega_1)\neq(\mathbf{k}_2,\omega_2)$, the mixed terms vanish under the usual space–time averaging, so the total exchange is just the sum of the per–mode contributions,
\begin{equation}
\int C_{E_j}(\mathbf{x},\mathbf{v})\,d\mathbf{v}
= j_{j,s}^{(1)}(\mathbf{x})\,E_j^{(1)}(\mathbf{x})
  + j_{j,s}^{(2)}(\mathbf{x})\,E_j^{(2)}(\mathbf{x}).
\end{equation}
However, in the case that the wavenumber and frequency are not distinct, such as entropy modes, which often have multiple degenerate solutions with $\omega_{\textrm{r}}=0$ as seen in Section~\ref{sec:weibelcomp}, one must consider the mixed term contributions even under space-time averaging. To illustrate this, we start with the definition of two real-valued, linearly superimposed waves
\begin{equation}
E_{j}(t)=\sum_{\alpha =1,2}\tfrac12\!\left(\hat E_{j,\alpha}\,e^{\,i[ \V{k}_\alpha\cdot \V{x} -\omega_{r \alpha}(\V{k})t]}
+\hat E_{j,\alpha}^{*}\,e^{-\,i[ \V{k}_\alpha\cdot \V{x} -\omega_{r \alpha}(\V{k})t]}\right)e^{\gamma_\alpha t},
\end{equation}
which has an associated superposition of perturbed distribution function eigenmodes, whose velocity derivative is
\begin{equation}
\frac{\partial f_{1,s}}{\partial v_j}(t)=\sum_{\alpha =1,2}\tfrac12\left(\frac{\partial \hat{f}_{1,s,\alpha}}{\partial v_j}\,e^{\,i[ \V{k}_\alpha\cdot \V{x} -\omega_{r \alpha}(\V{k})t]}
+\frac{\partial \hat{f}^{*}_{1,s,\alpha}}{\partial v_j}\,e^{-\,i[ \V{k}_\alpha\cdot \V{x} -\omega_{r \alpha}(\V{k})t]}\right)e^{\gamma_\alpha t}.
\end{equation}
Plugging these expressions into \eqref{eq:cei} and expanding, one finds that each cross term pairing produces an oscillatory factor of the form
\[
e^{\,i[(\V{k}_1-\V{k}_2)\cdot\V{x}-(\omega_{r1}-\omega_{r2})t]} \qquad\text{or} \qquad
e^{\,i[(\V{k}_1+\V{k}_2)\cdot\V{x}-(\omega_{r1}+\omega_{r2})t]},
\]
or their complex conjugates, which vanish when integrated over periodic boundary conditions in time unless the term in the exponent is zero. When the terms in the exponent are zero and the damping or growth rates are negligible, the mixed terms survive, showing that the complete expression for the linear superposition of the total energy of wave-modes is
\begin{eqnarray}
\big\langle C_{E_j,s}\big\rangle
&=& -\,q_s\,\frac{v_j^2}{2}\,\frac{1}{4}\,\Bigg[
\Bigg(
\frac{\partial \hat f_{1,s,1}}{\partial v_j}\,\hat E_{j,1}^{*}
+\frac{\partial \hat f^{*}_{1,s,1}}{\partial v_j}\,\hat E_{j,1}
+\frac{\partial \hat f_{1,s,2}}{\partial v_j}\,\hat E_{j,2}^{*}
+\frac{\partial \hat f^{*}_{1,s,2}}{\partial v_j}\,\hat E_{j,2}
\Bigg) \nonumber \\
&&+\,\delta(\mathbf k_1-\mathbf k_2)\,\delta(\omega_{r1}-\omega_{r2})\,
\Bigg(
\frac{\partial \hat f_{1,s,1}}{\partial v_j}\,\hat E_{j,2}^{*}
+\frac{\partial \hat f_{1,s,2}}{\partial v_j}\,\hat E_{j,1}^{*}
+\frac{\partial \hat f^{*}_{1,s,1}}{\partial v_j}\,\hat E_{j,2}
+\frac{\partial \hat f^{*}_{1,s,2}}{\partial v_j}\,\hat E_{j,1}
\Bigg) \nonumber \\
&&+\,\delta(\mathbf k_1+\mathbf k_2)\,\delta(\omega_{r1}+\omega_{r2})\,
\Bigg(
\frac{\partial \hat f_{1,s,1}}{\partial v_j}\,\hat E_{j,2}
+\frac{\partial \hat f_{1,s,2}}{\partial v_j}\,\hat E_{j,1}
+\frac{\partial \hat f^{*}_{1,s,1}}{\partial v_j}\,\hat E_{j,2}^{*}
+\frac{\partial \hat f^{*}_{1,s,2}}{\partial v_j}\,\hat E_{j,1}^{*}
\Bigg)
\Bigg] . \nonumber
\end{eqnarray}
}
Note that while the terms in the final row of the equation above vanish for propagating waves ($\omega_r \neq 0$) under time-averaging, they provide a non-vanishing contribution for non-propagating entropy modes where $\omega_r = 0$.

\section{Dimensionless Normalization}
\label{app:dimnormJetPlume}

\texttt{PLUME} computes all quantities in dimensionless form to ensure broad applicability across diverse plasma regimes. Consequently, \texttt{JET-PLUME} also evaluates all expressions in a dimensionless framework. In this section, we describe the dimensionless implementation of the additional terms introduced by \texttt{JET-PLUME}.

\subsection{Dimensionless Velocity Distribution Eigenfunctions}

Note that the coordinates in velocity space, whether in 2V gyrotropic velocity space \((v_\perp, v_\parallel)\) or in 3V Cartesian velocity space \((v_{\perp,1}, v_{\perp,2}, v_\parallel)\), are \emph{all} dimensionlessly normalized to the parallel thermal velocity of the corresponding species, \(w_{\parallel s}\). The normalized velocity coordinates are denoted with a hat and the species subscript, \emph{e.g.},
\begin{equation}
    \hat{v}_{\perp s} \equiv \frac{v_\perp}{w_{\parallel s}} \quad ; \quad \hat{v}_{\parallel s} \equiv \frac{v_\parallel}{w_{\parallel s}}.
    \label{eq:vel_norm}
\end{equation}
The exponential decay of the distribution at large velocities, critical to compute the field-particle correlations using \T{JET-PLUME}, is retained in the new dimensionless function $\hat{f}_{0s}(v_\perp,v_{\parallel})$, given by
\begin{equation}
  \overline{f}_{0s}(v_\perp,v_{\parallel}) \equiv  \frac{f_{0s}(v_\perp,v_{\parallel}) }{f_{00s}} =
   \exp \left[  - (\hat{v}_{\parallel s} -  \hat{V}_s )^2 - \frac{\hat{v}^2_{\perp s}}{ \aleph_s} \right]
    \quad ; \quad  f_{00s}=  \frac{n_{0,s}}{\pi^{3/2}w_{\parallel s} w_{\perp s}^2 }, 
  \label{eq:maxnorm}
\end{equation}
where we remind the reader that the perpendicular velocity coordinate $v_\perp$ is normalized by the species parallel thermal velocity $w_{\parallel s}$, as defined in \eqref{eq:vel_norm}. Note also that here we have defined an alternative normalization of the species parallel drift $V_s$ by the species parallel thermal velocity $w_{\parallel s}$,
\begin{equation}
 \hat{V}_s  \equiv \frac{V_s}{w_{\parallel s}}=  \overline{V}_s   \left(\frac{ \tau_s}{ \mu_s \beta_{\parallel R}} \right)^{1/2}.
 \label{eq:hatvs}
\end{equation}

We normalize the complex Fourier coefficient of the perturbed velocity distribution function for species $s$ in Equation~\eqref{eq:maxwellian} by 
\begin{eqnarray}
  \lefteqn{\overline{f}_{1s}(\V{k},\V{v},\omega) \equiv \frac{\hat{f}_{1s}(\V{k},\V{v},\omega)}{f_{00s}} = - i \left(\frac{\mu_s \tau_s} {\beta_{\parallel R}} \right)^{1/2} \frac{\overline{\hat{E}}_{\perp,1}}{Q_s}} \nonumber \\
  & & \times \sum_{m=-\infty}^{\infty}\sum_{n=-\infty}^{\infty}  \frac{J_m(b_s) e^{i(m-n)\phi}}{\widetilde{\omega} - \overline{k}_\parallel \hat{v}_{\parallel s} (\mu_s/\tau_s \aleph_R)^{1/2} - n \mu_s/Q_s}  
 \left\{ \frac{n J_n(b_s)}{b_s} \overline{U}_s \frac{\hat{E}_{\perp,1}}{\hat{E}_{\perp,1}}
      + i J'_n(b_s) \overline{U}_s  \frac{\hat{E}_{\perp,2}}{\hat{E}_{\perp,1}}
      +  J_n(b_s) \overline{W}_s \frac{\hat{E}_{\parallel}}{\hat{E}_{\perp,1}} \right\} \overline{f}_{0s},
      \label{eq:f1_norm}
\end{eqnarray}
where the constant $f_{00s}$ contains the dimensions of the bi-Maxwellian equilibrium with parallel drift $V_s$ given by Equations~\eqref{eq:maxwellian} and \eqref{eq:drift_maxwellian}. Here, the argument $b_s$ for the Bessel function of the first kind $J_n$ is expressed in our dimensionless parameters by 
\begin{equation}
    b_s =  \frac{Q_s}{(\mu_s \tau_s\aleph_R)^{1/2}}  \overline{k}_\perp \hat{v}_{\perp s}.
\end{equation} 
It is critical to note that while Equation~\eqref{eq:f1_norm} is expressed in terms of \(\hat{v}_{\parallel s}\) and \(\hat{v}_{\perp s}\), the expression results from algebraic manipulation and \emph{not} a change of variables. Therefore, in order to compute physical velocity moments, one must obtain the dimensional values of \(w_{\parallel s}\) and \(n_{0,s}\), even though the code outputs the perturbed distribution function on a dimensionless velocity grid. We address how to handle this subtlety in the following section, Appendix~\ref{app:dimensionalfix}.

The normalized supplementary functions  are defined by 
\begin{equation}
\overline{U}_s \equiv \frac{U_s w_{\parallel s}}{f_{0s}} 
\quad ; \quad
\overline{W}_s \equiv \frac{W_s w_{\parallel s}}{f_{0s}},
\end{equation}
effectively eliminating the exponential velocity dependence in these functions.
A couple of additional useful manipulations used in Equation~\eqref{eq:f1_norm} for the overall normalization and dimensionless representation of the resonant denominator are
\begin{equation}
 \frac{q_s \hat{E}_{\perp,1} }{m_s\Omega_R w_{\parallel s}} =
 \left(\frac{\mu_s \tau_s} {\beta_{\parallel R}} \right)^{1/2} \frac{\overline{\hat{E}}_{\perp,1}}{Q_s}
 \quad ; \quad
  \frac{\omega - k_{\parallel} v_{\parallel}- n \Omega_{s}}{\Omega_R} =
  \widetilde{\omega} - \overline{k}_\parallel \hat{v}_{\parallel s} (\mu_s/\tau_s \aleph_R)^{1/2} - n \mu_s/Q_s,
\end{equation}
where $\overline{\hat{E}}_{\perp,1} \equiv (c \hat{E}_{\perp,1})/(v_{A R} B_0)$.

The exponent of the bi-Maxwellian with a parallel drift may be expressed in our dimensionless parameters by
\begin{equation}
  -\frac{(v_{\parallel} - V_s)^2}{w_{\parallel s}^2} -\frac{v_\perp^2}{ w_{\perp s}^2}
  = - \left[\hat{v}_{\parallel s} - \overline{V}_s  \left(\frac{ \tau_s}{ \mu_s \beta_{\parallel R}} \right)^{1/2} \right]^2  -  \frac{\hat{v}^2_{\perp s} }{\aleph_s}.
\end{equation}

We can evaluate analytically the velocity derivatives of $f_{0s}$ in the supplementary functions $U_s$ and $W_s$ to obtain
\begin{equation}    
  U_s  = \frac{-2 v_\perp}{w_{\perp s}^2}  f_{0s} +  \frac{k_\parallel}{\omega}\left(
   \frac{-2 v_\perp(v_{\parallel} - V_s)}{w_{\parallel s}^2}  f_{0s} +  \frac{2 v_\perp v_{\parallel}}{w_{\perp s}^2}  f_{0s}\right)
\end{equation}
\begin{equation}
     W_s  =  \frac{-2 (v_{\parallel} - V_s)}{w_{\parallel s}^2}  f_{0s} - \frac{n \Omega_{s}}{\omega}
  \left(
   \frac{-2 (v_{\parallel} - V_s)}{w_{\parallel s}^2}  f_{0s} +  \frac{2 v_{\parallel}}{w_{\perp s}^2}  f_{0s}\right),
\end{equation}
and their expression in our normalized parameters becomes
\begin{equation}
  \overline{U}_s  = \frac{w_{\parallel s} }{ f_{0s}} U_s =
   -  2 \frac{ \hat{v}_{\perp s}}{\aleph_{s}}
   \left\{1 + \frac{ \overline{k}_{\parallel}}{\widetilde{\omega}} \left(\frac{ \mu_s }{ \tau_s\aleph_{R}}\right)^{1/2}
 \left[  \left(\aleph_{s} -1\right) \hat{v}_{\parallel s} - \aleph_{s}\overline{V}_s \left(\frac{ \tau_s}{ \mu_s \beta_{\parallel R}} \right)^{1/2}
     \right]     \right\},
\end{equation}
\begin{equation}
  \overline{W}_s  = \frac{w_{\parallel s} }{ f_{0s}}   W_s=
  2\left(  \frac{n\mu_s}{\widetilde{\omega}Q_s} -1 \right)\left[\hat{v}_{\parallel s} - \overline{V}_s \left(\frac{ \tau_s}{ \mu_s \beta_{\parallel R}} \right)^{1/2} \right]
  -2  \frac{n\mu_s}{\widetilde{\omega}Q_s \aleph_s} \hat{v}_{\parallel s}.
\end{equation}

In \eqref{eq:f1_norm}, the amplitude of the linear mode is characterized by
\begin{equation}
\overline{\hat{E}}_{\perp,1} \equiv \frac{c \hat{E}_{\perp,1}}{ v_{A R} B_0} = \left[  \frac{c \hat{E}_{\perp,1}}{ v_{A R} B_{\perp,2}}\right] 
\left( \frac{B_{\perp,2}}{B_0} \right)
\label{eq:fs1ampfact}
  \end{equation}
where the term in the brackets in the third form above is order unity for an \Alfven\ wave in large-scale MHD limit ($k \rho_R \ll 1$) when the plasma mass density is dominated by the reference species mass $m_R$; therefore, the second term in parentheses shows that this normalization is directly related to the amplitude of a linear \Alfven\ wave polarized in the $\hat{e}_y=\hat{e}_{\perp,2}$ direction. 

Critically, we note that \texttt{PLUME} normalizes the moments to $\hat{E}_{\perp,1}/B_0$, and thus the \emph{output} of \texttt{JET-PLUME} for the perturbed distribution function has the following normalization,
\begin{equation}
\check{f}_{1,s}(\V{k},\V{v},\omega) \equiv \frac{\overline{f}_{1s}(\V{k},\V{v},\omega)} {\hat{E}_{\perp,1}/B_0} = \frac{\hat{f}_{1s}(\V{k},\V{v},\omega)}{f_{00s} (\hat{E}_{\perp,1}/B_0)}.
\label{eq:f1sJPoutput}
\end{equation}
We opt to output quantities normalized to $\hat{E}_{\perp,1}/B_0$ as this is sufficient for the main use cases of \texttt{PLUME} and \texttt{JET-PLUME}, \emph{e.g.}, computing polarization angles, transport ratios, and velocity-space signatures.

\subsection{Obtaining Dimensional Quantities}
\label{app:dimensionalfix}

Assuming no numerical error, one can relate moments of $\check{f}_{1,s}$, produced by \texttt{JET-PLUME}, to the analytic eigenmodes computed by \texttt{PLUME} to ensure consistency throughout the code. However, because $\check{f}_{1,s}$ is evaluated on a dimensionless velocity grid, $\hat{v}_{j} = v_j / w_{\parallel,s}$, to preserve the dimensionless nature of the implementation, it is necessary to compute the product $w_{\parallel R} n_{0,R}$ when applying a variable substitution to integrate the perturbed distribution $\check{f}_{1,s}(\hat{v}_{\parallel,s} w_{\parallel,s}, \hat{v}_{\perp,s} w_{\perp,s}, \phi)$, where $\phi$ is the gyrophase angle, over its dimensionless grid. That is, the normalized \(n\)-th moment of the perturbed distribution function $\hat{f}_{1,s}$ takes the form:
\begin{eqnarray}
\frac{\mathbf{M}_s^{(n)}}{w^n_{\parallel,s}} &=& 
\frac{1}{w^n_{\parallel,s}}\int \mathbf{v}^{\otimes n} f_{1,s}(\mathbf{v})\, d^3v 
= \int \hat{\mathbf{v}}_s^{\otimes n} \frac{\hat{E}_{\perp,1}}{B_0} f_{00s}\,
\check{f}_{1,s}(\hat{v}_{\parallel,s} w_{\parallel,s}, \hat{v}_{\perp,s} w_{\perp s}, \phi)\, d^3\hat{v}
\nonumber \\
&&= \left(\frac{\mathsf{W}_s}{w_{\parallel,s}}\right)^{\otimes n}
\int \hat{\mathbf{v}}_s^{\otimes n} \frac{\hat{E}_{\perp,1}}{B_0} f_{00s}\,
\check{f}_{1,s}(\hat{v}_{\parallel,s}, \hat{v}_{\perp,s}, \phi)\, d^3\hat{v},
\nonumber \\
&\text{where}& \quad
\mathsf{W}_{i,s}
= w_{\perp s}\bigl(\delta_{i,\perp1} + \delta_{i,\perp2}\bigr)
+ w_{\parallel s}\delta_{i,\parallel},
\quad
w_{\parallel s} = w_{\parallel R}\sqrt{\frac{\mu_s}{\tau_s}},
\quad
w_{\perp s} = w_{\parallel R}\sqrt{\frac{\mu_s \aleph_s}{\tau_s}}.
\label{eq:usubmom}
\end{eqnarray}
with $w_{\parallel s}=w_{\perp,s}\sqrt{\aleph_s}$ and where $\hat{\mathbf{v}}^{\otimes n}$ yields a scalar, vector, or tensor depending on the order $n$, since $\mathbf{v}^{\otimes n}$ denotes the $n$-fold outer (tensor) product of the velocity vector with itself. Here $\delta_{i,j}$ represents the Kronecker delta. Note that the integral evaluated with arguments $(\hat{v}_{\parallel,s} w_{\parallel,s}, \hat{v}_{\perp,s} w_{\perp s}, \phi)$ is obtained via a variable transformation, whereas the final expression evaluated with purely dimensionless arguments $(\hat{v}_{\parallel,s}, \hat{v}_{\perp,s}, \phi)$ is achieved through algebraic manipulation. Upon appropriate manipulation of Equation~\eqref{eq:usubmom}, we can relate moments of $\check{f}_{1,s}(\hat{v}_{\parallel,s} w_{\parallel s}, \hat{v}_{\perp,s} w_{\perp s}, \phi)$ on the dimensionless grid to the normalized eigenmodes found in \texttt{PLUME} with
\begin{equation}
	\frac{U_{i,s}}{c \hat{E}_{\perp,1}/B_0} = \frac{n_{0,R}}{\pi^{3/2}}\frac{\hat{w}_{i,s}}{c} \sqrt{\tau_s^3 / \mu_s^3} \int  \hat{v}_i\check{f}_{1,s}(v_{\parallel,s},  v_{\perp,s}, \phi) \, d^{3}\hat{v}
	,
\end{equation}
and
\begin{equation}
	\frac{n_{1,s}}{n_{0,s} \hat{E}_{\perp,1}/B_0}  = \frac{1}{\pi^{3/2}}\int  \check{f}_{1,s}(v_{\parallel,s},  v_{\perp,s}, \phi) \, d^{3}\hat{v} 
	,
\end{equation}
being mindful of two key subtleties in these expressions. First, the dimensionless velocity arguments in $\check{f}_{1,s}$ arise from an algebraic scaling of the function rather than a formal change of integration variables, meaning the standard Jacobian factor must be explicitly managed when integrating. Second, we note the use of $n_{0,s} = f_{00,s} \int \hat{f}_{0,s} d^3v$ (from Equation~\eqref{eq:maxnorm}) to derive these expressions. This construction is necessary as our expressions are normalized to phase-space density, $f_{00s}=  n_{0,s}/(\pi^{3/2}w_{\parallel s} w_{\perp s}^2)$, which has an unwanted degree of freedom unless we introduce this additional expression.

To compute the quantity \(n_{0,R}\), we note the equality between the conductivity tensor derived from the susceptibility tensor (multiplied by the relevant field component) and the equivalent moment of \(\hat{f}_{1,s}\) weighted by the corresponding field contribution. Arbitrarily selecting the $(i,j) = (1,1)$ element of the susceptibility tensor for the reference species, we find 
\begin{equation}
-i \sqrt{\aleph_R}\widetilde{\omega} \chi_{R,11} \frac{\overline{w}_{\parallel R}}{\beta_{\parallel R}} \frac{\hat{E}_{\perp,1}}{\hat{E}_{\perp,1}} w_{\parallel R}^3= \frac{n_{0,R}}{\pi^{3/2}}  \int  \hat{v}_i\check{f}_{1,R;E_{\perp,2}=0,E_{\parallel}=0}(v_{\parallel,R},  v_{\perp,s}, \phi) \, d^{3}\hat{v},
\end{equation}
where $\chi_{s,ij}$ is the susceptibility tensor, which is calculated but not output in any typical current configuration of \texttt{PLUME} \citep{Klein:PlumeCodePaper:2025}. With the dimensional reference species parallel thermal velocity, $w_{\parallel R}$ we can trivially derive any arbitrary species parallel thermal velocity with $w_{\parallel s} = w_{\parallel R} \sqrt{\mu_s/\tau_s}$.

Here, numerical moments of $\check{f}_{1,s}$ are obtained using the Sokhotski–Plemelj theorem \citep{hunana2019introductory,plemelj1908riemannsche} to regulate singularities produced by the resonant denominator while capturing the residual. This procedure introduces error that depends on the location of the singularities in $\check{f}_{1,s}$ and the value of $i \epsilon_{SP}$, a small dimensionless parameter introduced by the Sokhotski–Plemelj theorem and added to the resonant denominator to compute the moments with regularization and inclusion of the residuals of the principal value. While \texttt{JET-PLUME} includes routines to compute these numerical moments for comparison against the analytic moments derived in \citet{Klein:PlumeCodePaper:2025}, this feature is disabled by default. In solar wind parameter regimes—the primary application of this code—the perturbed distribution $\check{f}_{1,s}$ frequently contains singularities within the integration domain, which can introduce substantial numerical errors.

\section{Numerical Workflow}
\label{app:workflow}

{\color{blue}
Here we summarize the numerical workflow implemented by \texttt{JET-PLUME}. The analytic expressions in the preceding appendices define the Fourier conventions, normalizations, and intermediate quantities used by the code.

\begin{enumerate}
    \item Specify the drifting bi-Maxwellian species, their species parameters, the and wavevector \(\mathbf{k}\).

    \item Use \texttt{PLUME} to compute the normalized linear eigenmode\citep{Klein:PlumeCodePaper:2025}:
    \begin{enumerate}
        \item construct the hot-plasma dielectric tensor for the specified equilibrium and wavevector;
        \item solve the dispersion relation for the complex eigenfrequency \(\omega=\omega_r+i\gamma\);
        \item evaluate the associated electromagnetic eigenfunction at that frequency;
        \item compute the corresponding density and bulk-velocity eigenfunctions for each species.
    \end{enumerate}

    \item Evaluate the linear perturbed distribution function \(\hat f_{1s}\) on a dimensionless velocity grid, retaining the separate \(E_{\perp,1}\)-, \(E_{\perp,2}\)-, and \(E_\parallel\)-driven response-channel contributions.

    \item Compute the corresponding FPC terms, including the total \(C_{E_j,s}\) and the channel-resolved pieces \(C_{E_j,s}^{(\perp,1)}\), \(C_{E_j,s}^{(\perp,2)}\), and \(C_{E_j,s}^{(\parallel)}\).

    \item Output the velocity-space energization signatures in gyrotropic or Cartesian reductions for comparison with simulations or observations.
\end{enumerate}
}

\end{widetext}

\bibliography{bibs/fpc_examples.bib,bibs/plumederiv.bib,bibs/random.bib}

@article{afshari2021importance,
  title={The importance of electron Landau damping for the dissipation of turbulent energy in terrestrial magnetosheath plasma},
  author={Afshari, AS and Howes, GG and Kletzing, CA and Hartley, DP and Boardsen, SA},
  journal={Journal of Geophysical Research: Space Physics},
  volume={126},
  number={12},
  pages={e2021JA029578},
  year={2021},
  publisher={Wiley Online Library}
}

@ARTICLE{Afshari:2024,
       author = {{Afshari}, A.~S. and {Howes}, G.~G. and {Shuster}, J.~R. and  {Klein}, K.~G. and  {McGinnis}, D. and {Martinovic}, M.~M. and  {Boardsen}, S.~A. {Brown}, C.~R. and {Huang}, R.~ and {Kletzing}, C.~A. and {Hartley}, D.~P.},
        title = "{Direct observation of ion cyclotron damping of turbulence in Earth’s magnetosheath plasma}",
  journal = {Nature Communications},
  publisher = {Nature Publishing Group},
  volume = {15},
  month = oct,
  pages = {7870},
  number = {1},
  year = 2024
}

@article{brown2022isolation,
  title={Isolation and phase-space energization analysis of the instabilities in collisionless shocks},
  author={Brown, Collin R and Juno, James and Howes, Gregory G and Haggerty, Colby C and Constantinou, Sage},
  journal={Journal of Plasma Physics},
  volume={89},
  number={3},
  pages={905890308},
  year={2023},
  publisher={Cambridge University Press}
}

@article{chen2019evidence,
  title={Evidence for electron Landau damping in space plasma turbulence},
  author={Chen, CHK and Klein, KG and Howes, Gregory G},
  journal={Nature communications},
  volume={10},
  number={1},
  pages={1--8},
  year={2019},
  publisher={Nature Publishing Group}
}

@article{horvath2020electron,
  title={Electron Landau damping of kinetic Alfv{\'e}n waves in simulated magnetosheath turbulence},
  author={Horvath, Sarah A and Howes, Gregory G and McCubbin, Andrew J},
  journal={Physics of Plasmas},
  volume={27},
  number={10},
  pages={102901},
  year={2020},
  publisher={AIP Publishing LLC}
}

@article{horvath2022observing,
  title={Observing particle energization above the Nyquist frequency: An application of the field-particle correlation technique},
  author={Horvath, Sarah A and Howes, Gregory G and McCubbin, Andrew J},
  journal={Physics of Plasmas},
  volume={29},
  number={6},
  pages={062901},
  year={2022},
  publisher={AIP Publishing LLC}
}

@article{howes2017diagnosing,
  title={Diagnosing collisionless energy transfer using field--particle correlations: Vlasov--Poisson plasmas},
  author={Howes, Gregory G and Klein, Kristopher G and Li, Tak Chu},
  journal={Journal of Plasma Physics},
  volume={83},
  number={1},
  year={2017},
  publisher={Cambridge University Press}
}

@article{howes2017prospectus,
  title={A prospectus on kinetic heliophysics},
  author={Howes, Gregory G},
  journal={Physics of plasmas},
  volume={24},
  number={5},
  pages={055907},
  year={2017},
  publisher={AIP Publishing LLC}
}

@ARTICLE{Howes:2024, 
title={The fundamental parameters of astrophysical plasma turbulence and its dissipation: non-relativistic limit}, 
volume={90}, 
DOI={10.1017/S0022377824001090}, 
number={5}, 
journal=jpp, 
author={Howes, Gregory G.}, 
year={2024}, 
pages={905900504}
}

@ARTICLE{Howes:2025,
       author = {{Howes}, Gregory G. and {Felix}, Alberto and {Brown}, Collin R. and {Haggerty}, Colby C. and {Juno}, James and {TenBarge}, Jason M. and {Wilson}, Lynn B. and {Caprioli}, Damiano},
        title = "{Velocity-space signatures of shock-drift acceleration at quasi-perpendicular collisionless shocks}",
      journal = {Physics of Plasmas},
         year = 2025,
        month = jun,
       volume = {32},
       number = {6},
          eid = {062904},
        pages = {062904},
          doi = {10.1063/5.0269528},
       adsurl = {https://ui.adsabs.harvard.edu/abs/2025PhPl...32f2904H}
}

@ARTICLE{Huang:2024,
       author = {{Huang}, Rui and {Howes}, Gregory G. and {McCubbin}, Andrew J.},
        title = "{The velocity-space signature of transit-time damping}",
      journal = {Journal of Plasma Physics},
         year = 2024,
        month = sep,
       volume = {90},
       number = {4},
          eid = {535900401},
        pages = {535900401},
          doi = {10.1017/S0022377824000667},
archivePrefix = {arXiv},
       eprint = {2401.16697},
 primaryClass = {physics.plasm-ph},
       adsurl = {https://ui.adsabs.harvard.edu/abs/2024JPlPh..90d5301H}
}

@article{klein2016measuring,
  title={Measuring collisionless damping in heliospheric plasmas using field--particle correlations},
  author={Klein, Kristopher G and Howes, Gregory G},
  journal={The Astrophysical Journal Letters},
  volume={826},
  number={2},
  pages={L30},
  year={2016},
  publisher={IOP Publishing}
}

@article{klein2017diagnosing,
  title={Diagnosing collisionless energy transfer using field--particle correlations: gyrokinetic turbulence},
  author={Klein, Kristopher G and Howes, Gregory G and TenBarge, Jason M},
  journal={Journal of Plasma Physics},
  volume={83},
  number={4},
  year={2017},
  publisher={Cambridge University Press}
}

@article{klein2020diagnosing,
  title={Diagnosing collisionless energy transfer using field--particle correlations: Alfv{\'e}n-ion cyclotron turbulence},
  author={Klein, Kristopher G and Howes, Gregory G and TenBarge, Jason M and Valentini, Francesco},
  journal={Journal of Plasma Physics},
  volume={86},
  number={4},
  year={2020},
  publisher={Cambridge University Press}
}

@article{li2019collisionless,
  title={Collisionless energy transfer in kinetic turbulence: field--particle correlations in fourier space},
  author={Li, Tak Chu and Howes, Gregory G and Klein, Kristopher G and Liu, Yi-Hsin and TenBarge, Jason M},
  journal={Journal of Plasma Physics},
  volume={85},
  number={4},
  year={2019},
  publisher={Cambridge University Press}
}

@article{juno2022phase,
  title={Phase-space energization of ions in oblique shocks},
  author={Juno, James and Brown, Collin R and Howes, Gregory G and Haggerty, Colby C and TenBarge, Jason M and Wilson Iii, Lynn B and Caprioli, Damiano and Klein, Kristopher G},
  journal={The Astrophysical Journal},
  volume={944},
  number={1},
  pages={15},
  year={2023},
  publisher={IOP Publishing}
}

@article{juno2021field,
  title={A field--particle correlation analysis of a perpendicular magnetized collisionless shock},
  author={Juno, James and Howes, Gregory G and TenBarge, Jason M and Wilson, Lynn B and Spitkovsky, Anatoly and Caprioli, Damiano and Klein, Kristopher G and Hakim, Ammar},
  journal={Journal of Plasma Physics},
  volume={87},
  number={3},
  year={2021},
  publisher={Cambridge University Press}
}

@article{mccubbin2022characterizing,
  title={Characterizing velocity--space signatures of electron energization in large-guide-field collisionless magnetic reconnection},
  author={McCubbin, Andrew J and Howes, Gregory G and TenBarge, Jason M},
  journal={Physics of Plasmas},
  volume={29},
  number={5},
  pages={052105},
  year={2022},
  publisher={AIP Publishing LLC}
}

@article{montag2022field,
  title={A field-particle correlation analysis of magnetic pumping},
  author={Montag, P and Howes, Gregory G},
  journal={Physics of Plasmas},
  volume={29},
  number={3},
  pages={032901},
  year={2022},
  publisher={AIP Publishing LLC}
}

@ARTICLE{Montag:2025,
       author = {{Montag}, P. and {Howes}, G.~G. and {McGinnis}, D. and {Afshari}, A.~S. and {Starkey}, M.~J. and {Desai}, M.~I.},
        title = "{MMS Observations of the Velocity-space Signature of Shock-drift Acceleration}",
      journal       = {The Astrophysical Journal Letters},
         year = 2025,
        month = feb,
       volume = {980},
       number = {2},
          eid = {L23},
        pages = {L23},
          doi = {10.3847/2041-8213/adb0b2},
archivePrefix = {arXiv},
       eprint = {2306.09061},
 primaryClass = {physics.space-ph},
       adsurl = {https://ui.adsabs.harvard.edu/abs/2025ApJ...980L..23M}
}

@article{schroeder2021laboratory,
  title={Laboratory measurements of the physics of auroral electron acceleration by Alfv{\'e}n waves},
  author={Schroeder, Jim WR and Howes, GG and Kletzing, CA and Skiff, F and Carter, TA and Vincena, S and Dorfman, S},
  journal={Nature communications},
  volume={12},
  number={1},
  pages={1--9},
  year={2021},
  publisher={Nature Publishing Group}
}

@article{howes2022revolutionizing,
  title={Revolutionizing Our Understanding of Particle Energization in Space Plasmas Using On-Board Wave-Particle Correlator Instrumentation},
  author={Howes, Gregory G and Verniero, Jaye L and Larson, Davin E and Bale, Stuart D and Kasper, Justin C and Goetz, Keith and Klein, Kristopher G and Whittlesey, Phyllis L and Livi, Roberto and Rahmati, Ali and others},
  journal={Frontiers in astronomy and space sciences},
  volume={9},
  pages={912868},
  year={2022},
  publisher={Frontiers Media SA}
}

@article{verniero2021determining,
  title={Determining threshold instrumental resolutions for resolving the velocity-space signature of ion Landau damping},
  author={Verniero, JL and Howes, GG and Stewart, DE and Klein, KG},
  journal={Journal of Geophysical Research: Space Physics},
  volume={126},
  number={5},
  pages={e2020JA028361},
  year={2021},
  publisher={Wiley Online Library}
}

@article{conley2023characterizing,
  title={Characterizing the velocity-space signature of electron Landau damping},
  author={Conley, Sarah A and Howes, Gregory G and McCubbin, Andrew J},
  journal={Journal of Plasma Physics},
  volume={89},
  number={5},
  pages={905890514},
  year={2023},
  publisher={Cambridge University Press}
}

@book{Fried:1961,
  title={II—Properties of Z},
  author={Fried, Burton D and Conte, Samuel D},
  journal={The Plasma Dispersion Function},
  year={1961},
  publisher={Elsevier}
}

@article{klein2015predicted,
  title={Predicted impacts of proton temperature anisotropy on solar wind turbulence},
  author={Klein, Kristopher G and Howes, Gregory G},
  journal={Physics of Plasmas},
  volume={22},
  number={3},
  pages={032903},
  year={2015},
  publisher={AIP Publishing LLC}
}

@book{Stix:1992,
  title={Waves in plasmas},
  author={Stix, Thomas H},
  year={1992},
  publisher={Springer Science \& Business Media}
}

@book{swanson2003plasma,
  title={Plasma waves},
  author={Swanson, Donald Gary},
  year={2003},
  publisher={CRC Press}
}

@ARTICLE{Klein:PlumeCodePaper:2025,
       author = {{Klein}, Kristopher G. and {Howes}, Gregory G. and {Brown}, Collin R.},
        title = "{PLUME: Plasma in a Linear Uniform Magnetized Environment}",
      journal = {Research Notes of the American Astronomical Society},
         year = 2025,
        month = apr,
       volume = {9},
       number = {4},
          eid = {102},
        pages = {102},
          doi = {10.3847/2515-5172/add1c2},
       adsurl = {https://ui.adsabs.harvard.edu/abs/2025RNAAS...9..102K}
}

@inproceedings{Hakim:2020b,
author = {Hakim, Ammar and Juno, James},
title = {Alias-Free, Matrix-Free, and Quadrature-Free Discontinuous Galerkin Algorithms for (Plasma) Kinetic Equations},
year = {2020},
isbn = {9781728199986},
publisher = {IEEE Press},
booktitle = {Proceedings of the International Conference for High Performance Computing, Networking, Storage and Analysis},
articleno = {73},
numpages = {15}
}

@article{Arnold:2011,
	Author = {{Arnold}, D.~N. and {Awanou}, G.},
	Doi = {10.1007/s10208-011-9087-3},
	Fjournal = {Foundations of Computational Mathematics. The Journal of the Society for the Foundations of Computational Mathematics},
	Journal = {Foundations of Comp. Math},
	Pages = {337--344},
	Subjclass = {65N30},
	Title = {{The serendipity family of finite elements}},
	Volume = 11,
	Year = 2011}

@article{cagas2017nonlinear,
  title={Nonlinear saturation of the Weibel instability},
  author={Cagas, Petr and Hakim, Ammar and Scales, Wayne and Srinivasan, Bhuvana},
  journal={Physics of Plasmas},
  volume={24},
  number={11},
  pages={112116},
  year={2017},
  publisher={AIP Publishing LLC}
}

@article{grassi2017electron,
  title={Electron Weibel instability in relativistic counterstreaming plasmas with flow-aligned external magnetic fields},
  author={Grassi, A and Grech, M and Amiranoff, Fran{\c{c}}ois and Pegoraro, F and Macchi, A and Riconda, C},
  journal={Physical Review E},
  volume={95},
  number={2},
  pages={023203},
  year={2017},
  publisher={APS}
}

@article{howes2006astrophysical,
  title={Astrophysical gyrokinetics: basic equations and linear theory},
  author={Howes, Gregory G and Cowley, Steven C and Dorland, William and Hammett, Gregory W and Quataert, Eliot and Schekochihin, Alexander A},
  journal={The Astrophysical Journal},
  volume={651},
  number={1},
  pages={590},
  year={2006},
  publisher={IOP Publishing}
}

@article{juno2018discontinuous,
  title={Discontinuous Galerkin algorithms for fully kinetic plasmas},
  author={Juno, James and Hakim, Ammar and TenBarge, Jason and Shi, Eric and Dorland, William},
  journal={Journal of Computational Physics},
  volume={353},
  pages={110--147},
  year={2018},
  publisher={Elsevier}
}

@article{klein2017applying,
  title={Applying Nyquist's method for stability determination to solar wind observations},
  author={Klein, Kristopher G and Kasper, Justin C and Korreck, KE and Stevens, Michael L},
  journal={Journal of Geophysical Research: Space Physics},
  volume={122},
  number={10},
  pages={9815--9823},
  year={2017},
  publisher={Wiley Online Library}
}

@article{klein2018majority,
  title={Majority of solar wind intervals support ion-driven instabilities},
  author={Klein, KG and Alterman, BL and Stevens, ML and Vech, D and Kasper, JC},
  journal={Physical review letters},
  volume={120},
  number={20},
  pages={205102},
  year={2018},
  publisher={APS}
}

@ARTICLE{Landau:1946,
   author = {{Landau}, L.~D.},
    title = "{On the Vibrations of the Electronic Plasma}",
  journal = {Journal of Physics},
     year = 1946,
   volume = 10,
    pages = {25}
}

@article{martinovic2021ion,
  title={Ion-driven instabilities in the inner heliosphere. I. Statistical trends},
  author={Martinovic, Mihailo M and Klein, Kristopher G and {\v{D}}urovcov{\'a}, Tereza and Alterman, Benjamin L},
  journal={The Astrophysical Journal},
  volume={923},
  number={1},
  pages={116},
  year={2021},
  publisher={IOP Publishing}
}

@article{numata2010astrogk,
  title={AstroGK: Astrophysical gyrokinetics code},
  author={Numata, Ryusuke and Howes, Gregory G and Tatsuno, Tomoya and Barnes, Michael and Dorland, William},
  journal={Journal of Computational Physics},
  volume={229},
  number={24},
  pages={9347--9372},
  year={2010},
  publisher={Elsevier}
}

@article{stockem2007relativistic,
  title={The relativistic filamentation instability in magnetized plasmas},
  author={Stockem, A and Lerche, I and Schlickeiser, R},
  journal={The Astrophysical Journal},
  volume={659},
  number={1},
  pages={419},
  year={2007},
  publisher={IOP Publishing}
}

@article{stockem2008suppression,
  title={Suppression of the filamentation instability by a flow-aligned magnetic field: testing the analytic threshold with PIC simulations},
  author={Stockem, Anne and Dieckmann, Mark E and Schlickeiser, Reinhard},
  journal={Plasma Physics and Controlled Fusion},
  volume={50},
  number={2},
  pages={025002},
  year={2008},
  publisher={IOP Publishing}
}

@article{zhao2022quantifying,
  title={Quantifying wave--particle interactions in collisionless plasmas: Theory and its application to the alfv{\'e}n-mode wave},
  author={Zhao, Jinsong and Lee, Louchuang and Xie, Huasheng and Yao, Yuhang and Wu, Dejin and Voitenko, Yuriy and Viviane, Pierrard},
  journal={The Astrophysical Journal},
  volume={930},
  number={1},
  pages={95},
  year={2022},
  publisher={IOP Publishing}
}

@article{shuster2021structures,
  title={Structures in the terms of the Vlasov equation observed at Earth’s magnetopause},
  author={Shuster, JR and Gershman, DJ and Dorelli, JC and Giles, BL and Wang, S and Bessho, N and Chen, L-J and Cassak, PA and Schwartz, SJ and Denton, RE and others},
  journal={Nature Physics},
  volume={17},
  number={9},
  pages={1056--1065},
  year={2021},
  publisher={Nature Publishing Group UK London}
}

@article{conley2024kinetic,
  title={The kinetic analog of the pressure--strain interaction},
  author={Conley, Sarah A and Juno, James and TenBarge, Jason M and Barbhuiya, M Hasan and Cassak, Paul A and Howes, Gregory G and Lichko, Emily},
  journal={Physics of Plasmas},
  volume={31},
  number={12},
  year={2024},
  publisher={AIP Publishing}
}

@article{coburn2024regulation,
  title={The Regulation of the Solar Wind Electron Heat Flux by Wave--Particle Interactions},
  author={Coburn, Jesse T and Verscharen, Daniel and Owen, Christopher J and Maksimovic, Milan and Horbury, Timothy S and Chen, Christopher HK and Guo, Fan and Fu, Xiangrong and Liu, Jingting and Abraham, Joel B and others},
  journal={The Astrophysical Journal},
  volume={964},
  number={1},
  pages={100},
  year={2024},
  publisher={IOP Publishing}
}

@article{verscharen2018alps,
  title={ALPS: the arbitrary linear plasma solver},
  author={Verscharen, Daniel and Klein, Kristopher G and Chandran, Benjamin DG and Stevens, Michael L and Salem, Chadi S and Bale, Stuart D},
  journal={Journal of Plasma Physics},
  volume={84},
  number={4},
  pages={905840403},
  year={2018},
  publisher={Cambridge University Press}
}

@article{bowen2022situ,
  title={In situ signature of cyclotron resonant heating in the solar wind},
  author={Bowen, Trevor A and Chandran, Benjamin DG and Squire, Jonathan and Bale, Stuart D and Duan, Die and Klein, Kristopher G and Larson, Davin and Mallet, Alfred and McManus, Michael D and Meyrand, Romain and others},
  journal={Physical Review Letters},
  volume={129},
  number={16},
  pages={165101},
  year={2022},
  publisher={APS}
}

@inproceedings{bowen2023data,
  title={Data-Driven Representations of Ion-Kinetic Distribution Functions},
  author={Bowen, Trevor A and Chandran, Benjamin DG and Klein, Kristopher G and Mallet, Alfred and Bale, Stuart D and Squire, Jonathan and Verniero, Jaye},
  booktitle={2023 XXXVth General Assembly and Scientific Symposium of the International Union of Radio Science (URSI GASS)},
  pages={1--4},
  year={2023},
  organization={IEEE}
}

@article{soto20232d,
  title={2D kinetic simulations of whistler wave generation by nonlinear scattering of lower-hybrid waves in turbulent plasmas},
  author={Soto-Chavez, A Rualdo and Crabtree, Chris and Ganguli, Gurudas and Fletcher, Alex C},
  journal={Physics of Plasmas},
  volume={30},
  number={10},
  year={2023},
  publisher={AIP Publishing}
}

@article{ganguli2010three,
  title={Three dimensional character of whistler turbulence},
  author={Ganguli, Gurudas and Rudakov, Leonid and Scales, Wayne and Wang, Joseph and Mithaiwala, Manish},
  journal={Physics of Plasmas},
  volume={17},
  number={5},
  year={2010},
  publisher={AIP Publishing}
}

@article{bernstein1957exact,
  title={Exact nonlinear plasma oscillations},
  author={Bernstein, Ira B and Greene, John M and Kruskal, Martin D},
  journal={Physical Review},
  volume={108},
  number={3},
  pages={546},
  year={1957},
  publisher={APS}
}

@article{johnson2001stochastic,
  title={Stochastic ion heating at the magnetopause due to kinetic Alfv{\'e}n waves},
  author={Johnson, Jay R and Cheng, CZ},
  journal={Geophysical research letters},
  volume={28},
  number={23},
  pages={4421--4424},
  year={2001},
  publisher={Wiley Online Library}
}

@article{chu2018lagrangian,
  title={A Lagrangian model for laser-induced fluorescence and its application to measurements of plasma ion temperature and electrostatic waves},
  author={Chu, Feng and Skiff, F},
  journal={Physics of plasmas},
  volume={25},
  number={1},
  year={2018},
  publisher={AIP Publishing}
}

@article{hunana2019introductory,
  title={An introductory guide to fluid models with anisotropic temperatures. Part 2. Kinetic theory, Pad{\'e} approximants and Landau fluid closures},
  author={Hunana, P and Tenerani, A and Zank, GP and Goldstein, ML and Webb, GM and Khomenko, E and Collados, M and Cally, PS and Adhikari, L and Velli, M},
  journal={Journal of Plasma Physics},
  volume={85},
  number={6},
  pages={205850603},
  year={2019},
  publisher={Cambridge University Press}
}

@article{plemelj1908riemannsche,
  title={Riemannsche funktionenscharen mit gegebener monodromiegruppe},
  author={Plemelj, Josip},
  journal={Monatshefte f{\"u}r Mathematik und Physik},
  volume={19},
  pages={211--245},
  year={1908},
  publisher={Springer}
}

@article{klein2021inferred,
  title={Inferred linear stability of parker solar Probe observations using one-and two-component proton distributions},
  author={Klein, KG and Verniero, JL and Alterman, B and Bale, S and Case, A and Kasper, JC and Korreck, K and Larson, D and Lichko, E and Livi, R and others},
  journal={The Astrophysical Journal},
  volume={909},
  number={1},
  pages={7},
  year={2021},
  publisher={IOP Publishing}
}

@article{mcmanus2024proton,
  title={Proton-and alpha-driven instabilities in an ion cyclotron wave event},
  author={McManus, Michael D and Klein, Kristopher G and Bale, Stuart D and Bowen, Trevor A and Huang, Jia and Larson, Davin and Livi, Roberto and Rahmati, Ali and Romeo, Orlando and Verniero, Jaye and others},
  journal={The Astrophysical Journal},
  volume={961},
  number={1},
  pages={142},
  year={2024},
  publisher={IOP Publishing}
}

@article{yang2022pressure,
  title={Pressure--strain interaction as the energy dissipation estimate in collisionless plasma},
  author={Yang, Yan and Matthaeus, William H and Roy, Sohom and Roytershteyn, Vadim and Parashar, Tulasi N and Bandyopadhyay, Riddhi and Wan, Minping},
  journal={The Astrophysical Journal},
  volume={929},
  number={2},
  pages={142},
  year={2022},
  publisher={IOP Publishing}
}

@article{cassak2022pressure,
  title={Pressure--strain interaction. I. On compression, deformation, and implications for Pi-D},
  author={Cassak, Paul A and Barbhuiya, M Hasan},
  journal={Physics of Plasmas},
  volume={29},
  number={12},
  year={2022},
  publisher={AIP Publishing}
}

@article{walters2023effects,
  title={The effects of nonequilibrium velocity distributions on Alfv{\'e}n ion-cyclotron waves in the solar wind},
  author={Walters, Jada and Klein, Kristopher G and Lichko, Emily and Stevens, Michael L and Verscharen, Daniel and Chandran, Benjamin DG},
  journal={The Astrophysical Journal},
  volume={955},
  number={2},
  pages={97},
  year={2023},
  publisher={IOP Publishing}
}

@article{juno2020noise,
  title={Noise-induced magnetic field saturation in kinetic simulations},
  author={Juno, J and Swisdak, MM and Tenbarge, JM and Skoutnev, V and Hakim, A},
  journal={Journal of Plasma Physics},
  volume={86},
  number={4},
  pages={175860401},
  year={2020},
  publisher={Cambridge University Press}
}

@article{mclaughlin2025spatially,
  title={Spatially resolved measurements of plasma ion velocity distributions in a dipole magnetic field},
  author={McLaughlin, Jacob W and Pette, Daniel V and Skiff, Fred N},
  journal={Physics of Plasmas},
  volume={32},
  number={3},
  year={2025},
  publisher={AIP Publishing}
}

@article{gilbert2024improving,
  title={Improving pulsed laser induced fluorescence distribution function analysis through matched filter signal processing},
  author={Gilbert, TJ and Steinberger, TE and Scime, EE},
  journal={Review of Scientific Instruments},
  volume={95},
  number={8},
  year={2024},
  publisher={AIP Publishing}
}

@article{bret2011robustness,
  title={Robustness of the filamentation instability as shock mediator in arbitrarily oriented magnetic field},
  author={Bret, Antoine and Alvaro, E Perez},
  journal={Physics of Plasmas},
  volume={18},
  number={8},
  year={2011},
  publisher={AIP Publishing}
}

@article{Klein_Verscharen_2025,
  author       = {K. G. Klein and D. Verscharen},
  title        = {The dielectric response of plasmas with arbitrary gyrotropic velocity distributions},
  journal      = {Physics of Plasmas},
  year         = {2025},
  volume       = {32},
  number       = {9},
  pages        = {092104},
  doi          = {10.1063/5.0286477},
}

@misc{Brown_JETPLUME_2026_misc,
  author       = {Brown, Collin R. and Howes, Gregory G. and Klein, Kristopher G.},
  title        = {{JET-PLUME}: A kinetic extension to the {PLUME} framework},
  year         = 2026,
  publisher    = {GitHub},
  howpublished = {\url{https://github.com/kgklein/PLUME/}},
  note         = {Includes the JET-PLUME extension developed for this work. Commit: \texttt{4271e40d893f19e4c7b9175d8c4613e1591d67b8}}
}

@ARTICLE{Klein:2026,
       author = {{Klein}, K.~G. and {Larson}, D. and {Livi}, R. and {Martinovi{\'c}}, M.~M. and {Rahmati}, A. and {Shankarappa}, N. and {Stevens}, M. and {Verscharen}, D. and {Whittlesey}, P.},
        title = "{Ion-Scale Wave Emission and Absorption for Non-Maxwellian Velocity Distributions in the Inner Heliosphere}",
  journal={Geophysical research letters},
         year = 2026,
        month = feb,
       volume = {53},
       number = {3},
          eid = {e2025GL118809},
        pages = {e2025GL118809},
          doi = {10.1029/2025GL118809},
       adsurl = {https://ui.adsabs.harvard.edu/abs/2026GeoRL..5318809K}
}

@book{gary1993theory,
  title={Theory of space plasma microinstabilities},
  author={Gary, S Peter},
  number={7},
  year={1993},
  publisher={Cambridge university press}
}

@book{brambilla1998kinetic,
  title={Kinetic theory of plasma waves: homogeneous plasmas},
  author={Brambilla, Marco},
  number={96},
  year={1998},
  publisher={Oxford University Press}
}

@article{schekochihin2009astrophysical,
  title={Astrophysical gyrokinetics: kinetic and fluid turbulent cascades in magnetized weakly collisional plasmas},
  author={Schekochihin, AA and Cowley, Steven Charles and Dorland, W and Hammett, GW and Howes, Gregory G and Quataert, E and Tatsuno, T},
  journal={The Astrophysical Journal Supplement Series},
  volume={182},
  number={1},
  pages={310--377},
  year={2009},
  publisher={The American Astronomical Society}
}

@article{Chandran:2010a,
  title={Perpendicular ion heating by low-frequency Alfv{\'e}n-wave turbulence in the solar wind},
  author={Chandran, Benjamin DG and Li, Bo and Rogers, Barrett N and Quataert, Eliot and Germaschewski, Kai},
  journal={The Astrophysical Journal},
  volume={720},
  number={1},
  pages={503--515},
  year={2010},
  publisher={The American Astronomical Society}
}

@article{Cerri:2021,
  title={On stochastic heating and its phase-space signatures in low-beta kinetic turbulence},
  author={Cerri, Silvio Sergio and Arzamasskiy, Lev and Kunz, Matthew W},
  journal={The Astrophysical Journal},
  volume={916},
  number={2},
  pages={120},
  year={2021},
  publisher={The American Astronomical Society}
}

\end{document}